\documentclass[iop,referee]{emulateapj-rtx4}

\usepackage{graphicx}  % needed for figures
\usepackage{dcolumn}   % needed for some tables
\usepackage{bm}        % for math
\usepackage{amsfonts,amsmath,amssymb,mathrsfs}
\usepackage{color}
\usepackage{hyperref}
\hypersetup{
  colorlinks=true,        % false: boxed links; true: colored links
  linkcolor=blue,         % color of internal links
  citecolor=cyan,         % color of links to bibliography
}

\usepackage{natbib,times}
\DeclareMathOperator\arctanh{arctanh}

\renewcommand{\apjl}{{ApJL}}

\newcommand{\rmn}[1]{{\mathrm{#1}}}
\newcommand{\ddt}[1]{{\frac{\rmn{d}#1}{\rmn{d}t}}}

\shorttitle{Electromagnetic emission from long-lived BNS
  merger remnants I}

\shortauthors{D. M. Siegel \& R. Ciolfi}

\begin{document}
\title{Electromagnetic emission from long-lived binary neutron star
  merger remnants I: formulation of the problem}

\author{Daniel M. Siegel\altaffilmark{1} and Riccardo Ciolfi\altaffilmark{2,3}}

\altaffiltext{1}{Max Planck Institute for Gravitational Physics (Albert
  Einstein Institute), Am M\"uhlenberg 1, D-14476 Potsdam-Golm, Germany}

\altaffiltext{2}{Physics Department, University of Trento, Via
  Sommarive 14, I-38123 Trento, Italy}

\altaffiltext{3}{INFN-TIFPA, Trento Institute for Fundamental Physics
  and Applications, Via Sommarive 14, I-38123 Trento, Italy}

\email{daniel.siegel@aei.mpg.de; riccardo.ciolfi@unitn.it}

\begin{abstract}
Binary neutron star (BNS) mergers are the leading model to explain the
phenomenology of short gamma-ray bursts (SGRBs), which are among the
most luminous explosions in the universe. 
Recent observations of long-lasting X-ray afterglows of SGRBs challenge standard
paradigms and indicate that in a large fraction of events a long-lived
neutron star (NS) may be formed rather than a black hole.
Understanding the mechanisms underlying these afterglows
is necessary in order to address the open questions concerning the nature of SGRB 
central engines. However, recent theoretical progress has been hampered 
by the fact that the timescales of interest for the afterglow emission are
inaccessible to numerical relativity simulations. 
Here we present a detailed model to bridge the gap between numerical 
simulations of the merger process and the relevant timescales for 
the afterglows, assuming that the merger results in a long-lived NS.
This model is formulated in terms of a set of coupled differential 
equations that follow the evolution of the post-merger system and
predict its electromagnetic (EM) emission in a self-consistent way,
starting from initial data that can be extracted from BNS merger simulations and 
taking into account the most relevant radiative processes.
Moreover, the model can accomodate the collapse of the remnant NS at any time 
during the evolution as well as different scenarios for the prompt SGRB 
emission. 
A second major reason of interest for BNS mergers is that they
are considered the most promising source of gravitational waves (GWs) for
detection with the advanced ground-based detector network LIGO/Virgo coming online this year. 
Multimessenger astronomy with joint EM and GW observations of the merger and post-merger 
phase can greatly enhance the scientific output of either type of
observation. However, the actual benefit depends on whether a suitable
EM counterpart signal to the GW emission can be identified (ideally
bright, isotropic, long-lasting, and associated with a high fraction
of BNS merger events). The model presented here allows us to search for
such counterparts, which carry the signature of a long-lived remnant
NS. The present paper is devoted to a detailed discussion of the
formulation and implementation of the model. In a companion paper, we
employ this model to predict the EM emission from $\sim\!10^{-2}$ to
$\sim\!10^7\,\text{s}$ after a BNS merger, considering a wide range of
physical parameters, and discuss the implications in the context of
SGRBs and multimessenger astronomy.
\end{abstract}
		

%%%%%%%%%%%%%%%%%%%%%%%%%%%%%%%%%%%%%%%%%		
\section{Introduction}
%%%%%%%%%%%%%%%%%%%%%%%%%%%%%%%%%%%%%%%%%
\label{sec:introduction}

The coalescence of binary neutron stars (BNS) represents the most promising 
source of gravitational waves (GWs) for the detection with ground-based interferometric 
detectors such as advanced LIGO and Virgo \citep{Harry2010,Accadia2011}. 
At the same time, BNS mergers are responsible for observable 
electromagnetic (EM) emission that carries complementary information on the source and 
that can be combined with the GW detection to significantly enlarge the scientific output.
As an independent observational channel, EM signals provide positional and temporal 
information that can enhance the search sensitivity of the GW detectors and lead to GW
detections (e.g., \citealt{Abadie2012b,Aasi2014,Williamson2014,Clark2014}). 
Inversely, EM follow-up observations triggered by a GW detection can confirm the 
astrophysical origin of the event (e.g., \citealt{Evans2012, Abadie2012a, Singer2014}),
provided a suitable, characteristic EM counterpart signal can be 
identified with the merger and/or post-merger phase of the evolution of the 
BNS system. In addition, EM counterparts can constrain physical properties of the
BNS merger remnant and its dynamics that cannot be probed by GW
observations (e.g., magnetic fields,
mass ejection). In the case of a confident GW detection, important
contraints on the source properties, the EM emission mechanisms, and
the energetics involved can be obtained even from a non-detection of
an EM counterpart. Moreover, detecting such EM signals could
significantly improve the sky localization 
of the source and thus the identification of the host galaxy, allowing
for two independent measurements of the redshift and the Hubble
constant (e.g.,
\citealt{Schutz1986,Metzger2012,Berger2014}). Furthermore, joint EM
and GW observations can reveal important information on when and how
short gamma-ray bursts (SGRBs) can be produced in BNS mergers. In
particular, they can verify or falsify the recently proposed
`time-reversal' scenario for SGRBs and possibly provide a very
accurate method to determine the lifetime of a remnant NS and to place strong
constraints on the unknown equation of state of nuclear matter at high
densities \citep{Ciolfi2015a,Ciolfi2015b}. With the advanced
LIGO/Virgo detector network starting their first science
runs later this year, such multimessenger astronomy will become reality in
the very near future.
The actual benefit of joint GW and EM observations, however, depends on our 
knowledge of the expected EM signals associated with BNS mergers and the 
underlying physical mechanisms, which are still poorly
understood. Therefore, a deeper understanding of these 
EM signals is urgently needed. 

BNS mergers are commonly considered the leading scenario to explain the phenomenology
of SGRBs (e.g.,
\citealt{Paczynski1986,Eichler1989,Narayan1992,Barthelmy2005a,Fox2005,Gehrels2005,Shibata2006b,Rezzolla2011,Paschalidis2015}). 
Currently, very strong evidence for
the association of SGRBs with BNS or neutron star--black hole (NS--BH)
mergers comes from the recent detection of possible radioactively powered
kilonova events
(e.g., \citealt{Li1998,Kulkarni2005,Rosswog2005,Metzger2010,Tanvir2013,Berger2013,Yang2015}). 
While this is still a matter of debate, joint EM and GW
observations will provide a powerful tool to unambiguously confirm the association of
SGRBs with compact object coalescence. 
For this reason, the prompt $\gamma$-ray
emission of SGRBs lasting less than $\approx
2\,\text{s}$ has so far been the prime target of studies on joint EM and GW
observations (e.g., \citealt{Evans2012,
  Abadie2012a,
  Singer2014,Abadie2012b,Aasi2014,Williamson2014,Clark2014}). 

Nevertheless, the SGRB prompt emission is thought to be
collimated and will thus be beamed away from the observer in most cases (see
\citealt{Berger2014} for an overview of observations to date; see also
Section~7 of \citealt{Siegel2015c}). 
During many years of observations, including the 10 years of operation of
the \textit{Swift} satellite \citep{Gehrels2004}, no SGRB with known
redshift has been detected within the sensitivity volume of advanced
LIGO/Virgo. Furthermore, joint EM and GW observations are so far based
on the assumption that the EM and the GW signals are characterized by
a relative time lag of at most a few seconds. However, the details of how the prompt
emission in SGRBs is generated still remain unclear, and the burst
could, e.g., be generated a long time ($\sim\!10^2-10^4\,\text{s}$)
after the merger
\citep{Ciolfi2015a,Ciolfi2015b}. In the latter case, joint
observations focusing on a
time window of a few seconds around the time of merger would miss the
SGRB. Likewise, focusing on a time lag of a few
seconds around a detected SGRB could lead to a non-detection of GW
emission. Hence, when assuming coincidence within a short time window
around the time of merger, one has to be cautious when drawing astrophysical
conclusions in either case. 
Finally, even if a SGRB is observed in
coincidence with a GW signal, such an observation alone will unlikely
be able to distinguish between a BNS and a NS--BH merger, as the SGRB
emission is expected to be very
similar for both progenitor models (see below). It is therefore important to
identify bright, long-lasting and highly isotropic EM counterparts 
that are produced in a high fraction of events and that can distinguish 
between a BNS and a NS--BH progenitor system.

A standard model to explain the generation of the SGRB prompt emission
in BNS and NS--BH mergers is an accretion powered relativistic
jet from a BH--torus system that is formed soon ($\lesssim
10-100\,\text{ms}$) after merger (e.g.,
\citealt{Narayan1992,Shibata2006b,Rezzolla2011,Paschalidis2015}). This
accretion process and the resulting energy release cease once the torus has 
been accreted on a timescale of at most one second, 
which is consistent with the typical timescale of the prompt SGRB emission
($\lesssim\!2\,\text{s}$). 
However, recent observations by the \textit{Swift} satellite have revealed long-lasting X-ray
afterglows in a large fraction of SGRB events that are indicative of
ongoing energy ejection on much longer timescales up to $\sim\!10^4\,\text{s}$
(\citealt{Rowlinson2013,Gompertz2013,Gompertz2014,Lue2015}). 
Even considering that the interaction of the jet with the interstellar 
medium might produce an afterglow signal lasting longer than the 
accretion timescale, persistent emission with a duration of $\sim\!10^4\,\text{s}$
remains difficult to explain within this BH--torus scenario (\citealt{Kumar2015} and referecnes
therein). A possible alternative is that a large fraction of BNS mergers lead to the formation 
of a stable or at least sufficiently long-lived NS rather than a BH--torus system (e.g.,
\citealt{Zhang2001,Metzger2008a,Rowlinson2010,Bucciantini2012,Rowlinson2013,
  Gompertz2013,Lue2015}). Such a long-lived NS can power ongoing energy 
ejection on the relevant timescales via loss of rotational energy. 
This is a clearly distinctive feature of BNS mergers as opposed to NS--BH mergers, 
which cannot produce a remnant NS. NS--BH mergers are thus challenged as a
progenitor model for at least a large class of SGRBs. 

The formation of a long-lived NS is indeed a very likely outcome of a BNS merger.
The merger product depends on the masses of the two progenitor NSs and the 
equation of state of nuclear matter at high densities, which is unknown. Recent
observations of high-mass NSs \citep{Demorest2010,Antoniadis2013}
indicate a maximum gravitational mass for stable NSs of $M_\text{TOV}\gtrsim
2\,\text{M}_\odot$. However, NSs with masses $M>M_\text{TOV}$ can be
centrifugally supported against gravitational collapse by uniform rotation up to the
mass-shedding limit, which is known as the supramassive regime. This
results in a maximum mass for uniformly rotating configurations of $M_\text{supra}\approx 1.2 M_\text{TOV}\gtrsim
2.4\,\text{M}_\odot$ \citep{Lasota1996}. Furthermore, the distribution
of NSs in binary systems is sharply peaked around
$1.3-1.4\,\text{M}_\odot$, with the first born NS slightly more
massive than the second one; this leads to a typical remnant NS mass
of $\approx\!2.3-2.4\,\text{M}_\odot$ when accounting for neutrino losses
and mass ejection \citep{Belczynski2008a}. 
Progenitors of lighter NSs
are much more abundant than progenitors for massive NSs, and
population synthesis calculations show that 99\% of all BNS mergers
should lead to a remnant mass between $\approx\!2.2-2.5\,\text{M}_\odot$
\citep{Belczynski2008a}. Hence, the most likely product of a BNS merger
should be a supramassive NS. Depending on $M_\text{TOV}$
there might also be a significant fraction of stable NSs. Furthermore, some
BNS mergers would lead to a slightly hypermassive NS (i.e., above the maximum mass 
supported by uniform rotation), which, however,
could still migrate to a supramassive configuration through subsequent
mass loss (cf.~Section~\ref{sec:phaseI}). Only a small fraction of BNS mergers
should promptly form a BH--torus system as assumed in the standard model. 

EM emission from a long-lived remnant NS when applied to
a large class of X-ray afterglows of SGRBs has so far been modeled in a
simple way as dipole spin-down emission from a uniformly rotating magnetar (e.g.,
\citealt{Rowlinson2013,Gompertz2013,Lue2015}). This model assumes an
instantaneous and direct conversion of spin-down luminosity
$L_\text{sd}$ into observed X-ray luminosity $L_X$ by some unspecified
process, $L_\text{sd}\propto L_X$, and consists of a simple analytically specified formula to
fit the X-ray lightcurves. In particular, it does not take into
account baryon pollution due to dynamical mass ejection and subsequent
neutrino and magnetically driven winds (e.g.,
\citealt{Hotokezaka2013a,Oechslin2007,Bauswein2013,Kastaun2015a,Dessart2009,Siegel2014a,Metzger2014c}),
which leads to a much more complex post-merger evolution (see,
however, \citealt{Metzger2008a,Bucciantini2012}). Based on a simple dynamical
model, \citet{Yu2013} and
\citet{Gao2015} have investigated EM emission from a
long-lived millisecond magnetar surrounded by an envelope of previously ejected
matter, finding a late-time brightening (termed ``magnetar-driven
macro-nova'' ) consistent with the optical and X-ray
(re)brightening of GRB 080503. Based on a refined
physical model, \citet{Metzger2014b} have investigated the evolution
of a stable millisecond magnetar in a similar setup and reported
late-time brightenings of the luminosity in the optical and X-ray
band, compatible with features observed in GRB 080503 and GRB
130603B (see Section~\ref{sec:discussion} for a comparison of these previous
models to ours). Accurately modeling the EM emission from a
long-lived remnant NS is challenging because of the complex physics
involved (baryon pollution, thermal effects, neutrino emission, strong
magnetic fields with complex structure, strong differential rotation)
and the wide range of timescales involved (from ms to days after the
merger). While the early post-merger phase and thus the generation of
the prompt SGRB emission can be probed by numerical relativity
simulations, such typical timescales for EM afterglows are
inaccessible to those simulations. On the other hand, semi-analytical
modeling has so far concentrated on computing EM emission in the
late-time regime ($\sim\!10^3-10^6\,\text{s}$ after merger;
\citealt{Yu2013,Gao2013,Metzger2014b,Gao2015}).

Here and in a companion Paper (\citealt{Siegel2015c}, henceforth Paper II), 
we consider the likely case of a long-lived NS remnant and provide a 
dynamical model to self-consistently compute the EM 
emission of the post-merger system based on some
initial data that can be extracted from a numerical relativity simulation tens of
milliseconds after the merger. This model thus bridges the gap between
the short timescales accessible to numerical relativity
simulations and the timescales of interest for EM
afterglow radiation as recorded by satellite missions like
\textit{Swift}. We note that our model does not include the EM
emission associated with the formation of a relativistic jet, which
can be added to the EM signal predicted here. The model is very general and should be applicable to
any BNS merger that leads to the formation of a long-lived NS. As we
have argued above, this should cover the vast majority of BNS merger
events. Therefore, our model represents an important tool to study and
identify promising EM counterparts of BNS mergers for coincident EM and GW
observations as discussed above and to investigate the nature of
long-lasting X-ray afterglows observed in a large fraction of SGRB
events. The present paper is devoted to a detailed discussion of the 
physical model and its numerical implementation. In Paper II, we
apply the model to a large number of possible long-lived 
BNS merger remnants, employing a wide range of physical input parameters.
Our results and their astrophysical implications are discussed in Paper II.

This paper is organized as follows. Section~\ref{sec:phenomenology}
describes the phenomenology underlying our evolution model. In
Section~\ref{sec:model_equations}, we provide a brief summary of
the model, which is formulated in terms of sets of highly coupled ordinary
differential equations. The subsequent section describes the ingredients
to these equations in detail. In Section~\ref{sec:numerics}, we discuss
some numerical aspects for integrating the model equations and, in
particular, present a scheme to reconstruct the observer lightcurves
and spectra including relativistic beaming, the relativistic Doppler
effect and the time-of-flight effect. Section~\ref{sec:discussion} is
devoted to discussion and conclusions. Several appendices are added
to streamline the discussion in the main part of the paper.

%%%%%%%%%%%%%%%%%%%%%%%%%%%%%%%%%%%%%%%%%
\section{Phenomenology}
%%%%%%%%%%%%%%%%%%%%%%%%%%%%%%%%%%%%%%%%%
\label{sec:phenomenology}

\begin{figure}[tb]
\centering
\includegraphics[width=0.53\textwidth]{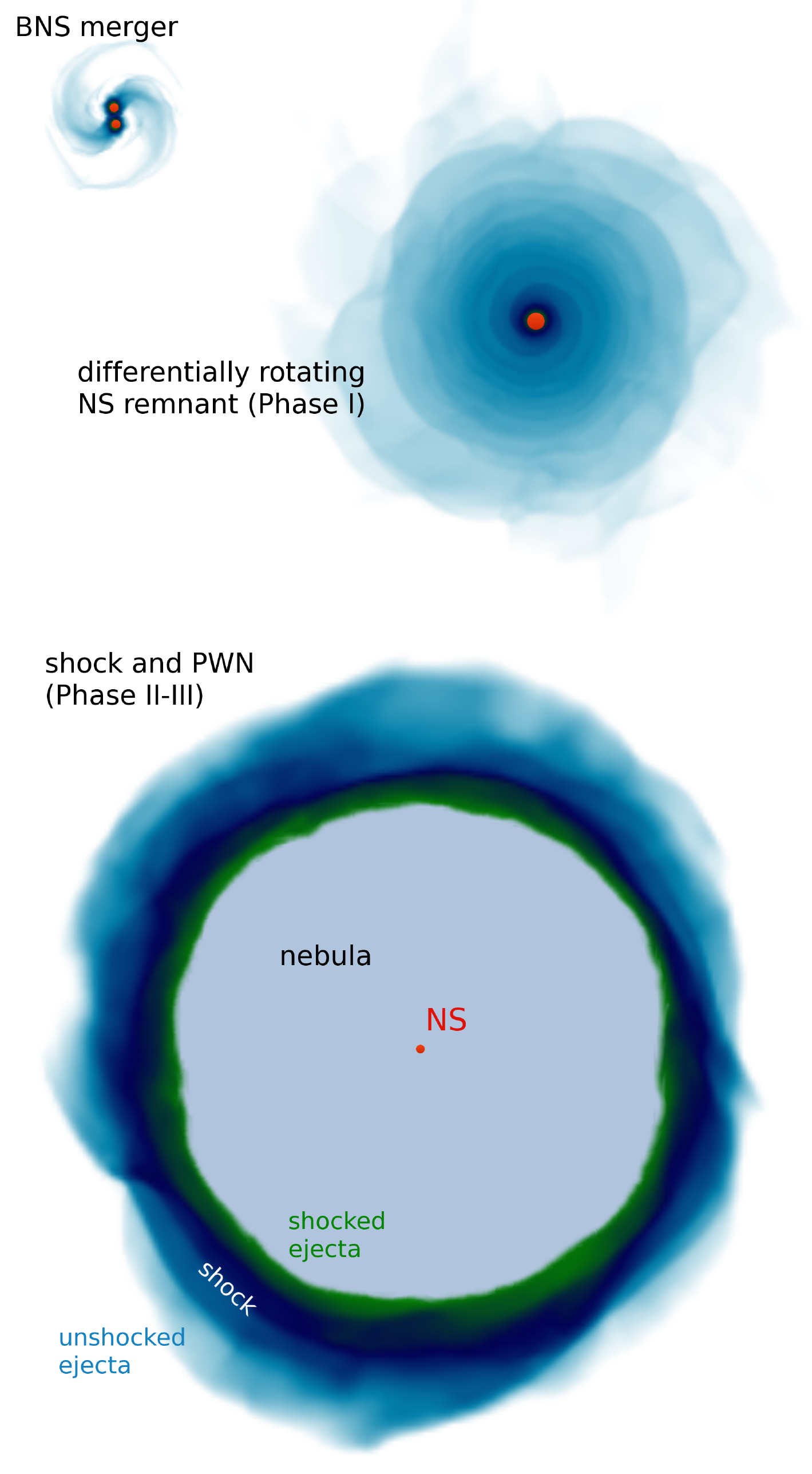}
 \caption{Evolution of the system according to the proposed scenario (with increasing spatial scale).  
   A BNS merger (top left) forms a differentially rotating NS that emits a baryon-loaded 
   wind (Phase I). The NS eventually settles down to uniform rotation
   and inflates a pulsar wind nebula (or simply `nebula')
   that sweeps up all the ejecta material into 
   a thin shell (Phase II). Spin-down emission from the NS continues while the nebula 
   and the ejecta shell keep expanding (Phase III).}
 \label{fig:overview}
\end{figure}

This section is intended to provide a conceptual outline of the
phenomenology our evolution model is built upon. 
We assume that a BNS merger leads to the formation of a long-lived
NS and divide the post-merger evolution into three phases: 
an early baryonic-wind phase (Phase I; $t\lesssim 1-10\,\text{s}$), a pulsar
wind shock phase (Phase II; $ t \sim 10\,\text{s}$), and a pulsar wind nebula (PWN) phase (Phase III; $\sim\!
10\,\text{s}\lesssim t \lesssim 10^{7}\,\text{s}$). 
These evolution phases are depicted in Figure~\ref{fig:overview}. 
Our model provides a self-consistent dynamical evolution once the physical properties 
of the system in the early post-merger phase are specified. These properties can be extracted 
or estimated from BNS merger simulations in general relativity.  
In particular, we start the evolution at a few to tens of milliseconds after the BNS merger, 
once a roughly axisymmetric state of the remnant NS has been reached and the strong 
GW emission characterizing the merger phase has been 
severely damped.

In our reference scenario, Phase I starts with a supramassive NS, which is a 
very likely outcome of a BNS merger (see Section~\ref{sec:introduction}). 
Nevertheless, the phenomenology described here also applies if the NS is 
stable or if its mass is only slightly above the hypermassive limit, provided 
that in the latter case the star enters the supramassive regime through 
additional mass loss before collapsing to a black hole (see below).

The newly-born NS is differentially rotating and endowed with 
strong magnetic fields (up to magnetar field strengths) due to 
compression of the two progenitor stars and
magnetic field amplification mechanisms acting during 
the merger, such as the Kelvin-Helmholtz instability 
(e.g., \citealt{Price2006,Zrake2013,Giacomazzo2015}). 
Moreover, further amplification occurs during Phase I, i.e., as long 
as differential rotation is active, via magnetic winding and possibly the 
magnetorotational instability (e.g., \citealt{Duez2006b,Siegel2013,Kiuchi2014}).
These latter mechanisms are likely to dominate the removal of differential 
rotation itself, which occurs on the 
timescale $t_\text{dr}$. In this case, $t_\text{dr}$ is roughly given by the Alf\'en
timescale and can be as long as $t_\text{dr}\sim 1-10\,\text{s}$ for the 
objects considered here, depending on the initial magnetic
field strength \citep{Shapiro2000}. 

At birth, the remnant star is already surrounded by material dynamically 
ejected during or shortly after the merger, with a total mass of up
to $M_\text{ej}\lesssim 10^{-3}\,\text{M}_\odot$ (e.g.,
\citealt{Hotokezaka2013a,Bauswein2013,Kastaun2015a}; see the
discussion in Section~\ref{sec:phaseI}). 
However, additional mass ejection is expected to take place 
during Phase I, which can even dominate over this early dynamical 
outflow in terms of total ejected mass. 
One main mass ejection mechanism results from magnetic field 
amplification in the stellar interior, which causes a 
build-up of magnetic pressure in the outer layers of the star. 
This pressure rapidly overcomes the gravitational binding at the stellar surface, 
launching a strong baryon-loaded magnetized wind
\citep{Siegel2014a,Siegel2015a}.
Furthermore, substantial mass loss can be caused by 
neutrino-induced winds over typical timescales for neutrino
cooling of $t_\nu \lesssim 1\,\text{s}$ \citep{Dessart2009}. 
For both mechanisms,
i.e., magnetically and neutrino-induced winds, we expect highly isotropic 
mass ejection. The material ejected from the NS
surface is typically hot with temperatures of up to tens of MeV
\citep{Siegel2014a,Kastaun2015a}, while further energy is carried 
by the wind in the form of a strong Poynting flux \citep{Siegel2014a}. 
These winds transport material outward at 
speeds of at most $v_\text{ej,in}\lesssim 0.1c$, creating a hot and
optically thick environment. EM
emission from these radially expanding winds is expected to be 
predominantly thermal, due to the very high optical depths at 
these early times.
However, because of the high optical depth, radiative energy loss is
still rather inefficient.

As differential rotation is being removed on the timescale $t_\text{dr}$, the
NS settles down to uniform rotation. Mass loss is suppressed and while 
the ejected matter keeps moving outward
the density in the vicinity of the NS is expected to drop on roughly
the same timescale. In the resulting essentially baryon-free
environment the NS can set up a pulsar-like magnetosphere.
Via dipole spin-down, the NS starts powering a highly relativistic, 
Poynting-flux dominated outflow of charged particles (mainly electrons and
positrons; see Section~\ref{sec:pulsar_properties}) or `pulsar wind'
at the expense of rotational energy. This occurs at a time $t=t_\text{pul,in}$ 
and marks the beginning of Phase II. 

\begin{figure*}[tb]
\centering
\includegraphics[width=0.38\textwidth]{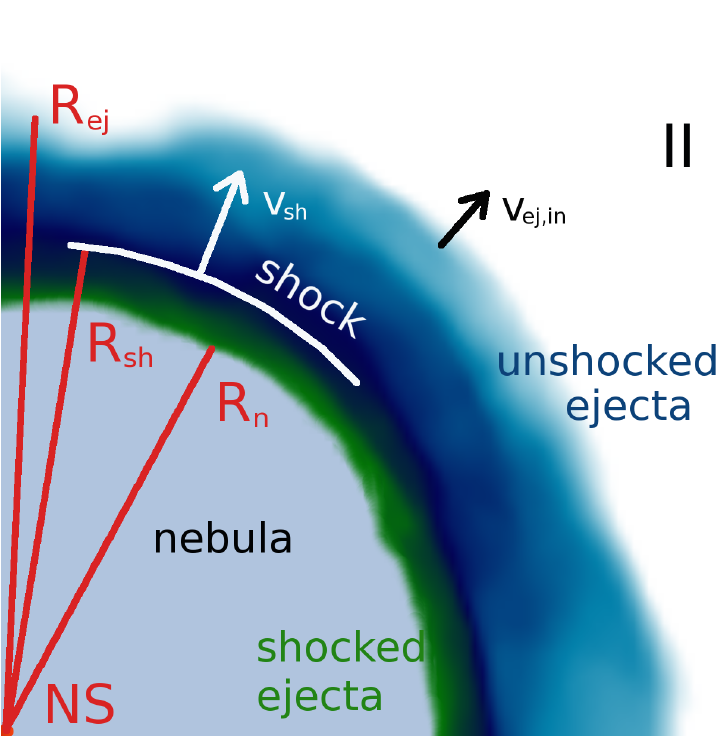}
\hspace{1.8cm}
\includegraphics[width=0.32\textwidth]{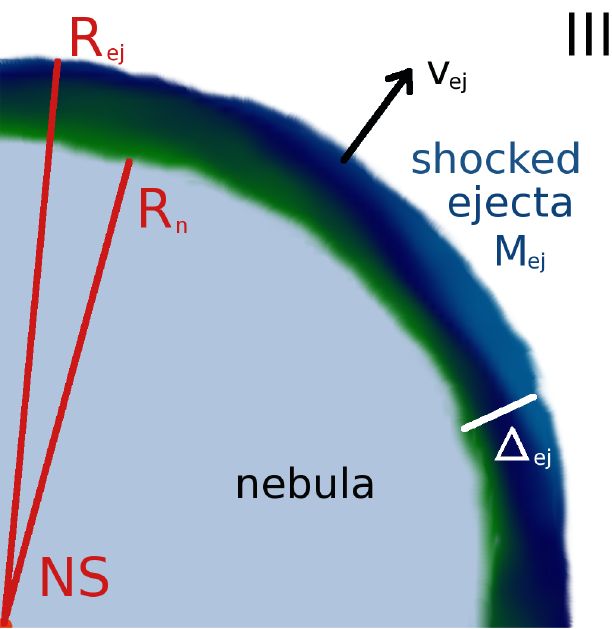}
\vspace{0.5cm}
 \caption{{\it Left}: Phase II of the evolution. The high pressure of
   the pulsar wind nebula (`nebula') powered by the spinning-down NS
   drives a strong shock through the ejecta with speed $v_\text{sh}$,
   thereby compressing, heating and accelerating the material. {\it
     Right}: Phase III of the evolution. The ejecta material of mass
   $M_\text{ej}$ has been entirely swept up by the shock into a thin
   layer of thickness $\Delta_\text{ej}$ moving outward at speed
   $v_\text{ej}$, while the spinning-down NS keeps injecting energy
   into the optically thick pulsar wind nebula.}
 \label{fig:schematic}
\end{figure*}

The pulsar wind inflates a PWN behind the less
rapidly expanding ejecta, a plasma of electrons, positrons and photons
(see Section~\ref{sec:phaseIII_PWN} for a
detailed discussion). As this PWN is highly overpressured with respect
to the confining ejecta envelope, it drives a strong hydrodynamical
shock into the fluid, which heats up the material upstream of the
shock and moves radially outward at relativistic
speeds, thereby sweeping up all the material behind the shock front into a thin
shell. During this phase the system is composed of a NS (henceforth ``pulsar'' in
Phase II and III) surrounded by an essentially baryon-free PWN and a
layer of confining ejecta material. The propagating shock front separates 
the ejecta material into an inner shocked part and an outer unshocked part 
(cf.~Figure~\ref{fig:overview} and \ref{fig:schematic}). While the shock front is moving outward
across the ejecta, the unshocked matter layer still emits thermal
radiation with increasing luminosity as the optical depth
decreases. Initially, the expansion of the PWN nebula is highly
relativistic and decelerates to non-relativistic speeds only when the
shock front encounters high-density material in the outer ejecta
layers. The total crossing time for the shock front is typically $\Delta t_\text{shock} = t_\text{shock,out} -
t_\text{pul,in} \ll t_\text{pul,in}$, where $t_\text{shock,out}$
denotes the time when the shock reaches the outer surface. At this break-out time, a
short burst-type non-thermal EM signal could be emitted that encodes 
the signature of particle acceleration at the shock front.

Phase III starts at $t=t_\text{shock,out}$. At this time, the entire ejecta
material has been swept
up into a thin shell of thickness $\Delta_\text{ej}$ (which we assume
to be constant during the following evolution) that moves outward with
speed $v_\text{ej}$ (cf.~Figure~\ref{fig:schematic}). In general, this speed is higher than the 
expansion speed of the baryon-loaded wind in Phase I ($v_\text{ej,in}$), 
as during shock propagation kinetic energy 
is deposited into the shocked ejecta. Rotational energy is extracted
from the pulsar via dipole spin-down and it is
reprocessed in the PWN via various radiative processes in analogy to 
pair plasmas in compact sources, such as active galactic nuclei (see
Section~\ref{sec:phaseIII_PWN} for a detailed
discussion). Radiation escaping from the PWN ionizes
the ejecta material, which thermalizes the radiation due to the
optical depth still being very high. Only at much later times
the ejecta layer eventually becomes transparent to radiation from the
nebula, which gives rise to a transition from predominantly thermal to
non-thermal emission spectra. We note that for reasons discussed in
Section~\ref{sec:stiffness_problem}, the total luminosity of the
system shows the characteristic $\propto t^{-2}$ behavior for dipole
spin-down at late times $t\gg t_\text{sd}$, where $t_\text{sd}$ is the
spin-down timescale. However, when restricted to individual frequency
bands, the late time behavior of the luminosity can significantly
differ from a $\propto t^{-2}$ power law.

As the NS is most likely not indefinitely stable against gravitational
collapse, it might collapse at any time during the evolution outlined
above (see Section~\ref{sec:collapse}). If the NS is supramassive, 
the collapse is expected to occur within timescales of the order of 
$\sim t_\text{sd}$, for the spin-down timescale
represents the time needed to remove a significant fraction of the rotational
energy from the NS and thus of its rotational support against collapse. For
typical parameters, the collapse occurs in Phase
III. However, if the NS is hypermassive at birth and does not migrate
to a supramassive configuration thereafter, it is expected to collapse
already in Phase I (i.e., on the timescale for removal of differential
rotation). Moreover, even a supramassive NS can collapse 
on timescales shorter than $\sim t_\text{sd}$, if the collapse is induced 
by a (magneto-)hydrodynamic instability. 

Assuming that a SGRB and its afterglows are produced by a BNS 
system merging into a long-lived NS, the prompt burst can be
associated either with the merger itself or with the delayed collapse of the remnant NS, as in the 
recently proposed `time-reversal' scenario 
(\citealt{Ciolfi2015a,Ciolfi2015b}; see \citealt{Rezzolla2015} for an
alternative proposal). Both scenarios can be 
accommodated by the framework of our model. In the former case, 
the lightcurves and spectra we predict after the time of merger 
represent a model for the observed afterglows. In the latter 
case, only the emission that follows the collapse should be directly 
compared with the observed afterglows.
Predictions for both scenarios are
discussed in detail in the companion paper (Paper II).

%%%%%%%%%%%%%%%%%%%%%%%%%%%%%%%%%%%%%%%%%
\section{Evolution equations}
%%%%%%%%%%%%%%%%%%%%%%%%%%%%%%%%%%%%%%%%%
\label{sec:model_equations}

According to the three main evolution phases discussed in
Section~\ref{sec:phenomenology}, our model consists of three sets of
coupled ordinary differential equations (ODEs), which we think
capture the main physical ingredients to describe the dynamical evolution and
EM emission of the
post-merger system. These
ODEs assume spherical symmetry and they are formulated in terms of
main evolution quantities, such as the
extent of certain dynamical structures (e.g., the bayon-loaded wind,
the PWN and the ejecta shell) and their associated energy
budgets. In summary, the evolution equations
we consider are as follows (see Table~\ref{tab:quantities} and
Figure~\ref{fig:schematic} for
definitions of variables):\\

\noindent{\bf Phase I: baryon-loaded wind}

\begin{eqnarray}
\ddt{R_\text{ej}} &=& v_\text{w}(R_\text{ej}(t),t) \label{eq:evoleqns_p1_Rej}\\
\ddt{E_\text{th}} &=& L_\text{EM}(t) + \ddt{E_\text{th,NS}} - L_\text{rad}(t) \label{eq:evoleqns_p1_Eth}
\end{eqnarray}

\vspace{8mm}

\noindent{\bf Phase II: pulsar wind shock}

\begin{eqnarray}
\ddt{R_\text{ej}} &=& v_\text{w}(R_\text{ej}(t),t) \label{eq:evoleqns_p2_Rej}\\
\ddt{R_\text{sh}} &=& v_\text{sh}(t) \label{eq:evoleqns_p2_shock} \\ 
\ddt{R_\text{n}} &=& \ddt{R_\text{sh}} - \ddt{\Delta_\text{sh}} \label{eq:evoleqns_p2_Rn}\\
\ddt{E_\text{th,sh}} &=& \ddt{E_\text{sh}} +
                        \ddt{E_\text{th,vol}} + \ddt{E_\text{PWN}} -
                         L_\text{rad,in}(t) \label{eq:evoleqns_p2_Esh} \\
\ddt{E_\text{th,ush}} &=& -\ddt{E_\text{th,vol}} - L_\text{rad}(t) \label{eq:evoleqns_p2_Eush}\\
\ddt{E_\text{th}} &=& \ddt{E_\text{th,sh}} + \ddt{E_\text{th,ush}}\\
\ddt{E_\text{nth}} &=&
                       -\frac{E_\text{nth}}{R_\text{n}}\ddt{R_\text{n}}
                       -  \ddt{E_\text{PWN}} \nonumber\\
                       &&+  L_\text{rad,in}(t) + \eta_\text{TS}[L_\text{sd}(t) + L_\text{rad,pul}(t)]
                      \label{eq:evoleqns_p2_Enth}\\
\ddt{E_B} &=& \eta_{B_\text{n}} [L_\text{sd}(t) + L_\text{rad,pul}(t)] \label{eq:evoleqns_p2_EB}
\end{eqnarray}

\vspace{4mm}

\noindent{\bf Phase III: pulsar wind nebula}

\begin{eqnarray}
\ddt{v_\text{ej}} &=& a_\text{ej}(t)  \label{eq:evoleqns_p3_vej}\\
\ddt{R_\text{ej}} &=& v_\text{ej}(t) + \frac{1}{2}a_\text{ej}(t)
\rmn{d}t\\
\ddt{R_\text{n}} &=& \ddt{R_\text{ej}} \label{eq:evoleqns_p3_Rn}\\
\ddt{E_\text{th}} &=& [1-f_\text{ej}(t)]\ddt{E_\text{PWN}} - L_\text{rad}(t) -
                      L_\text{rad,in}(t) \label{eq:evoleqns_p3_Eth}\\
\ddt{E_B} &=& \eta_{B_\text{n}} [L_\text{sd}(t) + L_\text{rad,pul}(t)] \label{eq:evoleqns_p3_EB}
\end{eqnarray}

Equations~\eqref{eq:evoleqns_p1_Rej} and \eqref{eq:evoleqns_p1_Eth} are
supplemented with a model to describe the baryonic wind emitted from
the NS during this phase (see Section~\ref{sec:wind_model}). Furthermore, Equations~\eqref{eq:evoleqns_p3_vej}--\eqref{eq:evoleqns_p3_EB} are
supplemented with Equations~\eqref{eq:LightmanI} and
\eqref{eq:LightmanII}, which model the radiative processes inside the
PWN and which need to be solved at every time step of
the evolution equations to determine the source term
$\rmn{d}E_\text{PWN}/\rmn{d}t$ and the emission spectrum of the nebula.

The above evolution equations are formulated in the rest frame of the
merger remnant (henceforth the ``lab frame''; see also
Appendix~\ref{app:frames}). The various
terms appearing in
Equations~\eqref{eq:evoleqns_p1_Rej}--\eqref{eq:evoleqns_p3_EB} are
motivated and discussed in detail in Section~\ref{sec:ingredients}.

\begin{table}[tb]
\caption{Definition of lab-frame quantities.}
\label{tab:quantities}
\centering
\begin{tabular}{cp{0.38\textwidth}}
\hline\hline
Quantity & Description \\
\hline
$t$ & time\\
$R_\text{ej}$ & outer radius of the ejected matter\\
$R_\text{n}$ & radius of the PWN (Phase II, III)\\
$R_\text{sh}$ & radial location of the pulsar wind shock front (Phase II)\\
$v_\text{w}(r,t)$ & velocity profile of the baryon-loaded wind (Phase
                    I, II; cf.~Section~\ref{sec:wind_model},
                    Equation~\eqref{eq:v_w_rt})\\
$v_\text{ej,in}$ & initial expansion speed of the baryonic ejecta
                   material (cf.~Section~\ref{sec:wind_model}, Equation~\eqref{eq:v_t})\\
$v_\text{sh}$ & velocity of the pulsar wind shock front (Phase II;
                cf.~Equation~\eqref{eq:v_sh})\\
$v_\text{ej}$ & velocity of the ejecta shell in Phase III (cf.~Equation~\eqref{eq:dvej_dt})\\
$a_\text{ej}$ & acceleration of the ejecta shell in Phase III (cf.~Equation~\eqref{eq:dvej_dt})\\
$\Delta_\text{sh}$ & radial thickness of the shocked ejecta shell in
                     Phase II (cf.~Section~\ref{sec:PhaseII_ejecta})\\
$\Delta_\text{ej}$ & radial thickness of the ejecta shell in Phase III (cf.~Section~\ref{sec:sec_PWN_p3})\\
$f_\text{ej}$ & function to smoothly `switch on/off'
                terms as the ejecta material becomes optically thin
                (cf.~Equation~\eqref{eq:f_ej})\\ 
$E_\text{th}$ & internal energy stored in the ejecta material, available
                to be emitted as thermal radiation\\
$E_\text{th,sh}$ & internal energy stored in the shocked ejecta matter
                  (Phase II)\\
$E_\text{th,ush}$ & internal energy stored in the unshocked ejecta matter
                  (Phase II)\\
$E_\text{nth}$ & internal energy of the PWN\\
$E_B$ & magnetic energy of the PWN\\
$\rmn{d}E_\text{th,NS}/\rmn{d}t$ & internal energy injected from the NS into the
                         baryon-loaded wind per unit time (Phase I;
                         cf.~Equation~\eqref{eq:Eth_NS})\\
$\rmn{d}E_\text{sh}/\rmn{d}t$ & energy deposited in the shocked ejecta
                                material by shock
                      heating per unit time (cf.~Equation~\eqref{eq:Esh})\\
$\rmn{d}E_\text{th,vol}/\rmn{d}t$ & total internal energy of unshocked ejecta in the
                    volume swept up by the shock front per unit time
                          (cf.~Equation~\eqref{eq:Eth_vol})\\
$\rmn{d}E_\text{PWN}/\rmn{d}t$ & total energy emitted by the PWN
                       per unit time (Phase II, III;
                       cf.~Equations~\eqref{eq:dE_PWN_dt} and \eqref{eq:dEPWN_dt})\\
$L_\text{EM}$ & EM energy deposited in the baryon-loaded
                wind per unit time (Phase
                I; cf.~Equation~\eqref{eq:L_EM})\\
$L_\text{rad}$ & luminosity of thermal radiation from the
                 outer surface of the ejected material
                 (cf.~Equations~\eqref{eq:Lrad_p1},
                 \eqref{eq:Lrad_p2}, and \eqref{eq:Lrad_phaseIII})\\
$L_\text{rad,in}$ & luminosity of thermal radiation from the ejected
                    material radiated toward the interior (Phase
                    II,III; cf.~Equations~\eqref{eq:Lradin_p2} and \eqref{eq:Lradin_phaseIII})\\
$L_\text{rad,pul}$ & luminosity of thermal radiation from
                    the NS surface (Phase II,III; cf.~Equation~\eqref{eq:Lrad_pul})\\
$L_\text{sd}$ & spin-down luminosity of the pulsar (Phase II, III; cf.~Equation~\eqref{eq:Lsd})\\
$\eta_{B_\text{n}}$ & fraction of the total pulsar wind power injected
                      as magnetic energy per unit time into the PWN
                      (cf.~Sections~\ref{sec:phaseII_PWN} and \ref{sec:phaseIII_PWN})\\
$\eta_\text{TS}$ & efficiency of converting pulsar wind power into random
              kinetic energy of accelerated particles in the PWN
              (cf.~Equations~\eqref{eq:evoleqns_p2_Enth}, \eqref{eq:le} and Sections~\ref{sec:phaseII_PWN},
                   \ref{sec:phaseIII_PWN})\\
\hline
\end{tabular}
\end{table}

%%%%%%%%%%%%%%%%%%%%%%%%%%%%%%%%%%%%%%%%%
 \section{Ingredients}
%%%%%%%%%%%%%%%%%%%%%%%%%%%%%%%%%%%%%%%%%
\label{sec:ingredients}

\subsection{Phase I: baryon-loaded wind}
\label{sec:phaseI}
Phase I starts with a differentially rotating NS that
generates a baryon-loaded wind, either due to a very strong
magnetization of the stellar interior \citep{Siegel2014a} and/or as the
result of neutrino emission from the stellar interior (e.g.,
\citealt{Dessart2009}). This wind can be treated as roughly
spherically symmetric at distances $r\gtrsim
R_\text{min}=30\,\text{km}$ (cf., e.g.,
\citealt{Siegel2014a,Siegel2015a}), which we define as
our inner spatial boundary of the evolution model.

Such a magnetically and/or neutrino-induced wind is likely to dominate
baryon pollution around the newly-formed NS on the timescales of
interest. Dynamical ejecta originating from the tidal tails during the
merger process will be ejected mostly into the equatorial plane and move away
from the merger site with high (mildly relativistic) velocities (e.g.,
\citealt{Davies1994,Rosswog2013,Oechslin2007,Hotokezaka2013a,Bauswein2013}). They
are thought to undergo r-process nucleosynthesis and to possibly power a
macronova (aka a kilonova; e.g.,
\citealt{Li1998,Metzger2010,Barnes2013,Piran2013,Tanaka2013}). The more
isotropic dynamical ejection originating from shock heating at
the contact interface during collision of the NSs and aided by radial
oscillations of the double-core structure of the newly-formed NS immediately after merger
\citep{Hotokezaka2013a,Bauswein2013,Kastaun2015a} only lasts for a few milliseconds
and is thus unlikely to dominate baryon pollution on the much longer
timescales relevant here. Such a
component can, however, be accounted for in our wind model by, e.g.,
tuning the mass ejection rate $\dot{M}$ (see below).

\subsubsection{Wind model}
\label{sec:wind_model}

In order to avoid performing expensive hydrodynamical simulations, we
describe the generation and evolution of the baryon-loaded wind by a
simple one-dimensional model. In particular, we assume that there are no
hydrodynamical shocks being formed in the wind. The assumption of
spherical symmetry is motivated by recent
three-dimensional magnetohydrodynamic simulations in general
relativity, which have shown that for the
most realistic magnetic field configurations, such a wind will be
highly isotropic \citep{Siegel2014a,Siegel2015a}.
Our simple approach is furthermore
justified by the fact that the exact details of
modeling this wind only have a minor influence on the final
predictions of our model, such as the lightcurves, as the pulsar
nebula shock eventually sweeps up all the material into a thin
shell in Phase II.

It is conceivable to assume that a fluid element of the wind ejected
from the NS at time $t$ moves
outward with a constant velocity $v(r,t) = v(t)$, once it has climbed up the
gravitational potential of the NS after a few tens of kilometers. Mass
conservation requires $v(r,t) = \dot{M}(t)/4\pi r^2 \rho(r,t)$, where
$\rho$ denotes the rest-mass density and $\dot{M}(t)$ the mass injection
rate into the wind. Hence, in order to satisfy $v(r,t) = v(t)$,
$\rho(r,t) = A(t)/r^2$ for matter ejected at time $t$. This assumption
is also supported by recent general-relativistic
magnetohydrodynamic simulations of magnetically driven winds that
yielded a $\rho\propto r^{-2}$ density profile and constant ejection
velocities for a constant $\dot{M}$ over the timescale of tens of
milliseconds \citep{Siegel2014a,Siegel2015a}.

While differential rotation is being removed on a timescale
$t_\text{dr}$, i.e., the gradient of the angular frequency decays as
\begin{equation}
  |\nabla \Omega| \propto \exp(- t/t_\text{dr}), \label{eq:nablaOmega}
\end{equation}
we also expect that the mass ejection rate, therefore the density profile, and the
ejection speed decay on roughly similar timescales:
\begin{eqnarray}
\dot{M}(t) &=& \dot{M}_\text{in} \exp(-\sigma_M t/t_\text{dr}), \label{eq:Mdot_t}\\
A(t) &=& A_\text{in} \exp(-\sigma_\rho t/t_\text{dr}), \label{eq:A_t}\\
v(t) &=& v_\text{ej,in} \exp(-\sigma_v t/t_\text{dr}), \label{eq:v_t}
\end{eqnarray}
where $\sigma_M,\sigma_\rho,\sigma_v$ are of the order of one. This
set of assumptions satisfies the `no-shocks-requirement' and
immediately yields for the outer radius of the wind:
\begin{equation}
  R_\text{ej}(t) = R_\text{min} + v_\text{ej,in} t. \label{eq:phaseI_Rej}
\end{equation}
Equations~\eqref{eq:Mdot_t}--\eqref{eq:v_t} only apply for a
magnetically driven wind \citep{Siegel2014a}. However,
typical values for $t_\text{dr}\lesssim 1\,\text{s}$ agree
well with typical neutrino cooling timescales, such that a
neutrino-driven wind can be additionally incorporated in the model by
tuning $\dot{M}_\text{in}$. We note that even for
moderate magnetic field strengths and realistic rotation periods of
the NS the magnetically driven wind can easily exceed mass loss rates
of a few in $10^{-3}\,\text{M}_\odot\,\text{s}^{-1}$ and thus dominate
over the other ejection mechanisms. If the neutrino cooling timescale
is very different from $t_\text{dr}$ the situation could be different. However, as
the pulsar wind shock eventually sweeps up all the ejecta material into a
thin shell, the following evolution will mostly depend on the total
amount of emitted material $M_\text{ej} \propto
\dot{M}_\text{in}t_\text{dr}$. Hence, by varying $t_\text{dr}$ and
$\dot{M}_\text{in}$ the above model for mass ejection can effectively also
accommodate a dominant neutrino-induced wind.

Simple scaling arguments can fix two of the free parameters
$\sigma_M,\sigma_\rho,\sigma_v$ that control the decay timescales
relative to $t_\text{dr}$. Magnetic winding in the stellar interior
converts rotational energy into magnetic energy (mostly into toroidal
field strength) at a rate $\propto
|\nabla\Omega|^2$ (e.g., \citealt{Duez2006b,Siegel2013}). This
magnetic energy is then available to be dissipated into kinetic and
EM energy of a wind driven by the built-up of magnetic
pressure in the stellar interior \citep{Siegel2014a}. Consequently,
\begin{equation}
  \frac{1}{2}\dot{M}v^2 + L_\text{EM} \propto |\nabla\Omega|^2, \label{eq:Erotdot}
\end{equation}
where $L_\text{EM}$ denotes the EM luminosity
corresponding to the Poynting flux carried by the wind (cf.~also Section~\ref{sec:ejecta_properties}). From
Equations~\eqref{eq:nablaOmega} and \eqref{eq:Erotdot} we conclude
that $L_\text{EM}\propto \exp(-\sigma_L t/t_\text{dr})$, with
$\sigma_L = 2$, and
\begin{equation}
  \sigma_M + 2\sigma_v = 2. \label{eq:sigma1}
\end{equation}
At $r = R_\text{min}$, we also need to satisfy $v(t) =
\dot{M}(t)/4\pi\rho(R_\text{min},t)R_\text{min}^2$. Using
Equations~\eqref{eq:Mdot_t}--\eqref{eq:v_t}, this yields
\begin{equation}
  \sigma_M = \sigma_\rho + \sigma_v \label{eq:sigma2}
\end{equation}
and $A_\text{in}=\dot{M}_\text{in}/4\pi v_\text{ej,in}$. According to
Equations~\eqref{eq:sigma1} and \eqref{eq:sigma2}, only
one $\sigma$-parameter can be chosen independently. The additional requirement of
$\sigma_M,\sigma_\rho,\sigma_v$ all being non-negative limits these
parameters to the following ranges:
\begin{eqnarray}
  \sigma_M &&\in [1, 2] \\
  \sigma_\rho &&\in [0.5, 2] \\
  \sigma_v &&\in [0, 0.5]. 
\end{eqnarray}

In order to compute the density profile $\rho(r,t)$ of the
wind resulting from the mass ejection model
\eqref{eq:Mdot_t}--\eqref{eq:v_t}, we first compute the total mass
contained in a volume of radius $r$ at time $t$,
\begin{eqnarray}
  m_\text{w}(r,t) &=& \int_{\bar{t}(r)}^{t} \dot{M}(t') \,\rmn{d}t' \label{eq:m_w_rt} \\
  &=& \frac{t_\text{dr}}{\sigma_M}\dot{M}_\text{in}\mskip-5mu\left[
    \exp\mskip-5mu\left(-\sigma_M\frac{\bar{t}(r,t)}{t_\text{dr}}\right)\mskip-5mu - \mskip-2mu\exp\mskip-5mu\left(-\sigma_M\frac{t}{t_\text{dr}}\right)  \right],\nonumber
\end{eqnarray}
where $\bar{t}(r,t)$ denotes the time a fluid element at $r$ and $t$ was
sent out from the inner boundary at $r=R_\text{min}$, i.e., $\bar{t}$ is determined by $r = v(\bar{t})(t-\bar{t})
+ R_\text{min}$, or, equivalently,
\begin{equation}
  (t-\bar{t})\exp\left(-\sigma_v\frac{\bar{t}}{t_\text{dr}}\right) -
  \frac{r - R_\text{min}}{v_\text{ej,in}} = 0. \label{eq:tbar}
\end{equation}
It is straightforward to show that Equation~\eqref{eq:tbar} has
exactly one root in $[0,t]$, such that $\bar{t}(r,t)$ is well
defined. At any time $t$ the density profile of the wind is then given
by
\begin{equation}
  \rho_{\text{w},t}(r) = \frac{1}{4\pi r^2} \frac{\partial
    m_\text{w}(r,t)}{\partial r}. \label{eq:rho_w_rt}
\end{equation}
The corresponding velocity profile is
\begin{equation}
  v_\text{w}(r,t) = v_\text{ej,in}
  \exp\left(-\sigma_v\frac{\bar{t}(r,t)}{t_\text{dr}}\right) \label{eq:v_w_rt}
\end{equation}
and the total amount of ejected mass at time $t$ is given by
\begin{eqnarray}
  M_\text{ej}(t) &=& m_\text{w}(R_\text{ej}(t),t) \nonumber\\ 
  &=& \frac{t_\text{dr}}{\sigma_M}\dot{M}_\text{in}\mskip-5mu\left[
    1 - \exp\mskip-5mu\left(-\sigma_M\frac{t}{t_\text{dr}}\right)  \right].
\end{eqnarray}

\subsubsection{Properties of the ejected matter}
\label{sec:ejecta_properties}

\paragraph{Injection of electromagnetic energy} The wind will be
endowed with a Poynting flux of luminosity (see \citealt{Siegel2014a}
and Section~\ref{sec:wind_model})
\begin{eqnarray}
  L_\text{EM}(t) \simeq && 10^{48} 
\left(\frac{\bar{B}}{10^{15}\,\mathrm{G}}\right)^2 \!
\left(\frac{R_\text{e}}{10^{6}\,\mathrm{cm}}\right)^{3} \!
\left(\frac{P_\text{c}}{10^{-4}\,\mathrm{s}}\right)^{-1}\nonumber\\
&&\times \exp\left(-\sigma_L  \frac{t}{t_\text{dr}}\right) \label{eq:L_EM}
\text{erg}\,\text{s}^{-1},
\end{eqnarray}
where $\sigma_L =2$. While $t_\text{dr}$ is a free parameter of our
model, the total magnetic field strength in the outer layers of the
NS, $\bar{B}$, the equatorial radius $R_\text{e}$, as well as the
central spin period $P_\text{c}$ of the differentially rotating NS can
be extracted from numerical relativity simulations. The wind itself is
highly turbulent in the vicinity of the NS \citep{Siegel2014a} and
magnetic dissipation will therefore be very effective. We assume that
this Poynting flux is dissipated and thermalized in the ejecta matter (due to the very high
optical depth) on the timescales of interest,
such that this energy is trapped and thus appears as a source term in
Equation~\eqref{eq:evoleqns_p1_Eth}.

\paragraph{Injection of thermal energy} The material ejected from the
NS surface as probed by numerical relativity simulations typically has
a very high specific internal energy $\epsilon_\text{ej,NS,in}$ that corresponds
to a temperature $T_\text{ej,NS,in}$ of the order of tens of MeV (e.g.,
\citealt{Siegel2014a,Kastaun2015a}). We therefore source the internal
energy of the ejected material in Equation~\eqref{eq:evoleqns_p1_Eth} by the
corresponding injection rate
\begin{equation}
\ddt{E_\text{th,NS}} =
\epsilon_\text{ej,NS,in}\dot{M}_\text{in}\exp\left(-\sigma_M\frac{t}{t_\text{dr}}-\frac{t}{t_{\nu}}\right), \label{eq:Eth_NS}
\end{equation}
where we have assumed that the NS matter cools down due to neutrino
cooling on a timescale $t_{\nu}$.

\paragraph{Thermal radiation from the wind} Due to the very high
optical depth of the wind material, the radiation
originating from the surface of the expanding wind will be
predominantly thermal in Phase I. This leads to radiative losses in Equation~\eqref{eq:evoleqns_p1_Eth} with
a luminosity
\begin{equation}
  L_\text{rad}(t) = 4 \pi R_\text{ej}^2(t) \sigma
  T_\text{eff}^4(t), \label{eq:Lrad_p1}
\end{equation}
where $\sigma$ denotes the Stefan-Boltzmann constant. The effective
temperature of the wind, $T_\text{eff}$, is given by
\begin{equation}
  T_\text{eff}^4(t)\simeq \frac{16}{3}\frac{T^4(t)}{\Delta\tau(t) + 1}. \label{eq:Teff_ph1}
\end{equation}
Here, 
\begin{equation}
  \Delta\tau (t) = \kappa \int_{R_\text{in}(t)}^{R_\text{ej}(t)}
  \rho_{\text{w},t}(r)\,\rmn{d}r \label{eq:delta_tau_w}
\end{equation}
is the optical depth of the wind with associated diffusion timescale
\begin{equation}
  t_\text{diff}(t) = \frac{R_\text{ej}(t)-R_\text{in}(t)}{c} [\Delta\tau(t) + 1]. 
\end{equation}
Adding unity here and in the definition of $T_\text{eff}$ takes a possible transition to the
optically thin regime into account. $\kappa$ is the opacity of
the ejecta material and $R_\text{in}(t)$ denotes the effective inner
radius of the ejected material, defined by the condition 
\begin{equation}
  \kappa \int_{R_\text{min}}^{R_\text{in}(t)}
  \rho_{\text{w},t}(r)\,\rmn{d}r = 1.
\end{equation}
We describe the wind material as a mixture of ideal gas with adiabatic index
$\Gamma_\text{ej} = 4/3$ and radiation. The temperature $T(t)$ of the
wind in Equation~\eqref{eq:Teff_ph1} at time $t$ is thus obtained as the root of the
following equation:
\begin{equation}
  T(t)^4 +
  \frac{1}{\Gamma_\text{ej}-1}\frac{k_\text{B}}{m_\text{p}a}\bar{\rho}_\text{w}(t)T(t)
  - \frac{E_\text{th}(t)}{aV_\text{ej}(t)} = 0, \label{eq:T_p1}
\end{equation}
where $k_\text{B}$ is the Boltzmann constant, $m_\text{p}$ denotes the
proton mass, $a$ is the radiation constant, $V_\text{ej}(t) = (4/3)\pi
[R_\text{ej}^3(t)-R_\text{in}^3(t)]$ the volume of the ejecta material,
and $\bar{\rho}_\text{w}(t)=[m_\text{w}(R_\text{ej}(t),t) -
m_\text{w}(R_\text{in}(t),t)] / V_\text{ej}(t)$ its mean density.

\subsection{Phase II: pulsar wind shock}

\subsubsection{Pulsar properties}
\label{sec:pulsar_properties}
\paragraph{`Ignition'} As differential rotation is being removed the
NS settles down to uniform rotation and mass ejection is suppressed
according to the wind model discussed in
Section~\ref{sec:wind_model}. This eventually creates the conditions
to build up
a pulsar magnetosphere. As a criterion to `switch on' the pulsar, we employ
$\rho(R_\text{min},t)=(A_\text{in}/R_\text{min}^2)\exp(-\sigma_\rho
t/t_\text{dr}) < \rho_\text{crit}$, which translates into a time
\begin{equation}
  t_\text{pul,in} =
  -\frac{t_\text{dr}}{\sigma_\rho}\ln\!\left(\frac{\rho_\text{crit}R_\text{min}^2}{A_\text{in}}\right). \label{eq:t_pulin}
\end{equation}
For the critical density we typically choose the rest-mass density for
Iron ions corresponding to the Goldreich-Julian charge density,
$\rho_\text{crit}\sim \rho_\text{GJ}[\text{Fe}]\sim 6\times
10^{-6}\,\text{g}\,\text{cm}^{-3}$, assuming typical magnetic fields of
$10^{15}\,\text{G}$ and rotational periods of the pulsar of $\sim
1\,\text{ms}$ \citep{Goldreich1969}. We note, however, that as the density in the vicinity
of the NS decreases exponentially, varying the choice of
$\rho_\text{crit}$ even by orders of magnitude does not influence the
model evolution noticeably, as long as the value is sufficiently small.

\paragraph{Initial spin and magnetic field}
Once the NS has settled down to uniform rotation, it is expected to
maintain roughly the same level of magnetization on the timescales of
interest. In Phase I, strong toroidal fields have been built up
through magnetic winding and possibly instabilities like the
magnetorotational instability \citep{Duez2006b,Siegel2013,Siegel2014_MRIproc,Kiuchi2014}, such
that the strength of the dipolar component of the poloidal field at
the pole $B_\text{p}$ is only a fraction $\eta_{B_\text{p}}$ of the total magnetic field strength
in the outer layers of the NS,
\begin{equation}
  B_\text{p}=\eta_{B_\text{p}} \bar{B}. \label{eq:B_p}
\end{equation}
This dipole component is the relevant quantity for dipole spin-down emission (see
next paragraph). Knowing the initial rotational energy
$E_\text{rot,NS,in}$ of the differentially rotating NS from numerical
relativity simulations, we can infer the initial spin period of the
pulsar $P_\text{pul,in}$ at $t_\text{pul,in}$ from its initial
rotational energy $E_\text{rot,pul,in}$:
\begin{equation}
  E_\text{rot,pul,in} = E_\text{rot,NS,in} - E_\text{EM} -
  E_\text{kin,w} -  E_\text{rot,ej}. \label{eq:Erot_pul}
\end{equation}
Here,
\begin{equation}
  E_\text{EM} = \int_0^{t_\text{pul,in}} L_\text{EM}(t)\,\rmn{d}t
\end{equation}
is the rotational energy dissipated into Poynting flux during Phase I,
\begin{equation}
  E_\text{kin,w} =
  \frac{1}{2} 4\pi \int_{R_\text{min}}^{R_\text{ej}(t_\text{pul,in})}
  \rho_{\text{w},t_\text{pul,in}}(r) v_\text{w}^2(r,t_\text{pul,in}) r^2\,\rmn{d}r \label{eq:Ekin_w}
\end{equation}
is the rotational energy dissipated into kinetic energy of the
radially expanding wind in
Phase I, and
\begin{equation}
  E_\text{rot,ej} \sim
  \frac{M_\text{ej}(t_\text{pul,in})}{M_\text{NS,in}}
  E_\text{rot,NS,in} \label{eq:Erot_ej}
\end{equation}
approximates the (subdominant) amount of initial rotational energy carried away by
the ejected matter. The initial mass $M_\text{NS,in}$ of the NS at
birth is known from numerical relativity simulations. The
initial spin period is then given by
\begin{equation}
  P_\text{pul,in} = 2\pi \left(
    \frac{2E_\text{rot,pul,in}}{I_\text{pul}}\right)^{-\frac{1}{2}}, \label{eq:Ppul_in}
\end{equation}
where $I_\text{pul}$ is the moment of inertia of the
pulsar.

\paragraph{Spin-down luminosity}
Due to unipolar induction charged particles are continuously extracted from the
pulsar surface \citep{Goldreich1969}. This primary particle current initiates copious pair
production, which populates the magnetosphere
with a nearly force-free plasma and drives a highly relativistic,
magnetized (Poynting-flux dominated) outflow of particles (mostly
electrons and positrons), referred to
as a pulsar wind. The associated Poynting-flux luminosity is given
by \citep{Spitkovsky2006,Philippov2015}
\begin{equation}
  L_\text{sd,in} =
 \frac{
   (2\pi)^4}{4c^3}\frac{B_\text{p}^2R_\text{pul}^6}{P_\text{pul,in}^4
 }(k_0 + k_1\sin^2\chi), \label{eq:Lsd_in}
\end{equation}
where $k_0=1.0\pm0.1$, $k_1 = 1.1\pm0.1$, $\chi$ is the inclination
angle of the dipole component of the magnetic field with respect to
the rotation axis of the pulsar, and $R_\text{pul}\simeq R_\text{e}$ denotes the radius
of the pulsar. This Poynting flux is extracted from the
pulsar via the magnetic field at the expense of rotational energy
$E_\text{rot,pul}$. Observations of well studied objects like the
Crab pulsar suggest that only $\sim 1\%$ of the observed spin-down
energy is directly radiated away by the pulsar \citep{Buehler2014},
such that essentially the entire spin-down energy is carried by the
pulsar wind leaving the light cylinder. Hence, we can set
$\dot{E}_\text{rot,pul}=L_\text{sd}$, which results in 
slowing down the pulsar spin $P_\text{pul}(t) \propto
\sqrt{1+t/t_\text{sd}}$ and yields a spin-down luminosity of
\begin{equation}
  L_\text{sd}(t) = L_\text{sd,in}
  \left(1+\frac{t-t_\text{pul,in}}{t_\text{sd}} \right)^{-2}, \label{eq:Lsd}
\end{equation}
where
\begin{equation}
  t_\text{sd} = \frac{E_\text{rot,pul,in}}{L_\text{sd,in}} \label{eq:tsd}
\end{equation}
is the
spin-down timescale. In the following, we assume an aligned rotator,
i.e., $\chi = 0$, as a non-zero inclination angle would result in a
change of $L_\text{sd}$ of order unity, which can be absorbed,
e.g., in the parameter $\eta_{B_\text{p}}$ that sets the initial field strength
$B_\text{p}$ (see Equation~\eqref{eq:B_p}).

\paragraph{Emission of thermal energy}
Immediately after the BNS merger the NS matter is very hot with
temperatures of a few tens of MeV. It cools
down via neutrino emission on a timescale $t_{\nu}\lesssim
1\,\text{s}$. This thermal energy reservoir can be
tapped via ejection of material in Phase I (cf.~Equation~\eqref{eq:Eth_NS})
and by thermal radiation from the pulsar surface in Phase II and
III. Thermal radiation from the pulsar can help initiate
secondary-particle cascades based on the primary charged particles extracted from the pulsar
surface. The associated energy is thus reprocessed in the
magnetosphere and will be carried away by the pulsar wind. However,
given typical neutrino cooling timescales,
this thermal energy contribution will become
energetically negligible very soon with respect to, e.g., the spin-down luminosity. Therefore, we adopt a simple cooling model for the pulsar surface,
\begin{equation}
  T_\text{NS}(t) = T_\text{NS,in}\exp\!\left( - \frac{t}{t_{\nu}}\right),
\end{equation}
with associated thermal radiation of luminosity
\begin{equation}
  L_\text{rad,pul}(t) = 4\pi R_\text{pul}^2\sigma T_\text{NS}^4(t). \label{eq:Lrad_pul}
\end{equation}
Here, $T_\text{NS,in}$ is the initial typical surface temperature of
the NS corresponding to the specific internal energy
$\epsilon_\text{ej,NS,in}$ (cf.~Section~\ref{sec:ejecta_properties}). As the
thermal energy is reprocessed in the magnetosphere, we add
$L_\text{rad,pul}$ as a source term to the evolution
equation \eqref{eq:evoleqns_p2_Enth} and as an input for the nebula in
Phase III (cf.~Section~\ref{sec:phaseIII_PWN}).

\subsubsection{Simple model for the expanding pulsar wind nebula}
\label{sec:phaseII_PWN}
The newly-formed pulsar wind (Section~\ref{sec:pulsar_properties})
leaves the magnetosphere with relativistic velocities and inflates a
PWN behind the less rapidly expanding ejecta
material (e.g., \citealt{Kennel1984a}; see
Section~\ref{sec:phaseIII_PWN} for more details). The PWN is highly
overpressured with respect to the surrounding ejecta matter and thus
drives a strong hydrodynamical shock into the material (see
Section~\ref{sec:PhaseII_ejecta}), which, in turn, leads to a rapid
expansion of the PWN. The exact physical description of such a
highly dynamical PWN is complex. However, the main energetical
features that govern the overall dynamics of the system (which is what we are
interested in here) can be captured using a simple
approach inspired by recent dynamical PWN models (e.g.,
\citealt{Gelfand2009,Kotera2013,Metzger2014a,Metzger2014b}): we formulate the
energetics of the PWN in terms of a balance equation for
the total internal energy $E_\text{nth}$ of the nebula (photons and random kinetic energy
of the particles; cf.~Equation~\eqref{eq:evoleqns_p2_Enth}). The internal
energy of the PWN is sourced by the
spin-down luminosity $L_\text{sd}$ (Equation~\eqref{eq:Lsd}),
thermal radiation from the pulsar surface $L_\text{rad,pul}$
(Equation~\eqref{eq:Lrad_pul}) as discussed above (which can represent a significant
contribution at early times), and thermal
radiation emitted by the hot ejecta matter toward the interior with luminosity
$L_\text{rad,in}$ (Equation~\eqref{eq:Lradin_p2}). We
assume that only a fraction $\eta_\text{TS}$ of the spin-down
luminosity and the thermal radiation from the pulsar surface are
dissipated into random kinetic energy of the particles in the
nebula. This parameter reflects our incomplete knowledge
about the efficiency with which the bulk energy of the pulsar wind is
dissipated in the nebula (see Section~\ref{sec:phaseIII_PWN} for more
details).

Rapid expansion of the PWN occurs at the expense of $p\,\rmn{d}V$ work,
\begin{equation}
  \ddt{E_{p\rmn{d}V}} = -p_\text{n}\ddt{V_\text{n}} = - \frac{E_\text{nth}}{R_\text{n}}\ddt{R_\text{n}},
\end{equation}
where 
\begin{equation}
  p_\text{n} = \frac{E_\text{nth}}{3V_\text{n}} +
  \frac{B_\text{n}^2}{8\pi} \label{eq:p_n}
\end{equation}
is the pressure inside the nebula with volume $V_\text{n} = (4/3)\pi R_\text{n}^3$ and
radius $R_\text{n}$. Here we have assumed that the particles inside
the nebula are relativistic and collisionless
(cf.~Section~\ref{sec:phaseIII_PWN}), such that they form an ideal gas
with adiabatic index $4/3$. The second pressure contribution on the
right hand side of Equation~\eqref{eq:p_n} is caused by the
uniform and isotropic magnetic field that we assign
to the nebula. Its field strength $B_\text{n}$ at time $t$ is inferred
from the total magnetic energy
\begin{equation}
  E_B = \frac{B_\text{n}^2}{8\pi}V_\text{n} \label{eq:B_n}
\end{equation}
of the nebula, which we evolve
according to Equation~\eqref{eq:evoleqns_p2_EB}. The parameter
$\eta_{B_\text{n}}$ in Equation~\eqref{eq:evoleqns_p2_EB} controls the level of magnetization of the
nebula (see also Section~\ref{sec:phaseIII_PWN}). We note that the
efficiency parameters have to satisfy $\eta_\text{TS}+\eta_{B_\text{n}}\le 1$. 

Furthermore, the nebula loses
energy through irradiation of the surrounding ejecta matter. The
internal energy of the nebula is radiated away on the diffusion
timescale through
various radiative processes that we do not model in detail in Phase II (see,
however, Section~\ref{sec:phaseIII_PWN}). We may write the associated
luminosity as $L_\text{nth}(t) = E_\text{nth}(t)/t_\text{diff,n}(t)$, where
\begin{equation}
  t_\text{diff,n}(t) = \frac{R_\text{n}(t)}{c} [1 + \Delta\tau_\text{T}(t)] \label{eq:tdiffn_p2}
\end{equation}
is the photon diffusion timescale of the nebula due to Thomson
scattering of photons off thermal electrons and positrons. Here,
\begin{equation}
  \Delta\tau_\text{T}(t) = \sqrt{\frac{4Y\sigma_\text{T} L_\text{sd}(t)}{\pi R_\text{n}(t)
        m_\text{e}c^3}}
\end{equation}
is the optical depth to Thomson scattering, with $\sigma_\text{T}$ the Thomson cross section,
$m_\text{e}$ the electron mass, and $Y$ the pair yield of the
nebula \citep{Lightman1987}. Given the very high electron compactness parameter
$l_\text{e}\gg 1$ during Phase II (see Section~\ref{sec:phaseIII_PWN}), the pair yield is given by $Y\simeq 0.1$
\citep{Svensson1987,Lightman1987}. We have added the light crossing time in
Equation~\eqref{eq:tdiffn_p2} to account for a possible transition to
the optically thin regime.

Ionization of ejecta matter due to nebula
radiation can cause a non-zero albedo $\mathcal{A}$, which, in turn, can be
frequency dependent, such that the radiative losses of the nebula may
be written as (cf.~also \citealt{Metzger2014a,Metzger2014b})
\begin{equation}
  \ddt{E_\text{PWN}} = f_\text{beam}(t)\int [1 -
  \mathcal{A}(t,\nu)] L_\text{nth}(t,\nu)\,\rmn{d}\nu, \label{eq:dE_PWN_dt}
\end{equation}
with $\int L_\text{nth}(t,\nu)\,\rmn{d}\nu = L_\text{nth}(t)$. Due to
relativistic expansion of the PWN with velocity
$v_\text{n}=\rmn{d}R_\text{n}/\rmn{d}t$ (cf.~Equation~\eqref{eq:evoleqns_p2_Rn}), relativistic beaming will
prevent a sizable fraction $1-f_\text{beam}$ of photons to escape from the
nebula. Therefore, we have additionally included a factor
$f_\text{beam}(t) = 1-v_\text{n}^2(t) /c^2$ in Equation~\eqref{eq:dE_PWN_dt}. As
the radiative losses given by Equation~\eqref{eq:dE_PWN_dt} are thermalized in the surrounding ejecta
shell, the resulting light curve neither depends on the exact spectral
shape of nebula radiation $L_\text{nth}(t,\nu)$ nor on the spectral
shape of $\mathcal{A}(\nu,t)$. Therefore, we typically
set $\mathcal{A}(t,\nu) = \mathcal{A}(t)$ and we need not assume a spectral
shape for $L_\text{nth}(t)$.

\subsubsection{Shock dynamics and ejecta properties}
\label{sec:PhaseII_ejecta}
The high nebula pressure \eqref{eq:p_n} drives a
strong hydrodynamic shock into the ejecta matter. The shock front
(denoted by the radial coordinate $R_\text{sh}$) divides the ejecta
matter into shocked ejecta with internal energy $E_\text{th,sh}$ and
unshocked ejecta with internal energy $E_\text{th,ush}$ (cf.~Figure~\ref{fig:schematic}). The
PWN with radius $R_\text{n}$ is separated from the
shocked ejecta material by a contact discontinuity. We denote the
radial thickness of the shocked ejecta layer by $\Delta_\text{sh}=R_\text{sh}-R_\text{n}$.

\paragraph{Shock propagation}
In order to avoid performing hydrodynamical simulations, we
describe the shock dynamics by a simple model (Equations~\eqref{eq:evoleqns_p2_shock} and \eqref{eq:evoleqns_p2_Rn}). According to the
special-relativistic Rankine-Hugoniot conditions, the shock speed
as measured in the lab frame is given by (see Appendix~\ref{app:shock})
\begin{equation}
  v_\text{sh}(t) = \frac{v_\text{w}(R_\text{sh}(t),t) +
    v_\text{sh,R}(t)}{1+v_\text{w}(R_\text{sh}(t),t)v_\text{sh,R}(t)/c^2}, \label{eq:v_sh}
\end{equation}
where
\begin{equation}
  v_\text{sh,R} = \left\{
    \frac{p_\text{n}}{\rho_\text{R}\left[ 1 -
        \frac{\rho_\text{R}}{\rho_\text{L}} \left( 1 +
        \frac{\Gamma_\text{ej}}{\Gamma_\text{ej} -
          1}\frac{p_\text{n}}{\rho_\text{L}c^2}\right) +
      \frac{p_\text{n}}{\rho_\text{R}c^2}\right]}\right\}^{\frac{1}{2}} \label{eq:v_R}
\end{equation}
is the shock velocity in the frame comoving with the unshocked
fluid. Here, $\rho_\text{R}(t) = \rho_{\text{w},t}(R_\text{sh}(t))$ is
the density of the unshocked fluid at the shock front and
\begin{equation}
  \rho_\text{L} = \frac{\Gamma_\text{ej} +
      1}{\Gamma_\text{ej} - 1}\rho_\text{R}   \frac{1}{2} \left\{ 1 +
      \left[1+4\frac{p_\text{n}}{\rho_\text{R}c^2}\frac{\Gamma_\text{ej}}{(\Gamma_\text{ej}+1)^2}
      \right]^{\frac{1}{2}} \right\} \label{eq:rho_L}
\end{equation}
is the density of the shocked fluid at the shock front (see
Appendix~\ref{app:shock}). A special-relativistic framework for shock
propagation is required here due to the very low baryon densities in
the surrounding of the pulsar (cf.~Equation~\eqref{eq:t_pulin}) and
the very high pressure of the pulsar nebula (Equation~\eqref{eq:p_n}),
which result in highly relativistic initial shock propagation
speeds. The shock front decelerates as it propagates outward into higher
density regions further out from the pulsar and eventually slows down
to non-relativistic or mildly relativistic velocities once it reaches the outer surface of the ejecta
material. As the shock front is evolved using
Equation~\eqref{eq:evoleqns_p2_shock}, the thickness
$\Delta_\text{sh}$ of the shocked ejecta layer and thus $R_\text{n}$
(cf.~Equation~\eqref{eq:evoleqns_p2_Rn}) is adjusted such that
pressure equilibrium between the nebula and the shocked ejecta layer
is maintained. For a mixture of ideal gas with $\Gamma_\text{ej}= 4/3$
and radiation, this results in the following constraint:
\begin{equation}
  R_\text{n} = R_\text{sh}\frac{E_\text{th,sh}}{E_\text{nth}}. \label{eq:press_equil}
\end{equation}
This conditions defines $\rmn{d}\Delta_\text{sh}/\rmn{d}t$ and couples
Equations~\eqref{eq:evoleqns_p2_Rn} and \eqref{eq:evoleqns_p2_Enth} to
make them implicit evolution equations for $R_\text{n}$ and $E_\text{nth}$.

\paragraph{Shock related energetics}
The shocked and unshocked ejecta material are assigned internal
energies $E_\text{th,sh}$ and $E_\text{th,ush}$, respectively, which are evolved
separately (cf.~Equations~\eqref{eq:evoleqns_p2_Esh} and
\eqref{eq:evoleqns_p2_Eush}). As the shock propagates across the
ejecta matter internal energy is transferred from the unshocked to the
shocked fluid part, which amounts to
\begin{equation}
  \rmn{d} E_\text{th,vol} =
  E_\text{th,ush}\frac{(R_\text{sh}+\rmn{d}R_\text{sh})^3}{R_\text{ej}^3-R_\text{sh}^3} \label{eq:Eth_vol}
\end{equation}
during a time $\rmn{d}t$. Furthermore, as the fluid is swept up by the
shock it is heated at a rate
\begin{equation}
  \ddt{E_\text{sh}} = \Delta \epsilon \ddt{m_\text{sh}},  \label{eq:Esh}
\end{equation}
where
\begin{equation}
  \Delta \epsilon =
  \frac{1}{\Gamma_\text{ej}+1}\frac{p_\text{n}}{\rho_\text{R}}\frac{2}{1+\left[
      1 + \frac{4\Gamma_\text{ej}}{(\Gamma_\text{ej} + 1)^2}\frac{p_\text{n}}{\rho_\text{R}c^2}
    \right]^{\frac{1}{2}}} \label{eq:del_e}
\end{equation}
is the jump in specific internal energy across the shock front (see
Appendix~\ref{app:shock}) and 
\begin{equation}
  \ddt{m_\text{sh}} = 4\pi
  R_\text{sh}^2\rho_\text{R}\frac{1}{\sqrt{1-\frac{v_\text{w}^2(R_\text{sh},\cdot)}{c^2}}}[v_\text{sh}
    - v_\text{w}(R_\text{sh},\cdot)] \label{eq:dm_sh}
\end{equation}
is the amount of material swept up by the shock per unit time. The
latter expression is obtained by noting that the
amount of material in a spherical volume changes by
\begin{equation}
  \rmn{d}m = \frac{4\pi r^2\rho}{\sqrt{1-\frac{v^2}{c^2}}}
    \rmn{d}r - \frac{4\pi r^2\rho v}{\sqrt{1-\frac{v^2}{c^2}}} \rmn{d}t 
\end{equation}
if the radius of the volume $r$ is altered by $\rmn{d}r$ during a time
$\rmn{d}t$, assuming that the fluid is radially moving outward with
velocity $v$.

For further reference (see Section~\ref{sec:ejecta_p3}), we also note that the shock deposits kinetic
energy in the shocked ejecta material. The jump $\Delta e_\text{kin}$ in specific kinetic
energy across the shock front as measured in the lab frame is given by
\begin{equation}
  \frac{\Delta e_\text{kin}}{c^2} =
    \left[1-\left(\frac{v_\text{L}^\text{Lab}}{c} \right)^2\right]^{-\frac{1}{2}}
  - \left[1-\left(\frac{v_\text{R}^\text{Lab}}{c} \right)^2\right]^{-\frac{1}{2}},\label{eq:delta_ekin}
\end{equation}
where $v_\text{R}^\text{Lab}(t) = v_\text{w}(R_\text{sh}(t),t)$ is the
velocity of the wind in the unshocked fluid part at the shock front as
measured in the lab frame. Furthermore, 
\begin{equation}
  v_\text{L}^\text{Lab}(t) = \frac{v_\text{L}(t) +
    v_\text{sh}(t)}{1 + v_\text{L}(t) v_\text{sh}(t)/c^2} \label{eq:v_L_Lab}
\end{equation}
is the velocity of the shocked fluid at the shock front as measured in
the lab frame, with $v_\text{L}$ being the velocity of the shocked
fluid at the shock front in the frame comoving with the shock front (see Appendix~\ref{app:shock}).

\paragraph{Thermal emission}
As in Phase I, we assume that the ejecta matter consists of a mixture
of ideal gas and photons. The corresponding fluid temperatures
$T_\text{sh}$ and $T_\text{ush}$ are given by
\begin{equation}
   T_\text{sh,ush}^4 +
  \frac{1}{\Gamma_\text{ej}-1}\frac{k_\text{B}}{m_\text{p}a}\bar{\rho}_\text{sh,ush}T_\text{sh,ush}
  - \frac{E_\text{th,sh,ush}}{aV_\text{sh,ush}} = 0, \label{eq:T_p2}
\end{equation}
where $\bar{\rho}_\text{sh}(t) =
m_\text{w}(R_\text{sh}(t),t)/V_\text{sh}(t)$ and
$\bar{\rho}_\text{ush}(t) = [ m_\text{w}(R_\text{ej}(t),t) -
m_\text{w}(R_\text{sh}(t),t)]/V_\text{ush}$ are the mean densities of
the shocked and unshocked ejecta, respectively, with associated
volumes $V_\text{sh} = 4\pi (R_\text{sh}^3 - R_\text{n}^3)/3$ and
$V_\text{ush} = 4\pi (R_\text{ej}^3 - R_\text{sh}^3)/3$. The
effective temperatures are then defined by
\begin{equation}
  T^4_\text{eff,sh,ush}(t) \simeq \frac{16}{3}
  \frac{T_\text{sh,ush}^4(t)}{\Delta\tau_\text{sh,ush}(t) + 1}, 
\end{equation}
with optical depths
\begin{equation}
  \Delta\tau_\text{sh}(t) = \kappa \bar{\rho}_\text{sh}(t)
  \Delta_\text{sh}(t) \label{eq:delta_tau_sh}
\end{equation}
and
\begin{equation}
  \Delta\tau_\text{ush}(t) = \kappa
  \int_{R_\text{sh}(t)}^{R_\text{ej}(t)}
  \rho_{\text{w},t}(r)\,\rmn{d}r. \label{eq:delta_tau_ush}
\end{equation}
These effective temperatures give rise to thermal radiation emitted
from the inner surface of the shocked ejecta toward the interior with
luminosity
\begin{equation}
  L_\text{rad,in}(t) = 4\pi R_\text{n}^2(t) \sigma
  T^4_\text{eff,sh}(t) \label{eq:Lradin_p2}
\end{equation}
and from the outer surface of the ejected material with luminosity
\begin{equation}
  L_\text{rad}(t) = 4\pi R_\text{ej}^2(t) \sigma
  T^4_\text{eff,ush}(t). \label{eq:Lrad_p2}
\end{equation}
While the radiation associated with $L_\text{rad}$ leaves the system
(cf.~Equation~\eqref{eq:evoleqns_p2_Eush}),
$L_\text{rad,in}$ is reabsorbed by the nebula, its energy is reprocessed
and eventually reemitted into the ejecta material
(cf.~Equations~\eqref{eq:evoleqns_p2_Esh} and \eqref{eq:evoleqns_p2_Enth}).

\paragraph{Shock break-out}
At the time when the shock front reaches the
outer ejecta surface a short transient signal can be produced in
addition to the thermal radiation from the ejecta. This signal can
carry the signature of particle acceleration at the shock front and it can
have a non-thermal spectrum reaching higher maximum energies than the
background thermal emission; therefore, it could be observable in the
hard X-ray and gamma-ray bands. Within the time-reversal scenario
\citep{Ciolfi2015a}, this transient signal might provide a convincing
explanation for the early precursors observed up to $\sim
10^2\,\text{s}$ prior to some SGRBs \citep{Troja2010}.

\subsection{Phase III: pulsar wind nebula}
\label{sec:sec_PWN_p3}
Phase III starts at $t_\text{shock,out}$, when the shock front reaches
the outer surface of the ejecta material,
$R_\text{sh}(t_\text{shock,out}) =
R_\text{ej}(t_\text{shock,out})$. At this time the shock has swept up
all the ejecta material of mass $M_\text{ej}\equiv M_\text{ej}(t_\text{pul,in})$ into a thin shell of thickness
$\Delta_\text{ej}=\Delta_\text{sh}(t_\text{shock,out})$ (cf.~Figure~\ref{fig:schematic}), which we
assume to be constant during the following
evolution. Therefore, we set
$\rmn{d}R_\text{n}/\rmn{d}t=\rmn{d}R_\text{ej}/\rmn{d}t$
for the evolution of the PWN radius $R_\text{n}$ in Phase III
(cf.~Equation~\eqref{eq:evoleqns_p3_Rn}). The following two
subsections detail the physical description of the PWN and the
ejecta shell in this phase.

\subsubsection{Physics of the pulsar wind nebula}
\label{sec:phaseIII_PWN}
\paragraph{Basic picture and assumptions} PWNe are
traditionally studied in the context of supernovae-born pulsars (see
\citealt{Gaensler2006}, \citealt{Kargaltsev2015}, and \citealt{Buehler2014} for reviews). Only
very recently, PWNe have also
been considered in the BNS merger scenario
\citep{Bucciantini2012,Metzger2014b}. We study such a system here
motivated by the fact that
long-lived NSs turn out to be very likely outcomes of BNS mergers
(cf.~Section~\ref{sec:introduction}) and by the fact that the baryonic ejecta
(cf.~Section~\ref{sec:phaseI}) naturally provide the conditions for
such a nebula to form (such
as providing the necessary confining envelope). Although the exact physical
conditions for a PWN in a BNS merger scenario are different from those
in a supernova event, it is conceivable that qualitatively similar
morphologies and dynamics emerge. 
For Phase III, we develop a physical description for a PWN in the
context of BNS mergers inspired by studies of
supernovae and radiative processes in active galactic nuclei. While
spectra of ordinary PWNe from radio to gamma-ray
energies can be explained in terms of synchrotron and inverse Compton
emission (for the best studied case, the Crab nebula, see, e.g.,
\citealt{Kennel1984b,Atoyan1996,Volpi2008,Olmi2014,Porth2014}), the
BNS merger scenario requires a more
detailed treatment of radiative processes due to the
presence of the strong photon field from the confining hot ejecta
material. In order to achieve this and in order to avoid performing
expensive MHD simulations, we neglect spatial variations and adopt a
`one-zone' model for the PWN, in which the nebula is considered to be
homogeneous and isotropic. The resulting  description of a PWN we propose for
Phase III is much more detailed than in Phase II for two reasons.
First, Phase III is less dynamical and a quasi-stationary description of
the PWN can be adopted that facilitates a detailed treatment of radiative
processes. Second, as the ejecta material surrounding the PWN
becomes transparent to the nebula radiation at some point, radiation
from the nebula itself is not reprocessed and thermalized anymore and
its spectrum becomes directly observable. We therefore employ a
formalism that predicts the nebula spectrum in a
self-consistent way.

The newly formed pulsar loses rotational energy at a rate
$L_\text{sd}$ given by Equation~\eqref{eq:Lsd} through a highly
relativistic magnetized particle outflow, referred to as a pulsar wind
(cf.~Section~\ref{sec:pulsar_properties}). As in Phase II
(cf.~Section~\ref{sec:phaseII_PWN}), we also assume that thermal
radiation from the pulsar surface with luminosity
$L_\text{rad,pul}$ (cf.~Equation~\eqref{eq:Lrad_pul}) is deposited in
this pulsar wind; this can represent a
significant energy contribution at early times when the NS is still
very hot. The pulsar wind inflates a PWN behind the less rapidly
expanding ejecta material, a bubble of radiation and charged particles
(mainly electrons and positrons), which is separated from the pulsar wind
by a termination shock at a distance \citep{Gaensler2006}
\begin{equation}
  R_\text{TS} = \sqrt{ \frac{L_\text{sd}}{4\pi  \xi c p_\text{n}} } \ll R_\text{n}
\end{equation}
from the pulsar, where the ram pressure of the pulsar wind equals the
intrinsic pressure
$p_\text{n}$ of the nebula; $\xi$ denotes the fraction of a sphere
covered by the wind.

Magnetosphere models predict that the pulsar wind leaving the
magnetosphere at the light cylinder $R_\text{LC} =
cP_\text{pul}/2\pi$ is highly magnetized, $\sigma_\text{m} \gg 1$
(e.g., \citealt{Arons2012}). Here, $\sigma_\text{m}$ denotes the
magnetization parameter,
defined as the ratio of Poynting flux to particle energy flux (e.g.,
\citealt{Kennel1984a}). One-dimensional and two-dimensional
axisymmetric MHD models only reproduce the observed properties of the Crab
nebula well if $\sigma_\text{m}\ll 1$ just upstream of the termination
shock (e.g.,
\citealt{Kennel1984a,Kennel1984b,Atoyan1996,Volpi2008,Olmi2014}). Dissipating
the Poynting flux into kinetic
energy of the particle flow between the light cylinder and the wind
termination shock is known as the $\sigma$-problem. Recent
three-dimensional MHD simulations, however, suggest a solution the
$\sigma$-problem \citep{Mizuno2011,Porth2014}. The simulations by
\citet{Porth2014} are able to reproduce the
morphology of the Crab nebula even for stronger magnetizations of up
to $\sigma_\text{m}\simeq \text{few}$, thanks to magnetic dissipation
inside the PWN. Enhanced turbulence downstream of the termination
shock in the 3D simulations efficiently dissipates magnetic energy,
such that the magnetic energy of the nebula is only a very small
fraction of the internal energy despite a high magnetization of the
pulsar wind at the termination shock. This is in agreement with previous
observations and spectral modeling that infer the
magnetic energy of the Crab nebula to be less than $\approx 3\%$ of the
total energy budget (\citealt{Buehler2014}, \citealt{Porth2014}, and
references therein). In our model of the PWN, we control the level of magnetization of the
nebula by introducing a factor $\eta_{B_\text{n}}$, which specifies
the amount of inflowing magnetic energy per time in terms of the total
power of the pulsar wind: $\dot{E}_B = \eta_{B_\text{n}}
[L_\text{sd}(t) + L_\text{rad,pul}(t)]$
(cf.~Equations~\eqref{eq:evoleqns_p2_EB} and \eqref{eq:evoleqns_p3_EB}). The magnetic energy $E_B$
then defines an average magnetic field strength $B_\text{n}$ of the
nebula through Equation~\eqref{eq:B_n}.

At the wind termination shock the wind plasma is
decelerated and heated, and efficient conversion of flow energy into
particle acceleration is thought to take place. The accelerated particles, which
are characterized by non-thermal energy spectra, are then advected
with the flow while cooling down due to synchrotron
emission. Observations of the Crab nebula synchrotron emission
indicate that the conversion of pulsar wind energy into accelerated particles
is $\gtrsim 10\%$ (\citealt{Kennel1984a,Buehler2014,Olmi2015}; we
henceforth denote this efficiency by $\eta_\text{TS}$). While diffusive shock acceleration at the
termination shock is now
thought to be rather unlikely to occur in the Crab nebula and other
PWNe \citep{Arons2012,Buehler2014}, enhanced turbulence and magnetic
dissipation downstream of the termination
shock (as found in recent 3D MHD simulations) might provide the required non-thermal particle
acceleration site. While the processes of particle acceleration still
remain unclear, assuming that the pulsar wind consists of electrons
and positrons, and that these particles are reaccelerated into a
power-law spectrum at the termination shock, 1D, 2D, and 3D MHD models
of the Crab nebula have been able to reproduce
the observed photon spectra very well
\citep{Kennel1984b,Atoyan1996,Volpi2008,Camus2009,Olmi2014,Porth2014,Olmi2015}.

Following this approach we consider a pulsar wind consisting of
electrons and positrons and assume that these particles are injected into the nebula with a
power-law spectrum, although
more complex injection spectra can easily be accommodated by our
model. In view of our `one-zone' description of
the nebula, we assume that electrons and positrons are continuously
injected uniformly throughout the nebula, instead of
being injected at the termination shock and then being advected with
the flow. We specify the injection of particles by the following
dimensionless compactness parameter (in analogy to
\citealt{Guilbert1983,Svensson1987,Lightman1987}):
\begin{eqnarray}
  l_\text{e}(t) &\equiv& \frac{\sigma_\text{T}}{m_\text{e}c^3
    R_\text{n}(t)} \eta_\text{TS}[L_\text{sd}(t) + L_\text{rad,pul}(t)] \nonumber\\
  &=&\frac{4\pi \sigma_\text{T} R_\text{n}^2(t)}{3c}
  \int_{1}^{\gamma_\text{max}} Q(\gamma,t) (\gamma - 1)
  \,\rmn{d}\gamma, \label{eq:le}
\end{eqnarray}
where $Q(\gamma,t) = Q_0(t) \gamma^{-\Gamma_\text{e}}$ is the number of
particles injected per unit time per unit normalized energy $\gamma =
\epsilon/m_\text{e}c^2$, where $\epsilon$ denotes the particle energy, and per unit volume
of the nebula. The parameter $\eta_\text{TS}$ defines the
aforementioned efficiency of converting pulsar wind power into random
kinetic energy of accelerated particles. The power-law injection parameters
$\eta_\text{TS}$, $\gamma_\text{max}$, and $\Gamma_\text{e}$ are model
input parameters, which, in the case of observed PWNe such as the Crab,
are usually determined by comparing simulated emission with observational data.

In contrast to ordinary PWNe, where intrinsic photon sources inherent
to the pulsar-nebula system are absent and only background photons
from the cosmic microwave background (CMB) and potentially from local
dust and starlight are `injected' into the nebula, we need to take
additional photon sources into account in our description of the PWN
that generate a strong photon field. The hot shock-heated ejecta
matter confining the PWN injects thermal photons with luminosity $L_\text{rad,in}$
(cf.~Section~\ref{sec:ejecta_p3}, Equation~\eqref{eq:Lradin_phaseIII}).
For completeness, we also include a thermal input spectrum $L_\text{CMB}$
from the CMB (see below),
once the ejecta shell has become optically thin. We
specify the resulting photon injection into the nebula in terms of the
following dimensionless compactness parameter (in analogy to \citealt{Lightman1987}):
\begin{eqnarray}
l_\text{ph}(t) &\equiv& \frac{\sigma_\text{T}}{m_\text{e}c^3
    R_\text{n}(t)} [L_\text{rad,in}(t) + f_\text{ej}(t)L_\text{CMB}(t)] \nonumber\\
&=&\frac{4\pi \sigma_\text{T} R_\text{n}^2(t)}{3c}
  \int_{x_\text{min}}^{x_\text{max}} \dot{n}_0(x,t) x \,\rmn{d}x, \label{eq:lph}
\end{eqnarray}
where $\dot{n}_0$ denotes the combined number of photons injected per
unit time per unit dimensionless energy $x=h\nu/m_\text{e}c^2$ per
unit volume of the PWN by the aforementioned sources
(cf.~Equation~\eqref{eq:nd0}). Here, $h$ denotes the Planck
constant and $\nu$ is the photon frequency. Photon energies are
assumed to range between $x_\text{min}$ and $x_\text{max}$, which we
define in Section~\ref{sec:PWN_numerical}. The function $f_\text{ej}$
is designed to `switch on' the (subdominant) CMB contribution once the
ejecta material becomes optically thin (cf.~Equation~\eqref{eq:f_ej}).

Radiative processes in the PWN reprocess the injected particles and
photons and determine the radiative losses of the nebula that are
transferred to the surrounding ejecta (denoted by $\rmn{d}E_\text{PWN}/\rmn{d}t$ in
Equation~\eqref{eq:evoleqns_p3_Eth}) and the associated emergent
spectrum. In particular, they determine how spin-down energy of
the pulsar is transmitted to the ejecta material to be radiated away
from the system eventually. In contrast to ordinary PWNe, the photon
field in our case is typically very strong, such that, expressed in
terms of the compactness parameters defined above, $l_\text{ph}$ can
become comparable to $l_\text{e}$ or even larger. The photon and
particle spectra inside the nebula thus become highly coupled and the
computations become intrinsically non-linear. In order to determine
those spectra in a self-consistent way under the combined effects of
synchrotron losses, (inverse) Compton scattering off thermal and
non-thermal particles, pair production and
annihilation, and photon escape, we employ, extend and modify a
formalism developed for
pair plasmas in compact sources such as active galactic nuclei \citep{Lightman1987}.

\paragraph{Balance equations}
Given the setup described above, the physical processes determining the
PWN properties and emergent spectra are remarkably similar to those in
pair plasmas of active galactic nuclei. Theoretical models to compute detailed
emergent spectra for active galactic nuclei have been developed by many authors in the past (e.g.,
\citealt{Lightman1987,Coppi1992,Belmont2008}). Consistent with our
assumptions, we adopt the approach of \citet{Lightman1987} (henceforth
LZ87), but extend and modify it to, e.g., include synchrotron losses. We
shall briefly outline the formulation of the equations here with emphasis on the
modifications and refer the reader to LZ87 for more details on all
other aspects.

According to our `one-zone' model we assume that particles and photons are uniformly injected
throughout the spherical volume of the PWN with radius $R_\text{n}$ and
associated compactness parameters $l_\text{e}$ and $l_\text{ph}$ (Equations~\eqref{eq:le} and \eqref{eq:lph}).
At a time $t$ and assuming quasi-stationarity, the number densities
per unit energy of photons, $n(x)$, and non-thermal particles,
$N(\gamma)$, are determined by the following set of highly non-linear, coupled
integro-differential equations:
\begin{eqnarray}
  0 &=& \dot{n}_0 + \dot{n}_\text{A} + \dot{n}_\text{C}^\text{NT} +
  \dot{n}_\text{C}^\text{T} + \dot{n}_\text{syn} \nonumber\\
   &&\qquad \qquad \qquad - \frac{c}{R_\text{n}} n
  (\Delta\tau_\text{C}^\text{NT} + \Delta\tau_{\gamma\gamma}) -
  \dot{n}_\text{esc}, \label{eq:LightmanI}\\
  0 &=& Q(\gamma) + P(\gamma) + \dot{N}_\text{C,syn}(\gamma). \label{eq:LightmanII}
\end{eqnarray}

Photons at energy $x$ are produced at rates $\dot{n}_0$,
$\dot{n}_\text{A}$, $\dot{n}_\text{C}^\text{NT}$,
$\dot{n}_\text{C}^\text{T}$, and $\dot{n}_\text{syn}$ via injection
(cf.~Equation~\eqref{eq:nd0}), $\text{e}^\pm$-pair annihilation,
Compton scattering off non-thermal electrons and positrons
(cf.~Equation~(9) in LZ87), Compton scattering off thermal electrons
(cf.~Equations~(22) and (23) in LZ87; see below), and synchrotron cooling of
particles (cf.~Equation~\eqref{eq:ndsyn}), respectively. For the pair
annihilation term $\dot{n}_\text{A}$ we use
Equations~(7),\,(8),\,(11),\,(15)--(18) of
\citet{Svensson1983}. Photons at energy $x$ are lost due to Compton
scattering off non-thermal particles with optical depth
$\Delta\tau_\text{C}^\text{NT}$ (cf.~Equation~(10) in LZ87), due to $\text{e}^\pm$-pair creation with
optical depth $\Delta\tau_{\gamma\gamma}$ (cf.~Equation~(11) in LZ87), and by escaping from the nebula at
a rate $\dot{n}_\text{esc}$ (see Equation~(21) of LZ87). Additionally, photons at energy $x$ are lost
via Compton scattering off thermal particles (see below), which is accounted for
in $\dot{n}^\text{T}_\text{C}$. Photons not `absorbed' by one of these
processes are still impeded from escaping the nebula by Thomson
scattering off thermal particles (see below) with scattering depth
$\Delta\tau_\text{T}$ (cf.~Equation~(20) in LZ87).

The rate of change of particles at
Lorentz factor $\gamma$ is given by injection into the nebula $Q$
(cf.~Equation~\eqref{eq:le}), $\text{e}^\pm$-pair creation $P$ (cf.~Equation~(13) in
LZ87), and non-thermal Compton scattering and synchrotron losses, which
are denoted by $\dot{N}_\text{C,syn}$ (see below). Particles do not
escape from the nebula, i.e., they are
assumed to be trapped, e.g., by the magnetic field $B_\text{n}$ of the nebula. Once cooled
down to $\gamma\sim 1$ and before annihilating, particles are assumed to thermalize and to
form a distinct thermal population described by a Maxwell-Boltzmann
distribution with dimensionless temperature $\theta_\text{e}\equiv
k_\text{B}T_\text{e}/m_\text{e}c^2\ll 1$. The number density of this
thermal population is determined in terms of the solution to
Equations~\eqref{eq:LightmanI} and \eqref{eq:LightmanII} by the
requirement that pairs must be destroyed at the same
rate as they are created in steady state (cf.~Equation~(18) in LZ87). The temperature
of this population can be determined self-consistently by not allowing
any net energy transfer between particles and photons via thermal
Compton scattering (cf.~Equations~(27) and (28) in LZ87).

In our implementation, the photon injection term is written as
\begin{equation}
  \dot{n}_0 = \dot{n}_{0,\text{ej}} +
  f_\text{ej}\dot{n}_{0,\text{CMB}},  \label{eq:nd0}
\end{equation}
where
\begin{equation}
  \dot{n}_{0,\text{ej}}(x,t) = \frac{6\pi}{R_\text{n}(t)}
  \frac{m_\text{e}^3c^4}{h^3}\frac{x^2}{\exp\left(\frac{xm_\text{e}c^2}{k_\text{B}T_\text{eff}(t)}\right)
  - 1} \label{eq:nd0bb}
\end{equation}
represents the injection of thermal photons from the ejecta material
with effective temperature $T_\text{eff}=T_\text{eff,com}/(\zeta\gamma_\text{ej})^{1/4}$
(cf.~Section~\ref{sec:ejecta_p3}). The thermal spectrum $\dot{n}_{0,\text{CMB}}$ of the CMB
is defined in the same way with an effective temperature of
$T_\text{eff,CMB}=2.725\,\text{K}$ and it is `switched on' by the
function $f_\text{ej}$ once the ejecta material becomes optically thin
(cf.~Equation~\eqref{eq:f_ej}).

In contrast to LZ87, we additionally include effects of synchrotron
radiation (as do \citealt{Coppi1992} and \citealt{Belmont2008}). Here, we
briefly outline the way we include those effects via the terms
$\dot{N}_\text{C,syn}$ and $\dot{n}_\text{syn}$. In deriving
Equation~\eqref{eq:LightmanII}, we assumed that the particle
distribution $N(\gamma)$ can be described by the continuity equation
\begin{equation}
  \frac{\partial N}{\partial t} + \frac{\partial }{\partial
    \gamma}(\dot{\gamma}N) = P(\gamma) + Q(\gamma), \label{eq:N_continuity}
\end{equation}
with $\partial N/\partial t = 0$ in steady state and thus
$\dot{N}_\text{C,syn}=-\partial (\dot{\gamma}N)/\partial \gamma$. Equation~\eqref{eq:N_continuity} is an accurate
description in the case of synchrotron emission and Compton scattering
excluding the Klein-Nishina limit (as assumed here for simplicity;
\citealt{Blumenthal1970}). Assuming steady state,
Equation~\eqref{eq:N_continuity} can be integrated to give
\begin{equation}
  N(\gamma) = -\dot{\gamma}_\text{C,syn}^{-1}(\gamma) \int_{\gamma}^{\gamma_\text{max}} [P(\gamma') +
  Q(\gamma')]\,\rmn{d}\gamma', \label{eq:N_gam}
\end{equation}
where 
\begin{equation}
  \dot{\gamma}_\text{C,syn}(\gamma)=\dot{\gamma}_\text{C}(\gamma)
+ \dot{\gamma}_\text{syn}(\gamma) \label{eq:gam_dot}
\end{equation}
is the combined particle cooling
rate due to Compton scattering (excluding the Klein-Nishina limit) and synchrotron emission, with
\begin{equation}
  \dot{\gamma}_\text{C}(\gamma) = -\sigma_\text{T} c
  \left(\frac{4}{3}\gamma^2 -1\right) \int_0^{3/4\gamma} n(x)x\,\rmn{d}x
\end{equation}
(cf.~Equation~(7) in LZ87) and
\begin{equation}
  \dot{\gamma}_\text{syn}(\gamma) =
  -\frac{\sqrt{3}e^3B_\text{n}}{hm_\text{e}c^2}
  \int_{x_\text{min}}^{x_\text{max}} \mathcal{R}(x/x_\text{c})\,\rmn{d}x \label{eq:gammadot_syn}
\end{equation}
(cf.~Appendix~\ref{app:synchrotron},
Equation~\eqref{eq:app_gammadot_syn}). The synchrotron losses
\eqref{eq:gammadot_syn} result in a photon source term
\begin{equation}
  \dot{n}_\text{syn}(x)=\frac{\sqrt{3}e^3B_\text{n}}{hm_\text{e}c^2}
  \frac{1}{x}\int_1^{\gamma_\text{max}} N(\gamma)
  \mathcal{R}(x/x_\text{c})\,\rmn{d}\gamma \label{eq:ndsyn}
\end{equation}
(cf.~Appendix~\ref{app:synchrotron},
Equation~\eqref{eq:app_ndsyn}). We note that for the stationarity
assumption leading to Equation~\eqref{eq:N_gam} to hold in our
evolution scheme (Equations~\eqref{eq:evoleqns_p3_vej}--\eqref{eq:evoleqns_p3_EB}), the timescale for equilibration
of the particle distribution given by the total cooling timescale
$|\gamma/\dot{\gamma}_\text{C,syn}|$ has to be much smaller than any
other evolution timescale for all energies $\gamma$. We therefore
monitor this quantity to check the validity of the stationarity
assumption during the numerical evolution (see
Section~\ref{sec:timescales}). Finally, we note that effects of
synchrotron self-absorption have not been considered so far and that
they can be neglected for our purposes. In order to
check the validity of this assumption, we monitor the optical depth to synchrotron
self-absorption, which can be approximated by
\begin{align}
  \Delta\tau_\text{syn}(x) =& \frac{\sqrt{3}e^3h^2 B_\text{n}
    R_\text{n}}{8\pi
    m_\text{e}^4c^6}\frac{1}{x^2} \label{eq:tau_syn}\\
   &\times\int_1^{\gamma_\text{max}}
  \mskip-20mu N(\gamma)\left[\frac{\partial}{\partial\gamma}\mathcal{R}(x/x_\text{c}) +
    f(\gamma)\mathcal{R}(x/x_\text{c}) \right]\rmn{d}\gamma \nonumber
\end{align}
(cf.~Appendix~\ref{app:synchrotron},
Equation~\eqref{eq:app_tau_syn}). Typically, the nebula is optically
thin to synchrotron self-absorption at energies of interest (see Section~\ref{sec:synchrotron_monitor}).

Solving the coupled
set of Equations~\eqref{eq:LightmanI} and \eqref{eq:LightmanII}
numerically at time $t$ yields a
self-consistent set of spectra \{$n(x)$, $N(\gamma)$,
$\dot{n}_\text{A}(x)$, $\dot{n}_\text{C}^\text{NT}(x)$,
$\dot{n}_\text{C}^\text{T}(x)$, $\dot{n}_\text{syn}(x)$, $\Delta\tau_\text{C}^\text{NT}(x)$,
$\Delta\tau_{\gamma\gamma}(x)$, $\Delta\tau_\text{T}$, $\dot{n}_\text{esc}(x)$, $P(\gamma)$,
$\dot{N}_\text{C,syn}(\gamma)$\}. It is important to point out that
with the modifications described above, these equations still conserve
energy (see Appendix~\ref{app:energy_conservation}).

The internal energy $E_\text{nth}$ of the nebula in terms of the
solution to Equations~\eqref{eq:LightmanI} and \eqref{eq:LightmanII}
can be written as
\begin{eqnarray}
  E_\text{nth}(t) &=& \frac{4}{3}\pi
                      R_\text{n}(t)^3m_\text{e}c^2\left[
                      \int_{x_\text{min}}^{x_\text{max}} n(x,t)
                      x\,\rmn{d}x \right.\nonumber\\
                      &&+ \left. \int_1^{\gamma_\text{max}}N(\gamma)(\gamma -1)\,\rmn{d}\gamma\right],
\end{eqnarray}
which is typically dominated by the photon contribution (first term on
the right-hand side).
The pressure of the nebula is then given by (cf.~also
Equation~\eqref{eq:p_n})
\begin{equation}
  p_\text{n}(t) = \frac{E_\text{nth}(t)}{4\pi R_\text{n}^3(t)} +
  \frac{B_\text{n}^2}{8\pi}. \label{eq:p_n_pIII}
\end{equation}
Moreover, the
luminosity of escaping radiation from the nebula is given by
\begin{equation}
  L_\text{PWN}(x,t) = \frac{4}{3}\pi R_\text{n}(t)^3m_\text{e}c^2
  \dot{n}_\text{esc}(x,t) x, \label{eq:L_PWN}
\end{equation}
and thus we arrive at the desired expression for the PWN source term
in Equation~\eqref{eq:evoleqns_p3_Eth}:
\begin{equation}
  \ddt{E_\text{PWN}} \equiv L_\text{PWN}(t) =
  \int_{x_\text{min}}^{x_\text{max}} L_\text{PWN}(x,t)\,\rmn{d}x. \label{eq:dEPWN_dt}
\end{equation}

\subsubsection{Ejecta properties and emergent radiation}
\label{sec:ejecta_p3}

\paragraph{Shell kinematics} The initial speed $v_\text{ej}(t_\text{shock,out})$ of the ejecta
shell in Phase III is inferred from its kinetic
energy $E_\text{kin}$ at $t_\text{shock,out}$:
\begin{equation}
  v_\text{ej}(t_\text{shock,out}) = c\left[ 1 - \left( 1+
      \frac{E_\text{kin}(t_\text{shock,out})}{M_\text{ej}c^2}\right)^{-2}\right]^{\frac{1}{2}}. \label{eq:vej_tso}
\end{equation}
The total kinetic energy $E_\text{kin}(t_\text{shock,out}) = E_\text{kin,w} +
E_\text{sh,tot}$ of the ejecta matter is the sum of the kinetic energy
of the original wind material $E_\text{kin,w}$ (cf.~Equation~\eqref{eq:Ekin_w})
and the total amount of kinetic energy deposited into the shocked
material by the shock,
\begin{equation}
  E_\text{sh,tot} = \int_{t_\text{pul,in}}^{t_\text{shock,out}} \Delta
  e_\text{kin} \ddt{m_\text{sh}}\,\rmn{d}t
\end{equation}
(cf.~Equations~\eqref{eq:dm_sh} and \eqref{eq:delta_ekin}). For
typical sub-relativistic or at most mildly relativistic speeds $v_\text{ej,in}$
of the outer ejecta front in Phase I and II (cf.~Equation~\eqref{eq:phaseI_Rej}), the initial speed of
the ejecta shell in Phase III is also at most mildly
relativistic. 

However, further acceleration
according to (cf.~Appendix~\ref{app:frames}; Equation~\eqref{eq:evoleqns_p3_vej})
\begin{equation}
  \ddt{v_\text{ej}} = \gamma_\text{ej}^{-3}(t) \alpha_\text{ej}(t)
  \equiv a_\text{ej}(t) \label{eq:dvej_dt}
\end{equation}
can be significant if the nebula pressure $p_\text{n}$
(Equation~\eqref{eq:p_n_pIII}) is high. Here,
$\gamma_\text{ej}=(1-v_\text{ej}^2/c^2)^{-1/2}$ denotes the Lorentz factor
of the ejecta shell and $\alpha_\text{ej} = 4\pi R_\text{n}^2
p_\text{n} / M_\text{ej}$ is the acceleration of the ejecta shell in
the frame comoving with the shell. In order to define a comoving frame
we employ the fact that such a freely
expanding, spherically symmetric shell at a time $t$ can be described as a thin ring of
finite extent in a Milne universe. The transformation between comoving
and lab frame is thus achieved via the Milne universe metric. For
further details, we refer to Appendix~\ref{app:frames}.

\paragraph{Thermal emission}
As in Phase I and II, the ejecta matter consists of a mixture of
photons and ideal gas with adiabatic index $\Gamma_\text{ej} = 4/3$,
the temperature of which in the comoving frame, $T_\text{com}$, can be found by solving
\begin{equation}
  T_\text{com}^4 +
  \frac{1}{\Gamma_\text{ej}-1}\frac{k_\text{B}}{m_\text{p}a}\rho_\text{ej}T_\text{com}
  - \frac{E_\text{th}}{aV_\text{ej}} = 0. \label{eq:T_com}
\end{equation}
Henceforth $q_\text{com}$ refers to the value of a quantity $q$ as measured in the
frame comoving with the ejecta shell. In Equation~\eqref{eq:T_com} we
have used the fact that $E_\text{th}/V_\text{ej} =
E_\text{th,com}/V_\text{ej,com}$
(cf.~Appendix~\ref{app:frames}). Furthermore, $\rho_\text{ej} =
M_\text{ej}/V_\text{ej,com} = M_\text{ej}/\zeta V_\text{ej}$ is the
density of the ejecta shell, where $V_\text{ej} = (4/3)\pi
(R_\text{ej}^3-R_\text{n}^3)$ is its volume in the lab frame and
\begin{equation}
  \zeta = \frac{3}{2}\frac{\gamma_\text{ej}^2\beta_\text{ej} -
    \arctanh \beta_\text{ej}}{\gamma_\text{ej}^3\beta_\text{ej}^3},
\end{equation}
with $\beta_\text{ej}=v_\text{ej}/c$
(cf.~Appendix~\ref{app:frames}). The associated effective temperature
in the comoving frame is given by
\begin{equation}
  T_\text{eff,com}^4(t) \simeq
  \frac{16}{3}\frac{T_\text{com}^4(t)}{\Delta\tau_\text{ej,com}(t) +
    1}, \label{eq:Teff_com}
\end{equation}
where
\begin{equation}
  \Delta\tau_\text{ej,com}(t) = \kappa\rho_\text{ej}
  (t)\Delta_\text{ej}(t)\gamma_\text{ej}(t), \label{eq:delta_tau_ej_com}
\end{equation}
which gives rise to thermal emission from the outer and inner surface of
the ejecta shell with luminosities
\begin{equation}
  L_\text{rad}(t) = 4\pi R_\text{ej}^2(t)
  \frac{f_\text{beam}(t)}{\zeta(t)\gamma_\text{ej}(t)} \sigma
  T_\text{eff,com}^4(t) \label{eq:Lrad_phaseIII}
\end{equation}
and
\begin{equation}
   L_\text{rad,in}(t) = 4\pi R_\text{n}^2(t)
  \frac{1}{\zeta(t)\gamma_\text{ej}(t)} \sigma T_\text{eff,com}^4(t), \label{eq:Lradin_phaseIII}
\end{equation}
respectively (cf.~Appendix~\ref{app:frames}; Equation~\eqref{eq:evoleqns_p3_Eth}). In
Equation~\eqref{eq:Lrad_phaseIII}, we have again included a factor
$f_\text{beam} = 1-v_\text{ej}^2/c^2$ to account for relativistic
beaming (cf.~also Section~\ref{sec:phaseIII_PWN}). This concludes the
discussion of source terms in the main evolution equations
\eqref{eq:evoleqns_p1_Rej}--\eqref{eq:evoleqns_p3_EB} of our model.

\subsubsection{Transition to the optically thin regime}
\label{sec:f_ej}

Once $\Delta\tau_\text{ej,com}$
(cf.~Equation~\eqref{eq:delta_tau_ej_com}) approaches unity, the
ejecta material becomes transparent to radiation from the nebula. This
non-thermal
radiation is not absorbed by the ejecta material anymore and becomes
directly observable. In order to ensure a smooth
transition between the optically thick and thin regimes, we employ an
auxiliary function
\begin{equation}
  f_\text{ej}(t) = \left\{ \begin{array}{ll}
 0 & \text{if}\; \Delta\tau_\text{ej}(t) >\Delta\tau_\text{ej,thres}\\
 1 - \left(\frac{\Delta\tau_\text{ej}(t)}{\Delta\tau_\text{ej,thres}}
                             \right)^b & \text{if}\; \Delta\tau_\text{ej}(t)\le\Delta\tau_\text{ej,thres}\\
  \end{array} \right.\label{eq:f_ej}
\end{equation}
to gradually `switch on or off' terms in the evolution equations
during this transition. Here, 
\begin{equation}
  \Delta\tau_\text{ej} = \Delta\tau_\text{ej,com}
\sqrt{ \frac{1-v_\text{ej}/c}{1+v_\text{ej}/c} } \label{eq:delta_tau_ej_lab}
\end{equation}
approximates the optical depth of the ejecta material as seen from the lab frame
(cf.~\citealt{Abramowicz1991}), with $\Delta\tau_\text{ej,com}$
defined in Equation~\eqref{eq:delta_tau_ej_com}. We note that this
transition function and its parameters $b$ and
$\Delta\tau_\text{ej,thres}$ are somewhat arbitrary and chosen in such
a way that they do not influence the numerical evolution significantly other than
guaranteeing a smooth transition from the optically thick to the
optically thin regime. Typically, $\Delta\tau_\text{ej,thres}\simeq \text{few}$ and
$b >1$.

In particular, as the ejecta material does not absorb nebula radiation
anymore, the source term $\rmn{d}E_\text{PWN}/\rmn{d}t$
(cf.~Equation~\eqref{eq:dEPWN_dt}) needs to be removed from the
corresponding evolution equation
(cf.~Equation~\eqref{eq:evoleqns_p3_Eth}). Instead, the non-thermal
emission from the nebula
\begin{equation}
  L_\text{rad,nth}(t) = f_\text{ej}(t)\ddt{E_\text{PWN}} \label{eq:Lrad_nth}
\end{equation}
and its associated spectrum given by Equation~\eqref{eq:L_PWN} become directly
observable. Consequently, the ejecta matter
also becomes optically thin to photons form the CMB,
which are now able to diffuse into the nebula
(cf.~Equations~\eqref{eq:lph} and \eqref{eq:nd0}).

\subsection{Collapse to a black hole}
\label{sec:collapse}

Our model applies to hypermassive, supramassive, and stable remnant
NSs. If the NS is hypermassive at birth, the expected lifetime is of
the order of $t_\text{dr}\lesssim 1\,\text{s}$, unless it migrates to
the supramassive regime through substantial mass ejection on shorter
timescales. If the NS is supramassive, it can survive on much
longer spin-down timescales, although (magneto-)hydrodynamic
instabilities can cause an earlier collapse to a black hole. In our
model, the time of collapse $t=t_\text{coll}$ to a black hole is an input parameter that
can be adjusted to cover all possible scenarios.

If the NS collapses to a black hole during Phase I, we keep evolving
Equations~\eqref{eq:evoleqns_p1_Rej} and \eqref{eq:evoleqns_p1_Eth},
setting $L_\text{EM}$ and $\rmn{d}E_\text{th,NS}/\rmn{d}t$ to zero. As the
collapse proceeds on the dynamical timescale, which is of the order of
milliseconds, we consider the NS collapse at $t\sim
t_\text{dr}\lesssim 1\,\text{s}$ to be instantaneous as far
as the numerical evolution is concerned. The time of collapse is
parametrized by $f_\text{coll,PI}$ in units of $t_\text{dr}$,
$t_\text{coll}=f_\text{coll,PI}t_\text{dr}$. We also note that the wind
model (cf.~Section~\ref{sec:wind_model}) has to be adjusted by setting
$\dot{M}(t')$ to zero for $t'>t_\text{coll}$ in Equation~\eqref{eq:m_w_rt}.
The resulting EM emission from the system will be predominantly
thermal and it will reflect the gradual depletion of the energy reservoir of
the ejecta material acquired up to the time of collapse (see also
Paper II).

In what follows, we consider a collapse occurring in Phase
III.\footnote{Here we do not discuss the possibility of a collapse during
shock propagation (Phase II), as this phase is typically very short
compared to Phase I and III, and a collapse is thus unlikely to occur.}
In the case that the
observed prompt $\gamma$-ray emission of a SGRB is associated to this
collapse as in the recently proposed time-reversal scenario
\citep{Ciolfi2015a,Ciolfi2015b}, the resulting lightcurves and spectra
of our model after $t=t_\text{coll}$ correspond to the observed afterglow
radiation of the SGRB.

We parametrize $t_\text{coll}$ in terms of the spin-down timescale
$t_\text{coll} = f_\text{coll} t_\text{sd}$. Scenarios for
a wide range of values for $f_\text{coll}$ are explored in the companion paper (Paper
II). The collapse of the NS proceeds on the dynamical timescale, which is
of the order of milliseconds and it can thus be considered instantaneous at
times $t_\text{coll} > t_\text{pul,in}\gtrsim 1\text{s}$. At times $t
\lesssim \text{few}\times t_\text{sd}$, the typical cooling
timescales $|\gamma/\dot{\gamma}_\text{C,syn}|$ of the non-thermal particles in the nebula
(cf.~Equation~\eqref{eq:gam_dot}) are orders of magnitude smaller than
any other evolution timescale of our model thanks to the relatively
strong magnetic field
(cf.~Equation~\eqref{eq:gammadot_syn}; see also
Section~\ref{sec:timescales} and Section~4.2 of Paper II). Furthermore, due to this
efficient cooling their total number
density $N_\text{tot} =
\int N(\gamma,t)\,\rmn{d}\gamma\ll N_\text{th,tot}$ is orders of
magnitude smaller than the total number density $N_\text{th,tot}$ of thermalized
particles in the nebula. We can therefore assume that the non-thermal
particle population is instantaneously thermalized at $t=t_\text{coll}$. Furthermore, noting
that the optical depth to Thomson scattering $\Delta\tau_\text{T}$ is given
by $\Delta\tau_\text{T}=\sigma_\text{T}R_\text{n}N_\text{th,tot}$, we can
conclude that $\Delta\tau_\text{T}$ is not affected by this instantaneous
thermalization. Pair annihilation becomes increasingly
unlikely as the nebula further expands, and can partially be
compensated by pair creation induced by photons in the high energy tail of the
photon spectrum above the pair creation threshold. As a
consequence, the total number of
particles is roughly conserved, and the evolution of the Thomson scattering
depth for $t>t_\text{coll}$ is approximately given by
\begin{equation}
  \Delta\tau_\text{T}(t) = \Delta\tau_\text{T}(t_\text{coll}) \left[\frac{R_\text{n}(t_\text{coll})}{R_\text{n}(t)}\right]^2.
\end{equation}
Due to efficient particle cooling prior to $t=t_\text{coll}$ (see above), the
energy budget of those photons above
the pair creation threshold is typically orders of magnitude smaller
than the combined photon energy below the pair creation threshold at
$t_\text{coll}$. Furthermore, we expect photons to interact
with particles mostly via Thomson scattering, as photons that could
Compton-scatter off thermal particles are small in number. Noting that
Thomson scattering is elastic, we can deduce that for our purposes, the
spectral shape of the non-thermal photon spectrum at
$t=t_\text{coll}$ remains frozen thereafter as photons diffuse out of
the nebula on the diffusion timescale
\begin{equation}
  t_\text{diff,n}(t) = \frac{R_\text{n}}{c}[1 + \Delta\tau_\text{T}(t)].
\end{equation}
Here, we have again added the light crossing time of the nebula to
account for a transition to the optically thin regime. The luminosity
of the photons diffusing out of the nebula is approximately given by
\begin{eqnarray} 
 \ddt{E_\text{PWN}} &=& L_\text{PWN}(t) \label{eq:dEPWN_coll}\\
 &=& \{1-[1-f_\text{ej}(t)]\mathcal{A}(t)\} f_\text{beam}(t) \frac{E_\text{nth}(t)}{t_\text{diff,n}(t)},\nonumber
\end{eqnarray}
where, in analogy to Equation~\eqref{eq:dE_PWN_dt} the prefactor takes
into account the combined effects of
relativistic beaming and a non-zero albedo of the surrounding ejecta
material.\footnote{Prior to collapse in Phase III, the quasi-stationarity
  assumption in the present implementation of the radiative processes
  of the PWN (Section~\ref{sec:phaseIII_PWN}) prevent us from
  introducing these effects (see also Section~\ref{sec:discussion}).}
The total energy budget of the nebula thus evolves according to
\begin{equation}
  \ddt{E_\text{nth}} = -L_\text{PWN}(t) + L_\text{rad,in}(t), \label{eq:Enth_coll}
\end{equation}
where $L_\text{rad,in}$ is the thermal radiation from the surrounding
ejecta shell (see Equation~\eqref{eq:Lradin_phaseIII}), and we can
write the non-thermal radiation leaving the system as
\begin{equation}
  L_\text{rad,nth}(x,t) = f_\text{ej}(t) L_\text{PWN}(x,t_\text{coll})
  \frac{L_\text{PWN}(t)}{L_\text{PWN}(t_\text{coll})}. \label{eq:Lrad_nth_coll}
\end{equation}
In conclusion, after a collapse to a black hole, we evolve the system
of evolution equations in Phase III
assuming conservation of magnetic energy ($\rmn{d}E_B/\rmn{d}t = 0$),
employing Equation~\eqref{eq:dEPWN_coll} in
Equation~\eqref{eq:evoleqns_p3_Eth}, and extending the set of
Equations~\eqref{eq:evoleqns_p3_vej}--\eqref{eq:evoleqns_p3_EB} by
Equation~\eqref{eq:Enth_coll} to evolve the nebula properties instead of
solving Equations~\eqref{eq:LightmanI} and \eqref{eq:LightmanII}.

\section{Numerical procedure}
\label{sec:numerics}
In this section, we discuss several aspects related to the numerical
evolution of the model presented in the previous sections. After
briefly outlining the overall numerical procedure to integrate the
evolution equations
\eqref{eq:evoleqns_p1_Rej}--\eqref{eq:evoleqns_p3_EB} in
Sections~\ref{sec:wind_numerical}, \ref{sec:PWN_numerical}, and
\ref{sec:main_model_numerical}, we identify and define important
timescales to be monitored during the numerical evolution
(Section~\ref{sec:timescales}), discuss how to monitor the validity of neglecting
synchrotron self-absorption (Section~\ref{sec:synchrotron_monitor}), discuss how to overcome
problems related to stiffness in Equations~\eqref{eq:evoleqns_p1_Rej}--\eqref{eq:evoleqns_p3_EB}
(Section~\ref{sec:stiffness_problem}), and finally discuss how to
numerically compute the lightcurves as seen by a distant observer
including relativistic effects such as relativistic beaming, the time-of-flight effect, and
the relativistic Doppler effect (Section~\ref{sec:observer_lightcurve}).

\subsection{Hydrodynamic wind evolution}
\label{sec:wind_numerical}

Prior to evolving the main evolution equations
\eqref{eq:evoleqns_p1_Rej}--\eqref{eq:evoleqns_p3_EB}, the evolution of
the background fluid (the baryonic wind emitted during Phase I; see Section~\ref{sec:wind_model}) has to
be computed numerically. First, at any given time $t$ the function
$\bar{t}(r,t)$ is constructed solving Equation~\eqref{eq:tbar}
numerically for every radius $r$. Knowing $\bar{t}(r,t)$ immediately yields the
velocity profile $v_\text{w}(r,t)$ via Equation~\eqref{eq:v_w_rt} and
the mass profile $m_\text{w}(r,t)$ via Equation~\eqref{eq:m_w_rt}. By
finite differencing the mass function $m_\text{w}(r,t)$ the corresponding density profile
$\rho_\text{w,t}(r)$ is obtained (cf.~Equation~\eqref{eq:rho_w_rt}). We
typically employ a spatial grid ranging between $r=R_\text{min}=30\,\text{km}$ and
$R_\text{ej}(t)$ (cf.~Equation~\eqref{eq:phaseI_Rej}), which is
logarithmically spaced with a fixed number of points per
decade. Automatic mesh refinement (where needed) is used to resolve
the profiles, which become increasingly `sharp' as the material moves outward to larger length
scales. By employing this wind model the hydrodynamic evolution of the
wind material has essentially been reduced to
a numerical root finding problem. 

The free parameters of the model,
$\dot{M}_\text{in}$, $t_\text{dr}$, $\sigma_M$, and $v_\text{ej,in}$
are listed in Table~\ref{tab:model_parameters}. We note that numerical
relativity simulations can be employed to directly determine
$\dot{M}_\text{in}$ and $v_\text{ej,in}$. Furthermore, they can provide estimates on
$t_\text{dr}$. The only unconstrained parameter is $\sigma_M$, which
controls the shape of the wind density profiles at later
times. However, as the shock sweeps up all the material
into a thin shell in Phase II, the exact distribution of matter is not
important and does not influence the lightcurves and spectra of our
model significantly.

\subsection{PWN balance equations}
\label{sec:PWN_numerical}

At any time $t$ in Phase III, the balance equations
\eqref{eq:LightmanI} and \eqref{eq:LightmanII} for the photon and
non-thermal particle distribution in the PWN have to be solved. These
distributions then specify, e.g., the source term
$\rmn{d}E_\text{PWN}/\rmn{d}t$ (cf.~Equation~\eqref{eq:dEPWN_dt})
required in Equation~\eqref{eq:evoleqns_p3_Eth}. We solve the coupled
set of highly non-linear, integro-differential equations
\eqref{eq:LightmanI}--\eqref{eq:LightmanII} in analogy to the
multi-step, iterative method outlined in Appendix~B of LZ87. Only Step
1 of their method differs slightly from our Step 1 due to the
modifications we introduced to the physical description of the pair
plasma (cf.~Section~\ref{sec:phaseIII_PWN}). Therefore, we rediscuss this step here for
completeness and refer the reader to LZ87 for the other Steps. 

Step 1 in solving
Equations~\eqref{eq:LightmanI}--\eqref{eq:LightmanII} consists of
neglecting thermal Comptonization (i.e., setting
$\dot{n}_\text{C}^\text{T}=0$ in Equation~\eqref{eq:LightmanI}) and proceeds
as follows. 
\begin{itemize}
  \item[(i)] Given an approximate photon spectrum $n(x)$ compute
    the corresponding particle distribution
    $N(\gamma)$ using Equation~\eqref{eq:N_gam}, where the integral
    over $P(\gamma)$ is computed using Equation~(13) of
    LZ87.

    \item[(ii)] Compute 
      \begin{equation*}
        \qquad A(x) \equiv \frac{R_\text{n}}{c}(\dot{n}_{0} + \dot{n}_\text{A}
        + \dot{n}_\text{C}^\text{NT}+\dot{n}_\text{syn})
      \end{equation*}
      using $N(\gamma)$ from (i), Equation~\eqref{eq:nd0}, Equations~(7),\,(8),\,(11),\,(15)--(18) of
\citet{Svensson1983}, Equation~(9) of LZ87, and
Equation~\eqref{eq:ndsyn},
      \begin{equation*}
        \qquad B(x) \equiv
        \frac{1}{1+
\frac{1}{3}\tau_\text{KN}(x)f(x)} + \Delta\tau_\text{C}^\text{NT}
      \end{equation*}
      using Equations~(21) and (10) of LZ87 together with $N(\gamma)$
      from (i). Furthermore, compute
      \begin{equation*}
        C(x) \equiv 0.2 R_\text{n}\sigma_\text{T}\frac{1}{x} 
      \end{equation*}
      as well as $D(x)\equiv A(1/x)$, $E(x)\equiv B(1/x)$, and
      $F(x)\equiv C(1/x)$. We note that, with these definitions, all
      coefficient functions $A,B,C,D,E,F$ are non-negative.

      \item[(iii)] Writing Equation~\eqref{eq:LightmanI} as $n(x) = A(x) / [B(x) + C(x) n(1/x)]$ and
        $n(1/x) = D(x)/[E(x) + F(x) n(x)]$ leads to a quadratic
        equation for $n(x)$,
        \begin{equation*}
          BFn^2(x) + (BE+CD-AF)n(x) - AE = 0,
        \end{equation*}
        which has only one physical root,
        \begin{align}
          \qquad n(x) =& -\frac{(BE+CD-AF)}{2BF} \nonumber\\
               &+ \frac{\sqrt{(BE+CD-AF)^2+4BFAE}}{2BF}. \label{eq:nx_new}
        \end{align}
        Equation~\eqref{eq:nx_new} is now used to update $n(x)$.
\end{itemize}

Initializing $n(x)$ by $(R_\text{n}/c) [\dot{n}_{0,\text{ej}} +
f_\text{ej}\dot{n}_{0,\text{CMB}}]$, we iterate on Steps (i)--(iii) until
for successive iterations $j-1$ and $j$ we have
\begin{equation}
  \epsilon \equiv\lVert n(x)_j - n(x)_{j-1}\rVert_\text{max} < \epsilon_\text{tol}.
\end{equation}
Here $\lVert u(x)\rVert_\text{max}=\text{max}_{x\in X}\vert
u(x)\rvert$ for a function $u(x)$ on the dimensionless photon energy
domain $X=[x_\text{min},x_\text{max}]$ and typically $\epsilon_\text{tol}
\le 0.01$. As $\gamma_\text{max}$ (cf.~Equation~\eqref{eq:le}) is the
maximum attainable photon energy, we set
$x_\text{max}=\gamma_\text{max}$. In order to cover the frequency
range down to the radio band, we typically set
$x_\text{min}=10^{-18}$. Both the particle energy and the photon
energy domains are logarithmically
spaced with typically 15 points per decade.

\subsection{Evolving the main model}
\label{sec:main_model_numerical}

After having computed the hydrodynamic evolution of the baryonic wind
(Section~\ref{sec:wind_numerical}), its density profile
$\rho_\text{w,t}(r)$ at any point in time and space is known and the
main equations of our model
\eqref{eq:evoleqns_p1_Rej}--\eqref{eq:evoleqns_p3_EB} can be
integrated. The time grid
$T=[t_\text{min},t_\text{max}]$ for this evolution is logarithmically spaced with
typically 50 points per decade. Additional mesh
refinement is used in order to properly resolve the propagation of the
relativistic shock front through the ejecta material in Phase II. In
our model, we associate the
time of merger of the BNS system with $t=0$, but start the
numerical evolution from an appropriate final checkpoint of a
numerical relativity simulation, corresponding to tens of
milliseconds after merger, and evolve it over the timescales
of interest until, e.g., $t_\text{max}\sim 10^7\text{s}$. It is important to point out that
numerical relativity simulations of BNS mergers can
be employed to determine most of the input parameters of our model,
which are listed in Table~\ref{tab:model_parameters}. Given those
initial parameters read off or estimated from a simulation, the
following evolution according to our model is
a self-consistent prediction based on these initial conditions.

\begin{table}[tb]
\caption{Model input parameters. Most of these parameters can be
  extracted from (or at least estimated/constrained using) numerical
  relativity simulations of BNS mergers.}
\label{tab:model_parameters}
\centering
\begin{tabular}{cp{0.38\textwidth}}
\hline\hline
Parameter & Description \\
\hline
$\dot{M}_\text{in}$ & initial mass-loss rate of the NS
                      (cf.~Section~\ref{sec:wind_model}) \\
$t_\text{dr}$ & timescale for removal of differential rotation from
                the NS  (cf.~Section~\ref{sec:wind_model})\\
$\sigma_M$ & ratio of $t_\text{dr}$ to the timescale for decrease of the mass-loss
             rate (cf.~Section~\ref{sec:wind_model})\\
$v_\text{ej,in}$ & initial expansion speed of the baryonic ejecta
                   material (cf.~Section~\ref{sec:wind_model})\\
$\bar{B}$ & magnetic field strength in the outer layers of the NS
            (cf.~Equation~\eqref{eq:L_EM})\\
$\eta_{B_\text{p}}$ & dipolar magnetic field strength of
                      the pulsar in units of $\bar{B}$ (cf.~Equation~\eqref{eq:B_p})\\
$E_\text{rot,NS,in}$ & initial rotational energy of the NS (cf.~Equation~\eqref{eq:Erot_pul})\\
$P_\text{c}$ & initial central spin period of the NS
               (cf.~Equation~\eqref{eq:L_EM})\\
$R_\text{e}$ & equatorial radius of the NS
               (cf.~Equation~\eqref{eq:L_EM})\\
$M_\text{NS,in}$ & initial mass of the NS
                   (cf.~Equation~\eqref{eq:Erot_ej})\\
$I_\text{pul}$ & moment of inertia of the pulsar
                 (cf.~Equation~\eqref{eq:Ppul_in})\\
$\epsilon_\text{ej,NS,in}$ & initial specific internal energy of the
                             material ejected from the NS surface
                             (cf.~Equation~\eqref{eq:Eth_NS})\\
$\kappa$ & opacity of the ejecta material
           (cf.~Equations~\eqref{eq:delta_tau_w},
           \eqref{eq:delta_tau_sh}, \eqref{eq:delta_tau_ush}, and
           \eqref{eq:delta_tau_ej_com})\\
$\mathcal{A}(\nu,t)$ & frequency and time-dependent albedo of the
                       ejecta shell in Phase II and III
                       (cf.~Sections~\ref{sec:phaseII_PWN} and \ref{sec:collapse})\\
$t_\nu$ & neutrino-cooling timescale
          (cf.~Equation~\eqref{eq:Eth_NS})\\
$\eta_{B_\text{n}}$ &  fraction of the total pulsar wind power injected
                      as magnetic energy per unit time into the
                      PWN (cf.~Equations~\eqref{eq:evoleqns_p2_EB},
                      \eqref{eq:evoleqns_p3_EB}, 
                      and Sections~\ref{sec:phaseII_PWN},
                      \ref{sec:phaseIII_PWN}).\\
$\eta_\text{TS}$ & efficiency of converting pulsar wind power into random
              kinetic energy of accelerated particles in the PWN
              (cf.~Equations~\eqref{eq:evoleqns_p2_Enth},
              \eqref{eq:le} and Sections~\ref{sec:phaseII_PWN},
                   \ref{sec:phaseIII_PWN})\\
$\gamma_\text{max}$ & maximum Lorentz factor for non-thermal
                        particle injection into the PWN
                      (cf.~Section~\ref{sec:phaseIII_PWN}, Equation~\eqref{eq:le})\\
$\Gamma_\text{e}$ & power-law index of the
                                      non-thermal spectrum for particle
                                      injection into the PWN (cf.~Section~\ref{sec:phaseIII_PWN}, Equation~\eqref{eq:le})\\
$f_\text{coll}$ & (only in the collapse scenario,
                  Section~\ref{sec:collapse}) parameter specifying the
                  time of collapse of the NS in units of the spin-down
                  timescale (collapse during Phase III) or in units of
                  $t_\text{dr}$ (``$f_\text{coll,PI}$'', collapse during Phase I)\\
\hline
\end{tabular}
\end{table}

\subsubsection{Phase I}

In order to evolve Equations~\eqref{eq:evoleqns_p1_Rej} and
\eqref{eq:evoleqns_p1_Eth} from time $t$ to $t+\Delta t$ we need to
compute the source terms on the right-hand sides. This can be done in the
following way:

\begin{itemize}

  \item[(i)] Compute $v_\text{w}(R_\text{ej}(t),t)$ using the
    precomputed baryonic wind model (Section~\ref{sec:wind_numerical}) or Equation~\eqref{eq:phaseI_Rej}.

  \item[(ii)] Employ the previously generated wind profiles
    $m_\text{w}(r,t)$ and $\rho_\text{w}(t)$ to compute the optical
    depth of the ejecta material (Equation~\eqref{eq:delta_tau_w}) and to numerically solve for
    the temperature of the ejecta material using
    Equation~\eqref{eq:T_p1}. This yields the source term
    $L_\text{rad}(t)$ (cf.~Equations~\eqref{eq:Lrad_p1} and \eqref{eq:Teff_ph1}).

  \item[(iii)] Compute the remaining two source terms using
    Equations~\eqref{eq:L_EM} and \eqref{eq:Eth_NS}.

\end{itemize}

This phase ends when $t=t_\text{pul,in}$
(Equation~\eqref{eq:t_pulin}) and Phase
II starts. If the NS collapses to a black hole during Phase I, we keep evolving
Equations~\eqref{eq:evoleqns_p1_Rej} and \eqref{eq:evoleqns_p1_Eth},
setting $L_\text{EM}$ and $\rmn{d}E_\text{th,NS}/\rmn{d}t$ to zero
(cf.~Section~\ref{sec:collapse}).

\subsubsection{Phase II}

At $t=t_\text{pul,in}$, the initial rotational energy of the pulsar is
computed using Equation~\eqref{eq:Erot_pul} and its initial spin
period is then inferred from Equation~\eqref{eq:Ppul_in}. This allows
us to compute the initial spin-down luminosity
(Equation~\eqref{eq:Lsd_in}) and the spin-down timescale
(Equation~\eqref{eq:tsd}), which together define the spin-down
luminosity at later times (Equation~\eqref{eq:Lsd}).

In order to evolve the evolution equations
\eqref{eq:evoleqns_p2_Rej}--\eqref{eq:evoleqns_p2_EB} from time $t$ to
$t+\Delta t$, one can proceed as follows:

\begin{itemize}
  
\item[(i)] Compute $v_\text{w}(R_\text{ej}(t),t)$ using the
    precomputed baryonic wind model (Section~\ref{sec:wind_numerical})
    or Equation~\eqref{eq:phaseI_Rej}.

  \item[(ii)] Compute $\rmn{d}E_\text{th,vol}$ and $v_\text{sh}(t)$
    from Equations~\eqref{eq:Eth_vol} and \eqref{eq:v_sh},
    respectively. Use $v_\text{sh}(t)$ and the wind profiles to
    evaluate shock heating according to Equation~\eqref{eq:Esh}.

  \item[(iii)] Compute the source terms $L_\text{sd}(t)$,
    $L_\text{rad,pul}$, and $\rmn{d}E_\text{PWN}/\rmn{d}t$ using
    Equations~\eqref{eq:Lsd}, \eqref{eq:Lrad_pul}, and \eqref{eq:dE_PWN_dt}.

  \item[(iv)] Find the temperatures of the shocked and unshocked
    ejecta parts by using the precomputed wind profiles
    $m_\text{w}(r,t)$, $\rho_\text{w}(t)$ and solving
    Equation~\eqref{eq:T_p2}. This yields the optical depths
    (Equations~\eqref{eq:delta_tau_sh}, \eqref{eq:delta_tau_ush}) and
    luminosities \eqref{eq:Lradin_p2} and \eqref{eq:Lrad_p2}.

  \item[(v)] Evolve all quantities to $t+\Delta t$, except for
    $R_\text{n}$ and $E_\text{nth}$.

  \item[(vi)] Define
    \begin{align}
      \qquad E_\text{nth}(t+\Delta t) &= E_\text{nth}(t) +
                                 \ddt{E_\text{nth}}\Delta t\\
      R_\text{n}(t+\Delta t) &= R_\text{sh}(t+\Delta
                               t)\frac{E_\text{th,sh}(t+\Delta
                               t)}{E_\text{nth}(t+\Delta t)}
    \end{align}
    and iterate until both quantities have converged to some desired
    accuracy. The second condition ensures pressure equilibrium
    between the nebula and the shocked ejecta layer
    (cf.~Equation~\eqref{eq:press_equil}).

\end{itemize}

Phase II ends when $R_\text{sh}(t+\Delta t) > R_\text{ej}(t+\Delta t)$.

\subsubsection{Phase III}

Once the shock has reached the outer ejecta layers, Phase III
begins. The initial speed of the shocked ejecta layer is calculated from
Equation~\eqref{eq:vej_tso}. Evolving
Equations~\eqref{eq:evoleqns_p3_vej}--\eqref{eq:evoleqns_p3_EB} from
time $t$ to $t+\Delta t$ can be
accomplished by the following steps:

\begin{itemize}

  \item[(i)] Compute $L_\text{sd}$ and $L_\text{rad,pul}$ using
    Equations~\eqref{eq:Lsd} and \eqref{eq:Lrad_pul}.

  \item[(ii)] Compute the acceleration of the ejecta shell according
    to Equation~\eqref{eq:dvej_dt}.

  \item[(iii)] Find the temperature of the ejecta shell by solving
    Equation~\eqref{eq:T_com}. This yields the optical depth
    (Equations~\eqref{eq:delta_tau_ej_com}) and the
    luminosities $L_\text{rad}$ and $L_\text{rad,in}$
    (Equations~\eqref{eq:Lrad_phaseIII} and
    \eqref{eq:Lradin_phaseIII}).

   \item[(iv)] Solve Equations~\eqref{eq:LightmanI} and
     \eqref{eq:LightmanII} as described in
     Section~\ref{sec:PWN_numerical} to find the source term $\rmn{d}E_\text{PWN}/\rmn{d}t$.

\end{itemize}

In the case the NS collapses to a black hole (see
Section~\ref{sec:collapse}), Step (i) is omitted and
Equation~\eqref{eq:Enth_coll} is added as another evolution equation to
Equations~\eqref{eq:evoleqns_p3_vej}--\eqref{eq:evoleqns_p3_EB}. Furthermore,
Step (iv) is replaced by
\begin{itemize}
  \item[(iv)'] Find the source term $\rmn{d}E_\text{PWN}/\rmn{d}t$
    using Equation~\eqref{eq:dEPWN_coll}.
\end{itemize}

This concludes our discussion of the overall procedure to evolve Equations~\eqref{eq:evoleqns_p1_Rej}--\eqref{eq:evoleqns_p3_EB}.

\subsection{Timescales}
\label{sec:timescales}

As our evolution model in Phase III is built upon the assumption of
quasi-stationarity as far as radiative processes in the PWN are concerned,
we need to monitor several timescales during the numerical evolution in order to assess the validity
of this assumption. For these diagnostic purposes, we define the
following timescales in Phase III:
\begin{eqnarray}
  \tau_\text{e}(t) &=& \frac{l_\text{e}}{\dot{l}_\text{e}},\\
  \tau_\text{ph}(t) &=& \frac{l_\text{ph}}{\dot{l}_\text{ph}},\\
  \tau_\text{c}(\gamma, t) &=& \left|
                               \frac{\gamma}{\dot{\gamma}_\text{C,syn}}\right|,\\
  \tau_\text{l}(t) &=& \frac{R_\text{n}}{c},
\end{eqnarray}
where $\gamma$ denotes the Lorentz factor of a particle in the PWN and
$l_\text{e}$, $l_\text{ph}$, and $\dot{\gamma}_\text{C,syn}$ are
defined by Equations~\eqref{eq:le}, \eqref{eq:lph}, and
\eqref{eq:gam_dot}, respectively.

In order for the stationarity assumption regarding the particle
distribution to hold, which allowed us to integrate
Equation~\eqref{eq:N_continuity} to obtain Equation~\eqref{eq:N_gam}, the
timescale for equilibration of the particle distribution given by the
cooling timescale $\tau_\text{c}$ has to be much smaller than any
other timescale involved in the problem. In particular, it has to be
smaller than the timescale for change of the photon distribution. Thus
we have the requirements that
\begin{eqnarray}
  \tau_\text{c}(\gamma,t) &\ll \tau_\text{ph}(t) \mskip20mu &\forall\, t,\gamma,\\
  \tau_\text{c}(\gamma,t) &\ll \tau_\text{e}(t) \mskip20mu &\forall\, t,\gamma.
\end{eqnarray}
Here, we have assumed that the timescales for change of the photon and
particle distributions inside the nebula are approximately given by
the timescales $\tau_\text{ph}$ and $\tau_\text{e}$ for change of the
injected spectra. As we shall discuss in the companion paper (Paper II), we typically
find that these conditions are very well satisfied across the entire parameter
space in absence of significant acceleration of the ejecta shell after $t=t_\text{shock,out}$
and except for very late times $t\gtrsim 10^{7}\,\text{s}$ when the
nebula has grown in size, the magnetic field strength has decreased, and
synchrotron cooling has become less efficient. However, we
are not interested in the evolution at these late times.

For the stationarity assumption regarding the photon spectrum to hold
(i.e., for Equation~\eqref{eq:LightmanI} to hold), the timescale for
the photon spectrum to change has to be much larger than the timescale
for equilibration inside the nebula. As the positronic plasma inside
the PWN is relativistic, its sound speed is close to the speed of
light. We can thus assume that the nebula plasma adjusts to changing
exterior conditions essentially on the
timescale $\tau_\text{l}$. Therefore, one also has the requirement that
\begin{equation}
  \tau_\text{l}(t) \ll \tau_\text{ph} (t)\mskip20mu \forall\,t.
\end{equation}
 As long as further acceleration of the ejecta shell after
 $t=t_\text{shock,out}$ is not significant, we typically find that
 this condition is well satisfied during the numerical evolution as we
 shall discuss in more detail in the companion paper (Paper II).

\subsection{Synchrotron self-absorption}
\label{sec:synchrotron_monitor}

In our model of the PWN in Phase III, we have neglected synchrotron
self-absorption. In order to check the validity of this assumption
during the numerical evolution of our model, we monitor the
dimensionless photon energy
$x_{\Delta\tau_\text{syn}=1}(t)$ defined by
\begin{equation}
  \Delta\tau_\text{syn}(x_{\Delta\tau_\text{syn}=1}(t),t) = 1,
\end{equation}
where $\Delta\tau_\text{syn}(x,t)$ is the approximate optical depth of
the nebula to synchrotron self-absorption at time $t$
(cf.~Equation~\eqref{eq:tau_syn}). This dimensionless energy separates
the optically thick and thin regimes of the photon spectrum and lies
well below X-ray frequencies (typically by many orders of magnitude) as we
shall discuss in the companion paper (Paper II). Therefore, the part
of the spectrum that we are mostly interested in, i.e., at X-ray and
$\gamma$-ray energies, is unaffected by effects of synchrotron self-absorption.

\subsection{Stiffness problem}
\label{sec:stiffness_problem}

At some point in the evolution (in Phase III), the photon diffusion timescale of
the ejecta shell 
\begin{equation}
  t_\text{diff,ej} = \frac{R_\text{ej}-R_\text{n}}{c}(\Delta\tau_\text{ej}+1)
\end{equation}
becomes comparable to the temporal
resolution $\Delta t$ and Equation~\eqref{eq:evoleqns_p3_Eth} becomes
stiff. Here, $\Delta\tau_\text{ej}$ denotes the optical depth as seen
from the lab frame (cf.~Equation~\eqref{eq:delta_tau_ej_lab}). This
problem can be noticed in the following way. Assuming
$\Delta\tau_\text{ej,com}\gg 1$, a radiation
dominated gas, such that $T_\text{com}\simeq
E_\text{th}/aV_\text{ej}$ (cf.~Equation~\eqref{eq:T_com}), and using
Equations~\eqref{eq:Teff_com} and \eqref{eq:Lrad_phaseIII} it is
straightforward to show that the thermal emission from the ejecta
shell scales as $L_\text{rad} \propto
E_\text{th}/t_\text{diff,ej}$. The photon diffusion timescale itself scales
as $t_\text{diff,ej}\propto R_\text{ej}^{-2}$
(cf.~Equation~\eqref{eq:delta_tau_ej_com}). Hence, as $R_\text{ej}$ is
monotonically increasing with time, at some point $t_\text{diff,ej}\approx
\Delta t$ and Equation~\eqref{eq:evoleqns_p3_Eth} becomes stiff.

Fortunately, there is an easy way to bypass this problem. As $L_\text{rad} \propto
E_\text{th}/t_\text{diff,ej}$, this term will eventually deplete the
energy reservoir $E_\text{th}$ of the ejecta shell until the
shell enters an asymptotic regime in which the amount of injected
energy equals the amount of emitted energy on the photon diffusion
timescale. In other words, this regime is defined by setting
$\rmn{d}E_\text{th}/\rmn{d}t = 0$ in
Equation~\eqref{eq:evoleqns_p3_Eth} on a timescale $t\sim
t_\text{diff,ej}$. At any time in this asymptotic regime, the total
internal energy $E_\text{th}$ of the ejecta shell is then given by
\begin{equation}
  E_\text{th}(t) = [1-f_\text{ej}(t)]\ddt{E_\text{PWN}}
  t_\text{diff,ej}(t) \label{eq:E_th_eff}
\end{equation}
and a separate evolution equation for $E_\text{th}$ is not needed
anymore. 

In practice, the grid spacing $\Delta t$ has to be adjusted such that
this asymptotic regime is reached well before $\Delta t \approx
t_\text{diff,ej}$. Once this is achieved, we typically switch to the
asymptotic regime during the numerical evolution at
$t_\text{res}$, defined by $t_\text{diff,ej}(t_\text{res}) =
2 \Delta t$. By definition of this regime, $L_\text{rad}$ and
$L_\text{rad,in}$ scale as $\rmn{d}E_\text{PWN}/\rmn{d}t$. Therefore,
assuming that the thermal photons from the ejecta shell dominate the
photon injection $l_\text{ph}$ of the PWN
(cf.~Equation~\eqref{eq:lph}), $l_\text{ph}$ is expected to scale as
$l_\text{e}$. As the PWN conserves energy (total luminosity of
injection equals the luminosity of emitted radiation; see
Appendix~\ref{app:energy_conservation}) we hence conclude that
\begin{equation}
  \ddt{E_\text{PWN}} \propto (L_\text{sd} + L_\text{rad,pul}). \label{eq:dEPWN_eff}
\end{equation}
In order to avoid numerical runaway instabilities due to the fact that
$\rmn{d}{E_\text{PWN}}/\rmn{d}t$ in Equation~\eqref{eq:E_th_eff}
depends on the solution to Equations~\eqref{eq:LightmanI} and
\eqref{eq:LightmanII}, which, in turn, depends on the injected
photons, i.e., on $\rmn{d}{E_\text{PWN}}/\rmn{d}t$, we employ
Equation~\eqref{eq:dEPWN_eff} in Equation~\eqref{eq:E_th_eff} and
calibrate the prefactor at $t_\text{res}$ to ensure continuity over
$t=t_\text{res}$. This calibration, however, is only valid for
$t>t_\text{res}$ if also the dynamics of the shell have reached an asymptotic
regime, i.e., if no further acceleration occurs such that, e.g., the
beaming factor does not depend on time. In practice, a grid spacing
can always be set up to satisfy this constraint as well.

\begin{figure}[tb]
  \includegraphics[width=0.48 \textwidth]{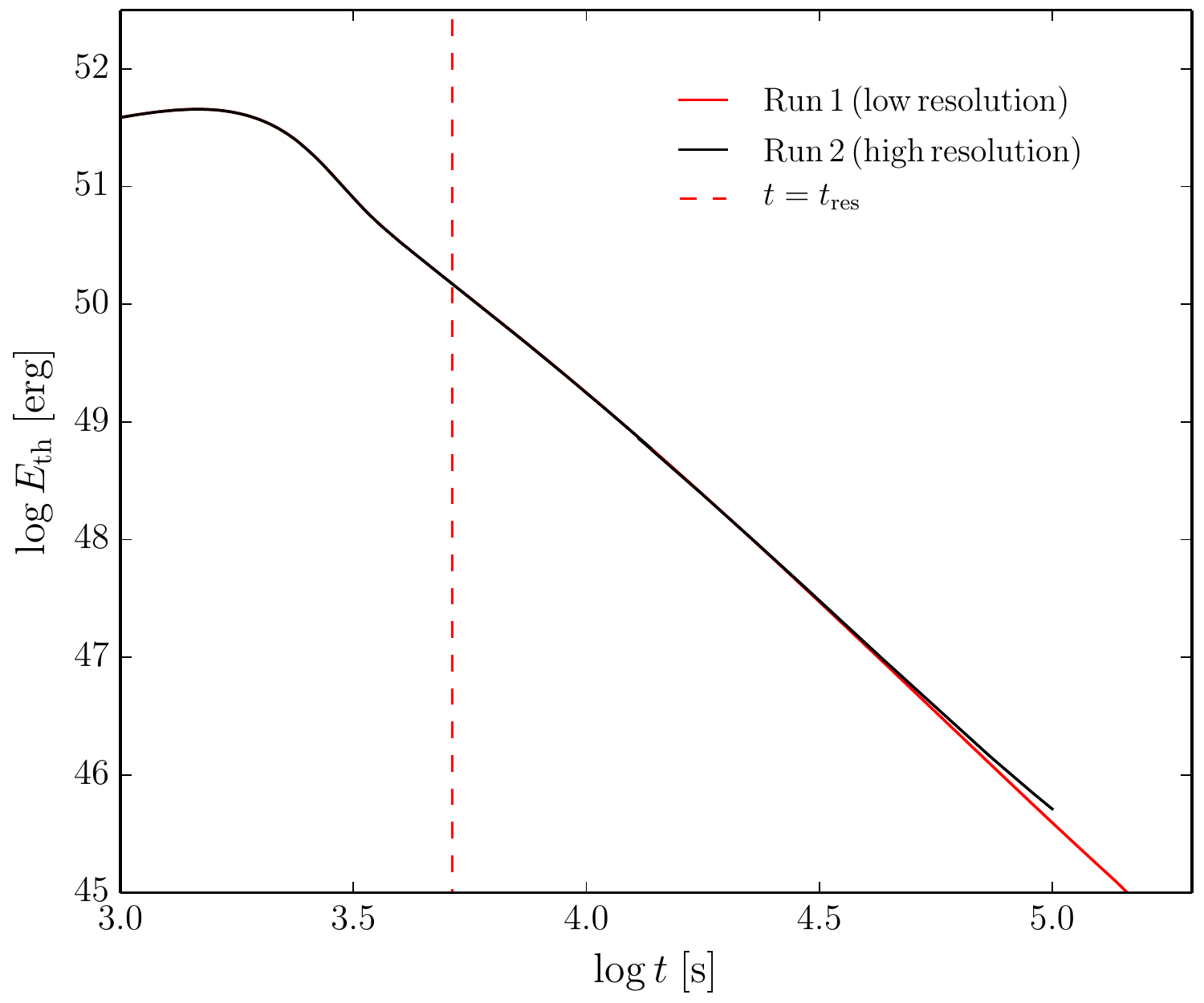}
  \caption{Evolution of the internal energy $E_\text{th}$ of the ejecta
  shell in Phase III according to a run using the evolution scheme
  described in this section (Run 1) and a run using additional mesh
  refinement to evolve Equation~\eqref{eq:evoleqns_p3_Eth} up to late
  times (Run 2). The dashed red line indicates the time of transition $t=t_\text{res}$
  to the modified evolution scheme for Run 1.}
  \label{fig:effective_evol}
\end{figure}

Figure~\ref{fig:effective_evol} compares the internal energy
$E_\text{th}$ of a typical model case evolved using the scheme outlined above (Run
1) with
the corresponding result for a simulation in which a power-law refinement
$\Delta t \propto t_\text{diff,ej}\propto R_\text{ej}^{-2}$ was
implemented (Run 2). For the latter run, the refinement guaranteed
$\Delta t \ll t_\text{diff,ej}$ at all times and thus allowed us to integrate
Equation~\eqref{eq:evoleqns_p3_Eth} for all times. The agreement
between the two runs is
remarkably good and indicates that the scheme described above well
captures the evolution of the system at later times than
$t_\text{res}$. Remaining discrepancies at late times $t\sim
10^{5}\,\text{s}$ are due to numerical errors as
the ever decreasing time step in Run 2 results in increasing accuracy
for the time integrations of
Equations~\eqref{eq:evoleqns_p3_vej}--\eqref{eq:evoleqns_p3_EB}. Finally,
we note that runs with power-law refinement are computationally very
expensive, such that comparisons like the one presented here cannot be
carried out routinely and we will thus not add this comparison to the
set of routine checks (see Sections~\ref{sec:timescales} and
\ref{sec:synchrotron_monitor}) that we shall discuss in more
detail in the companion paper (Paper II).

\subsection{Observer lightcurve reconstruction}
\label{sec:observer_lightcurve}

This section is devoted to a discussion of how the
lightcurve and spectra as seen by a distant observer can be reconstructed, including
relativistic effects such as the relativistic Doppler effect, the
time-of-flight effect, and relativistic beaming. As the effective temperature of blackbody
radiation is altered by the relativistic Doppler factor and enters the
luminosity to the fourth power, even mildly relativistic ejecta shell
velocities of $v_\text{ej}\sim 0.1c$ can have a significant influence
on the lightcurve. Furthermore, as the PWN and the ejecta shell expand to large
radii, radiation reaching the observer from different locations on the
surface of the expanding sphere can have appreciable delays. These
delays are particularly important in dynamical situations, in which the ejecta
shell is accelerated or in which the NS collapses to a black hole and
induces an abrupt change in the radiation escaping
from the system. Finally,
relativistic beaming is influential in the sense that it normalizes
the total luminosity by selecting an effective surface area of the
expanding sphere that emits in the direction of the observer. In so
doing, it also affects the maximum delay concerning the time-of-flight
effect and must therefore be taken into account.
Consequently, it is important to consider these relativistic effects
when predicting observer lightcurves and spectra with our model.

We numerically compute the lightcurves and spectra as seen by a
distant observer including the
relativistic effects described above by decomposing the radiation
originating from the surface area of the half-sphere facing the
observer into individual energy packages
released at time $t$ during a time $\Delta
t$ from ring-shaped portions of surface area of equal distance to the
observer. These
packages are emitted at time $t$ and then recollected by the observer
according to their time of flight and their Doppler-shifted frequency. To
this end, we define a coordinate system centered about the NS, with its
$z$-axis in direction of the remote observer and $\theta$ denoting
the polar angle. At any time $t$ let us consider two half spheres
facing the observer, defined by $\theta \in [0,\pi/2]$ and radii
$R_\text{ej}(t)$ and $R_\text{n}(t)$, respectively. Due to
relativistic beaming, only radiation
originating from the regions $\theta \in [0,\theta_\text{max}(t)]$ will
eventually reach the observer \citep{Rees1966}, where
\begin{equation}
  \theta_\text{max}(t) =  \arccos\left( \frac{v_\text{ej}(t)}{c}\right).
\end{equation}
This area is then further decomposed into rings of constant $\theta$,
defined by a grid $\Theta =  [0, \ldots, \theta_k,
\ldots,\theta_\text{max}(t)]$, with $n_\theta$ points and spacing $\Delta
\theta_k$ that is equidistant in the limb angle $\mu = \cos\theta$.

Moreover, given a time grid
$T=[t_\text{min},\ldots,t_i,\ldots,t_\text{max}]$ for the evolution of
Equations~\eqref{eq:evoleqns_p1_Rej}--\eqref{eq:evoleqns_p3_EB}
appropriate to properly resolve the shock dynamics during Phase II and
to overcome the stiffness problem
(cf.~Section~\ref{sec:stiffness_problem}), we define a time grid for
the observer $T'=[t'_\text{min}=t_\text{min},\ldots,t'_j,\ldots,t'_\text{max}=t_\text{max}]$ by
\begin{align}
  t'_0 &= t'_\text{min},\\
  t'_{j+1} &=t'_j + N_{t'}\frac{v_\text{ej,in}t'_j}{c n_\theta}, \label{eq:t_obs}
\end{align}
where $N_{t'} < 1$. In order for the observer to resolve the time-of-flight effect, i.e.,
to distinguish between the various energy contributions from different
locations on the emitting surface area, the
grid spacing $\Delta t'$ at time $t'$ has to be smaller than the light
crossing time of the half sphere divided by the number of individual
rings. Since
\begin{equation}
  \Delta t'_j = N_{t'}\frac{v_\text{ej,in} t'_j}{c n_\theta} < \frac{R_\text{ej}(t'_j)}{c n_\theta},
\end{equation}
the grid spacing defined by Equation~\eqref{eq:t_obs} guarantees that
the time-of-flight effect is properly resolved. 

We also define a receiver energy grid $E'(T',X')$ for the remote
observer, where $X'$ denotes the grid in dimensionless energy
$x'=h\nu'/m_\text{e}c^2$, which is typically similarly spaced as the
corresponding one used in the evolution of the main model
(cf.~Section~\ref{sec:PWN_numerical}). The energy contribution of a
ring of width $d\theta$ at
$\theta$ to the total energy emission during a time $\rmn{d}t$ per
unit frequency as measured by the distant observer is given by
\begin{equation}
  \frac{\rmn{d}E'}{\rmn{d}\nu'} = 4\pi I'(\nu') \rmn{d}\sigma
  \cos\theta_\text{com} \rmn{d}t, \label{eq:dEdnu_sph}
\end{equation}
where $I'(\nu')$ is the specific intensity as seen by the observer,
$\rmn{d}\sigma = 2\pi R^2_\text{ej/n}\sin\theta\rmn{d}\theta$ is the
surface area of the ring, and $\theta_\text{com}$ given by
\begin{equation}
  \tan\theta_\text{com} =\sqrt{1-v_\text{ej}^2/c^2}
  \frac{\sin\theta}{\cos\theta - v_\text{ej}/c}
\end{equation}
defines the direction of the observer as seen in the local comoving frame
of the expanding surface. We note that the effective emitting surface
area $\rmn{d}\sigma\cos\theta_\text{com}$ approaches zero as $\theta$
reaches $\theta_\text{max}$. We have already multiplied by
$4\pi$ in Equation~\eqref{eq:dEdnu_sph} to account for an isotropic source.

According to the relativistic Doppler effect applied to a
photon of frequency $\nu$ emitted from the surface at a point
specified by an angle $\theta$ in direction of the remote observer, 
\begin{equation}
D\equiv \frac{\nu'}{\nu} = \frac{x'}{x}
  = \frac{\sqrt{1-v_\text{ej}^2/c^2}}{1-(v_\text{ej}/c)\cos\theta}. \label{eq:D}
\end{equation}
Noting that $I/\nu^3$ is Lorentz invariant, where $I$ is the specific
intensity, we have $I'=D^3 I$. Therefore,
for the blackbody emission of the ejecta material,
\begin{equation}
  I'_\text{th}(\nu') = \frac{2 h \nu'^3}{c^2}
  \left[\exp\left(\frac{h\nu'}{Dk_\text{B}T_\text{eff,com}}\right) -
    1\right]^{-1}, \label{eq:Ith}
\end{equation}
where $T_\text{eff,com}$ is given by Equation~\eqref{eq:Teff_com}. We
note that the effective temperature as seen by the distant observer is thus
$T'_\text{eff}=D T_\text{eff,com}$. Since the associated luminosity scales as
$L'\propto D^4 T_\text{eff,com}^4$, small deviations of $D$ from unity
can already have an appreciable influence on the observer
lightcurve. Furthermore, as long as the nebula is optically thick,
i.e., $\Delta\tau_\text{T}>1$, the escaping photons originate from the
outer surface layer at $r=R_\text{n}$ and we can set
\begin{equation}
  I'_\text{nth}(x') = D^3 \frac{L_\text{rad,nth}(x'/D,t)}{4\pi
    R_\text{n}^2} \label{eq:Inth}
\end{equation}
for the non-thermal radiation, with
$L_\text{rad,nth}(x,t)$ being the non-thermal luminosity of the
nebula (cf.~Equations~\eqref{eq:Lrad_nth} and
\eqref{eq:Lrad_nth_coll}). When beaming is already encoded in
$L_\text{rad,nth}(x,t)$ itself (cf.~Section~\ref{sec:collapse}), the
right-hand side of Equation~\eqref{eq:Inth} has to be divided by
$f_\text{beam}$ to solely take the geometric time-of-flight effect
into account (through the subdivision
into individual rings up to $\theta=\theta_\text{max}$).

When the nebula is optically thin, i.e., $\Delta\tau_\text{T}<1$, the
escaping photons are emitted uniformly and isotropically throughout
the volume of the nebula. In this case, we slice the nebula
volume $V_\text{n}$ into spherical shells of radius $r\in
R=[R_\text{min},R_\text{n}]$ and thickness $\rmn{d}r$ and
subdivide each shell into rings of constant $\theta_\text{vol}\in\Theta_\text{vol}=[0,\pi]$ and width
$\rmn{d}\theta_\text{vol}$. The contribution of such volume elements $\rmn{d}V= \rmn{d}\sigma
\rmn{d}r = 2\pi r^2\sin\theta_\text{vol} \rmn{d}\theta_\text{vol} \rmn{d}r$ to the total energy
emission during a time $\rmn{d}t$ per unit frequency as measured by
the distant observer can then be written as
\begin{equation}
   \frac{\rmn{d}E'}{\rmn{d}x'} = D^3L_\text{rad,nth}(x'/D,t)
   \frac{\rmn{d}V}{V_\text{n}} \rmn{d}t. \label{eq:dEdnu_seg}
\end{equation}
Here, $D$ is defined as in Equation~\eqref{eq:D} in terms of the local velocity $v(r)=v_\text{ej} r/R_\text{n}$.

With these definitions and expressions at hand,
we compute the radiation emitted at time $t$ during a time $\Delta t$ as received by the remote
observer according to the following steps: 

\begin{itemize}
  \item[(i)] First, compute the arrival times of the energy packages
    for all emitting rings and spherical segments, i.e., for all
    $\theta\in\Theta$, $\theta_\text{vol}\in\Theta_\text{vol}$, and
      $r\in R$:
\begin{align}
  t'_\text{arr,th}(\theta) &= \frac{R_\text{min}}{c} + t
  - \frac{R_\text{ej}(t)}{c}\cos\theta\\
t'_\text{arr,nth,surf}(\theta) &= \frac{R_\text{min}}{c} + t
  - \frac{R_\text{n}(t)}{c}\cos\theta\\
\qquad t'_\text{arr,nth,vol}(\theta_\text{vol},r) &= \frac{R_\text{min}}{c} + t
  - \frac{r}{c}\cos\theta_\text{vol}
\end{align}
This definition of arrival times ensures that a photon emitted at
$t=t_\text{min}$ from the outer surface of the baryonic wind at $R_\text{ej}(t_\text{min})=R_\text{min}$ and
$\theta = 0$ reaches the observer at $t'=t_\text{min}$.

\item[(ii)] Second, compute the energy spectrum $\Delta E'_\text{th}(\Theta,X')$
  that is generated by thermal radiation from the surface of the
  ejecta matter using Equations~\eqref{eq:dEdnu_sph} and \eqref{eq:Ith}:
\begin{multline}
  \qquad \Delta E'_\text{th}(\theta_k,\nu'_l) = 8\pi^2 R_\text{ej}^2 I'_\text{th}(\nu'_l)\\
  \times\sin\theta_k\cos[\theta_\text{com}(\theta_k)] \Delta\theta_k \Delta t. \nonumber
\end{multline}

\item[(iii)] In analogy to the thermal radiation in Step (ii), use
  Equations~\eqref{eq:dEdnu_sph}, \eqref{eq:Inth}, and \eqref{eq:dEdnu_seg} to compute
  the energy contributions generated by the non-thermal radiation
  from the PWN once the ejecta shell has become optically thin:
\begin{align}
  \Delta E'_\text{nth,surf}(\theta_k,x'_l) &= (1-f_\text{n}) 8\pi^2
                                              R_\text{n}^2
                                              I'_\text{nth}(x'_l)\nonumber\\
   \times &\sin\theta_k\cos[\theta_\text{com}(\theta_k)] \Delta\theta_k \Delta t,\nonumber\\
  \qquad\Delta E'_\text{nth,vol}(\theta_{\text{vol},k},x'_l, r_m) &=
                                                           f_\text{n}\frac{3}{2}D^3L_\text{rad,nth}(x'_l/D)\nonumber\\
 \times &\frac{r_m^2}{R_\text{n}^3}\sin\theta_{\text{vol},k}
  \Delta\theta_{\text{vol},k}\Delta r_m \Delta t.\nonumber
\end{align}
The function $f_\text{n}(t)$ is defined
in terms of $\Delta \tau_\text{T}$ in analogy to $f_\text{ej}$ in
Equation~\eqref{eq:f_ej}. It is used here to guarantee a smooth
transition between the optically thick and thin regimes of the nebula,
corresponding to surface and volume nebula emission, respectively.

\item[(iv)] Use the mapping of angles to arrival times from Step (i)
  to put the energy packages from Steps (ii) and (iii) into the correct time and frequency bins
  of the observer energy grid $E'(T',X')$.
\end{itemize}

Once the evolution of
Equations~\eqref{eq:evoleqns_p1_Rej}--\eqref{eq:evoleqns_p3_EB} has
been accomplished up to $t=t_\text{max}$ and all energy packages have
been sent and received, the energy array of the observer, $E'(T',X')$,
may be divided by the time steps $\Delta t'_j$ to obtain the
corresponding observer luminosity 
\begin{equation}
  L_\text{obs}(t'_j,x'_l) = E'(t'_j,x'_l)/\Delta t'_j.
\end{equation}
This quantity can then be used to compute
detailed predictions for the luminosity as seen by a remote observer in
specific wavelength bands, such as in the $\gamma$-ray, X-ray,
UV, optical, and radio bands (see Paper II).

%%%%%%%%%%%%%%%%%%%%%%%%%%%%%%%%%%%%%%%%%
\section{Discussion and Conclusion}
%%%%%%%%%%%%%%%%%%%%%%%%%%%%%%%%%%%%%%%%%
\label{sec:discussion}

In this paper, we have presented
a dynamical model to describe the
post-merger evolution of a BNS system and its EM
emission. Our model assumes that the merger of two NSs leads to the
formation of a  (hypermassive, supramassive, or stable) NS which does not
collapse to a black hole on timescales of at least tens of
milliseconds after merger. As we have argued
(cf.~Section~\ref{sec:introduction}), such a long-lived NS is a very
likely possibility, such that the model should be applicable to a
large fraction of BNS merger events. In contrast to a black
hole promptly formed after merger and surrounded by a short-lived
accretion disk, such long-lived
objects can power long-lasting EM emission from
$\gamma$-ray to radio energies that might be responsible for at least
part of the observed long-lasting afterglow emission observed in many
SGRBs. We refer to Paper
II for a detailed account on modeling X-ray afterglow lightcurves
with our model in the context of SGRBs. Moreover, this model
also represents an important tool to identify EM
counterparts associated with the GW signal of the
inspiral and merger of a BNS system (see Paper II for a detailed
discussion). The identification of such
EM counterparts is essential for performing joint
EM and GW observations. Joint
observations involving EM counterparts of the kind
discussed here (long-lasting and highly isotropic) can confirm the
association of the GW signal with a BNS merger
(i.e., distinguish form the NS--BH binary merger). With the advanced
LIGO/Virgo detector network
starting its first science run later this year, such multimessenger
astronomy will turn into exciting reality in the very near future.

Our model is formulated in terms of a set of highly coupled
differential equations, which provide a self-consistent evolution of
the post-merger system and its EM emission given some
initial data. Such initial data can be
extracted from a numerical relativity simulation of the merger and
early post-merger phase at a few to tens of milliseconds after the
merger, once a roughly axisymmetric state has been
reached. Our model allows us to
evolve the post-merger system over time and
lengthscales inaccessible to numerical relativity simulations,
typically up to $\sim\!10^7\,\text{s}$ after merger. It thus bridges
the gap between numerical relativity simulations of the merger process
and the timescales of interest for SGRB afterglow radiation. The model
evolves the system through three main
evolutionary phases: an early baryonic wind phase (Phase I), a pulsar
wind shock phase (Phase II), and a PWN phase (Phase
III). Furthermore, the possibility of collapse to a black hole during
any of the three phases is accounted for. Our model links the
evolution of the central engine directly to the observed afterglow
radiation, taking into account relativistic dynamics as well as an
accurate reconstruction of the observer lightcurve including
relativistic beaming, the relativistic Doppler and the time-of-flight effect.

\paragraph{Prompt SGRB emission and time-reversal scenario}
Our model makes no assumption on how and when the prompt $\gamma$-ray
emission of the SGRB itself is produced. It can accommodate both the
standard scenario (SGRB at the time of merger) and the recently
proposed time-reversal scenario (SGRB at the time of collapse of the
remnant NS; \citealt{Ciolfi2015a,Ciolfi2015b}). It can obviously also
accommodate the case in which no relativistic jet and thus no SGRB is
produced at all. 

In the context of magnetar models for
SGRBs, the prompt emission is assumed to be generated by an accretion
powered relativistic jet emerging from a NS--torus system shortly
($\sim\!\text{ms}$) after the time of merger (e.g.,
\citealt{Metzger2008a,Bucciantini2012,Gompertz2014,Metzger2014b,Gao2015}). However,
a generic feature
of a newly-born NS after a BNS merger will be strong baryon pollution
in its vicinity due to dynamical ejecta from the merger process and
neutrino and magnetically driven winds from its surface or possibly
from an accretion disk
(\citealt{Hotokezaka2013a,Oechslin2007,Bauswein2013,Kastaun2015a,Dessart2009,Siegel2014a,Metzger2014c};
see Section~\ref{sec:phaseI}). Such baryon pollution can not only
choke jets \citep{Nagakura2014,Murguia-Berthier2014}, but it is likely
to even prevent the generation of any relativistic outflow at
all. We note that numerical simulations of BNS mergers that lead to
the formation of a remnant NS have not found
indications for the generation of a relativistic jet
\citep{Giacomazzo2013}. 

In the time-reversal scenario, the SGRB is
generated at the time of collapse of the supramassive remnant NS. In
this case, the baryon-free PWN surrounding the collapsing NS does not threaten
the formation of an
accretion powered jet from the remaining BH--torus
system. \citet{Margalit2015} have recently
argued that the formation of an accretion disk following the collapse
of the supramassive NS is rather unlikely. However, numerical
simulations will be needed to further investigate this issue.

As there are still many open questions related to the formation of the
SGRB prompt emission in BNS mergers, our model is designed to be
general in the sense
that it does not make assumptions on the generation of the prompt
emission. In either scenario, it provides detailed predictions for the
intrinsic afterglow
emission emerging from the post-merger system that can be compared to
observations (see Paper II). In the time-reversal
scenario, it additionally predicts the EM emission prior to the
SGRB itself (see Paper II). If such emission is found, it would
represent strong evidence in favor of the time-reversal scenario (see
Section~7 of Paper II for a more detailed discussion and implications
for joint GW and EM observations).

\paragraph{Comparison to earlier work} 
Other authors, e.g., \citet{Yu2013} and \citet{Metzger2014b}, have
previously considered radiation from a similar physical
setup.\footnote{\citet{Gao2013} have also considered a similar
  scenario, but focus on computing afterglow radiation from the
  interaction of ejected mass with the ambient medium.} During the
time of writing also \citet{Gao2015} have
investigated EM emission in a similar context. All of the
aforementioned authors considered a setup that is qualitatively
similar to the phenomenology in Phase III of our model: a remnant NS
(previously born in a BNS merger), surrounded by a PWN that is confined by a shell
of matter. They present different models to evolve such a setup and to
predict thermal and non-thermal radiation emerging from such a
system. Based on a simple dynamical model, \citet{Yu2013} find thermal
emission peaking around $\sim\!10^4\!-\!10^5\,\text{s}$ after merger with
luminosities of $\sim\!10^{44}\!-\!10^{45}\,\text{erg}\,\text{s}^{-1}$, which
they term ``merger-nova''. Building on this model, \citet{Gao2015}
argue that such a merger-nova is consistent with the late-time
rebrightening observed at optical and X-ray wavelengths in GRB
080503. \citet{Metzger2014b} consider a more detailed physical
model, in particular, regarding the PWN physics, but without
relativistic dynamics. Furthermore, they implement a detailed
formalism to compute the ejecta opacities, the degree of ionization,
and the resulting albedo in a self-consistent way. As a result of a
more detailed physical description of the PWN and its interaction with
the ejecta layer, \citet{Metzger2014b} obtain a dimmer thermal optical/UV
transient peaking at a luminosity of
$\sim\!10^{43}\!-\!10^{44}\,\text{erg}\,\text{s}^{-1}$ on a timescale
of $\sim\!10^4\!-\!10^5\,\text{s}$ after merger and argue that the late-time
X-ray excess of GRB 130603B \citep{Fong2014} as well as the late-time optical
rebrightening and X-ray emission of GRB 080503 is consistent with
their model.

In contrast to previous work, we start the evolution shortly after the
BNS merger and introduce a first baryonic wind phase that we expect to
be generic of BNS mergers leading to the formation of a long-lived remnant
NS. Furthermore, we introduce the pulsar ignition and pulsar wind
shock phase (qualitatively similar to the initial phase of
\citealt{Metzger2014a}). As the pulsar wind shock is
propagating at relativistic speeds in our setup, we implement a
relativistic scheme to describe the propagation across the ejecta and
a detailed description of the energy transfer between the PWN, the
shock heated, and unshocked ejecta during this phase. In Phase III, we
implement a self-consistent model for the radiative processes occurring
in the PWN, including pair creation and annihilation, Compton
scattering, Thomson scattering, and synchrotron cooling. Moreover, we
employ a relativistic description of the dynamics in terms
of a Milne-universe model and develop a reconstruction of the observer
lightcurve and spectra taking into account the combined effects of
relativistic beaming, the relativistic Doppler effect, and the
time-of-flight effect. As a result of baryon pollution being a general
feature and the level of detail reached here, we expect our model to
be applicable to a much larger class of SGRBs than previously
thought. We explore this possibility in Paper II. In particular, we
point out that previous magnetar models, which are applied to large
classes of SGRB events (e.g.,
\citealt{Rowlinson2013,Gompertz2013,Gompertz2014,Lue2015}) neglect the
effects of surrounding ejecta material on the magnetar emission and
instead assume a direct and instantaneous conversion of spin-down
energy $L_\text{sd}$ into observed X-ray luminosity $L_\text{X}$ by
some unspecified process, $L_\text{sd}\propto L_\text{X}$. This typically
leads to very simple analytical fitting formulae for the X-ray
lightcurves, which is in sharp contrast to the self-consistent
dynamical evolution considered here.

\paragraph{Rayleigh-Taylor instability, kick velocities, and PWN jet}
Our model is one dimensional and highly idealized. As noted by
\citet{Metzger2014b}, at early times the high pressure of the PWN
pushing against the ejecta shell can trigger a Rayleigh-Taylor
instability. This would presumably cause some porosity in the ejecta envelope
thanks to which non-thermal radiation from the PWN
could escape from the system already at much earlier times. Such effects,
however, are difficult to take into account in a one dimensional model
like ours and are neglected here. 

Another deviation from spherical symmetry not accounted for by our
one-dimensional model could arise from the possible presence of a kick velocity of
the newly-formed NS. A large kick velocity could qualitatively alter
the evolutionary scenario considered here.

Furthermore, based on axisymmetric magnetohydrodynamic simulations,
\citet{Bucciantini2012} show that for sufficiently high spin-down
energies, the built-up of strong toroidal magnetic field in the PWN
interior can drive a bipolar jet through the ejecta shell and might
even disrupt it entirely. Such jet breakouts were employed by
\citet{Bucciantini2012} to model SGRBs with extended
emission. However, recent three-dimensional simulations indicate that
such jet breakouts are an intrinsic effect of two-dimensional simulations, and
that the kink instability in three dimensions destroys the polar jets
and dissolves them into the PWN \citep{Porth2014}. Therefore, we
exclude such catastrophic events for our model.

\paragraph{Future improvements}
Our model already reflects a certain degree of detail and
sophistication, but many aspects can be
improved in future work. A more accurate description of radiative
transfer through the ejecta shell and the PWN is required to predict
more accurate lightcurves and spectra, which would be desirable for a
more detailed comparison with observational data. Furthermore, the
present implementation of a self-consistent modeling of the radiative
processes occurring in the PWN requires the assumption of
quasi-stationarity (as far as those processes are concerned). In the
future, it would be desirable to develop a time-dependent formalism,
which would then allow us to include a self-consistent computation of
a time and frequency-dependent albedo of the ejecta material (similar
to \citealt{Metzger2014b}). Moreover, such a time-dependent formalism
would more accurately
describe further acceleration of the ejecta shell in Phase III, it
would allow us to include the associated $p\rmn{d}V$ work done by the
nebula, and it would provide a more accurate model of the PWN emission
during the transient phase following the collapse of the remnant
NS. Finally, we note that with different initial data and
some modifications, our model could also be employed to investigate
EM afterglows of long GRBs.

%\bigskip
\acknowledgements
We thank B.~F.~Schutz, W.~Kastaun, and B.~D.~Metzger for valuable
discussions. R.~C.~acknowledges support from MIUR FIR Grant No.~RBFR13QJYF. 
 
%%%%%%%%%%%%%%%%%%%%%%%%%%%%%%%%%%%%%%%%%

\bibliographystyle{apj}
\bibliography{aeireferences}

%%%%%%%%%%%%%%%%%%%%%%%%%%%%%%%%%%%%%%%%%

\appendix

\section{Relativistic Rankine-Hugoniot conditions}
\label{app:shock}
This appendix is devoted to deriving Equations~\eqref{eq:v_sh}, \eqref{eq:v_R},
\eqref{eq:rho_L}, and \eqref{eq:del_e} from the special-relativistic
Rankine-Hugoniot conditions (e.g., \citealt{Taub1948,Marti1994}),
thereby also rederiving some of the expressions in \citet{Taub1948}
using our notation and conventions. The Rankine-Hugoniot
conditions were obtained in the special-relativistic case by
\citet{Taub1948} and relate the fluid properties across a shock wave
imposing continuity of the mass and energy-momentum flux:
\begin{align}
  [\rho u^\mu] n_\mu  &= 0, \label{eq:RanHug_1or}\\
  [T^{\mu\nu}] n_\mu &= 0. \label{eq:RanHug_2or}
\end{align}
Here, $[f] = f_\text{R}-f_\text{L}$ relates the values of a quantity
$f$ on one side and the other side of the discontinuity surface with normal vector $n^\mu$.
We choose a frame in which the shock is at rest and adopt Cartesian
coordinates such that $n^\mu = (0,1,0,0)$. Furthermore, we assume an
ideal fluid with rest-mass density $\rho$, pressure
$p$, specific internal energy $\epsilon$, four-velocity
$u^\mu=(\gamma,\gamma u,0,0)$, where $\gamma = (1-u^2)^{-1/2}$, and energy-momentum tensor
$T^{\mu\nu}=\rho(c^2 + \epsilon + p/\rho)u^\mu u^\nu +
p\eta^{\mu\nu}$, where $\eta^{\mu\nu} = \text{diag}(-1,1,1,1)$ is the
Minkowski metric. Finally, we assume an ideal gas equation of state,
$p = (\Gamma -1)\rho\epsilon$, where $\Gamma$ denotes the adiabatic
index. Under these assumptions, the Rankine-Hugoniot conditions
(Equations~\eqref{eq:RanHug_1or} and \eqref{eq:RanHug_2or}) are
written as (cf.~also \citealt{Taub1948})
\begin{eqnarray}
  \frac{\rho_\text{L}u_\text{L}}{\sqrt{1-u_\text{L}^2}} &=&
  \frac{\rho_\text{L}u_\text{R}}{\sqrt{1-u_\text{L}^2}}\equiv m, \label{eq:RanHug_1}\\
  \rho_\text{L} \left(c^2 + \frac{\Gamma}{\Gamma
  -1}\frac{p_\text{L}}{\rho_\text{L}}\right)\frac{u_\text{L}}{1-u_\text{L}^2}
  &=& \rho_\text{R} \left(c^2 + \frac{\Gamma}{\Gamma
  -1}\frac{p_\text{R}}{\rho_\text{R}}\right)\frac{u_\text{R}}{1-u_\text{R}^2},
  \label{eq:RanHug_2}\\
  p_\text{R} - p_\text{L} &=& m^2c^2\left[ \frac{1}{\rho_\text{L}} \left(1 + \frac{\Gamma}{\Gamma
  -1}\frac{p_\text{L}}{\rho_\text{L}c^2}\right) - \frac{1}{\rho_\text{R}} \left(1 + \frac{\Gamma}{\Gamma
  -1}\frac{p_\text{R}}{\rho_\text{R}c^2}\right)   \right]. \label{eq:RanHug_3}
\end{eqnarray}
We note that for a mixture of radiation and ideal gas with $\Gamma =
4/3$ (i.e., the ejecta matter considered in our model), the
Rankine-Hugoniot conditions take exactly the same form, with $p_\text{L}$
and $p_\text{R}$ being replaced by the respective sums of the radiation and fluid pressures.

Squaring Equation~\eqref{eq:RanHug_2} and subtracting
Equation~\eqref{eq:RanHug_3} multiplied by $m^2[\ldots + \ldots]$ yields (where $[\ldots + \ldots]$ denotes the square bracket on the right hand
side of Equation~\eqref{eq:RanHug_3} with a $+$ sign separating
the two terms; cf.~also Equation~(7.8) in \citealt{Taub1948}):
\begin{equation}
  c^2\left[\left( 1 + \frac{\Gamma}{\Gamma -
        1}\frac{p_\text{L}}{\rho_\text{L}c^2}\right)^2 - \left( 1 + \frac{\Gamma}{\Gamma -
        1}\frac{p_\text{R}}{\rho_\text{R}c^2}\right)^2 \right] =
  (p_\text{L} - p_\text{R}) \left[ \frac{1}{\rho_\text{L}}\left( 1 + \frac{\Gamma}{\Gamma -
        1}\frac{p_\text{L}}{\rho_\text{L}c^2}\right) - \frac{1}{\rho_\text{R}}\left( 1 + \frac{\Gamma}{\Gamma -
        1}\frac{p_\text{R}}{\rho_\text{R}c^2}\right) \right]. \label{eq:rho_L_intermediate}
\end{equation}
This equation can be read as a quadratic equation for $\rho_\text{L}$
in terms of $\rho_\text{R}$, $p_\text{L}$, and $p_\text{R}$. In the
following, we assume a strong shock, i.e., 
\begin{equation}
  p_\text{L}\gg p_\text{R}, \label{eq:strong_shock}
\end{equation}
and a non-relativistic fluid ahead of the shock, i.e.,
\begin{equation}
  p_\text{R}/(\rho_\text{R}c^2) \ll 1. \label{eq:non-rel_fluid}
\end{equation} 
These assumptions are very well satisfied for the pulsar wind shock in Phase II (see
Section~\ref{sec:PhaseII_ejecta}), given the non-relativistic
unshocked ejecta ahead of the shock front and the very high nebula
pressure $p_\text{n}$ (Equation~\ref{eq:p_n}) that equals the pressure
$p_\text{L}$ of the shocked ejecta according to the pressure
balance condition \eqref{eq:press_equil}.
With these assumptions, we obtain Equation~\eqref{eq:rho_L} from Equation~\eqref{eq:rho_L_intermediate}:
\begin{equation}
 \rho_\text{L} = \frac{\Gamma +
      1}{\Gamma - 1}\rho_\text{R}   \frac{1}{2} \left\{ 1 +
      \left[1+4\frac{p_\text{L}}{\rho_\text{R}c^2}\frac{\Gamma}{(\Gamma
          +1)^2}
      \right]^{\frac{1}{2}} \right\}. \label{eq:rho_L_app}
\end{equation}

With the help of Equation~\eqref{eq:RanHug_1}, Equation~\eqref{eq:RanHug_3} can be written as
\begin{equation}
  \frac{u_\text{R}^2}{1-u_\text{R}^2} =
  \frac{\frac{p_\text{L}}{\rho_\text{R}c^2} \left( 1-
      \frac{p_\text{R}}{p_\text{L}} \right)}{1 -
    \frac{\rho_\text{R}}{\rho_\text{L}} + \frac{\Gamma}{\Gamma -
      1}\left( \frac{p_\text{R}}{\rho_\text{R}c^2} -
      \frac{\rho_\text{R}}{\rho_\text{L}}\frac{p_\text{L}}{\rho_\text{L}c^2}
    \right)} = \frac{\rho_\text{R}^2}{\rho_\text{L}^2}\frac{u_\text{L}^2}{1-u_\text{L}^2},
\end{equation}
from which using Equations~\eqref{eq:strong_shock} and
\eqref{eq:non-rel_fluid} we obtain:
\begin{eqnarray}
  |u_\text{R}| &=& \left\{ \frac{p_\text{L}}{\rho_\text{R}c^2} \left[ 1 -
      \frac{\rho_\text{R}}{\rho_\text{L}}\left( 1 + \frac{\Gamma}{\Gamma -
      1}\frac{p_\text{L}}{\rho_\text{L}c^2}\right) + \frac{p_\text{L}}{\rho_\text{R}c^2}
    \right]^{-1}\right\}^{\frac{1}{2}}, \\
     |u_\text{L}| &=& \left\{ \frac{p_\text{L}}{\rho_\text{L}c^2} \left[ \frac{\rho_\text{L}}{\rho_\text{R}} - 1 - \frac{1}{\Gamma -
      1}\frac{p_\text{L}}{\rho_\text{L}c^2} \right]^{-1}\right\}^{\frac{1}{2}}.
\end{eqnarray}
In our setup, $u_\text{R}<0$, i.e., the fluid ahead of the shock is moving toward the shock
in the rest frame of the shock front. As seen from the frame comoving with the fluid ahead of the shock, the
shock velocity $v_\text{sh,R}$ is then given by
$v_\text{sh,R}=|u_\text{R}|c$ (Equation~\eqref{eq:v_R}). In order to
obtain the shock speed in the lab frame, we apply a
Lorentz transformation and arrive at
Equation~\eqref{eq:v_sh}. Accordingly, we have $u_\text{L}> 0$ and the
velocity of the fluid behind the shock as seen from the frame comoving
with the shock front is given by $v_\text{L}=u_\text{L}c$ (this
quantity is needed in Equation~\eqref{eq:v_L_Lab}).

For the jump in specific internal energy across the shock, we have
\begin{equation}
  \frac{\Delta\epsilon}{c^2} = \frac{\epsilon_\text{L} -
    \epsilon_\text{R}}{c^2} = \frac{1}{\Gamma -1}\left(
    \frac{p_\text{L}}{\rho_\text{L}c^2} - \frac{p_\text{R}}{\rho_\text{R}c^2}\right).
\end{equation}
Using Equations~\eqref{eq:strong_shock}, \eqref{eq:non-rel_fluid}, and
\eqref{eq:rho_L_app} this results in
\begin{equation}
  \Delta \epsilon = \frac{1}{\Gamma + 1}
  \frac{p_\text{L}}{\rho_\text{R}} \frac{2}{1  + \left[ 1 +
     \frac{4\Gamma}{(\Gamma + 1)^2}\frac{p_\text{L}}{\rho_\text{R}c^2} \right]^{\frac{1}{2}}},
\end{equation}
which is the desired expression in Equation~\eqref{eq:del_e}.

\section{Transformation between comoving and lab frame}
\label{app:frames}
This appendix defines the lab and comoving frames used
in Phase III of our evolution model to describe the expansion of the
ejecta shell (cf.~Section~\ref{sec:ejecta_p3}) and it provides transformations for
some quantities required by our model. The Milne universe
metric has recently been employed to describe an expanding homogeneous GRB
fireball \citep{Li2007,Li2013}. We follow
this approach and apply it to the expansion of a thin ejecta
shell. Some of the expressions that we derive here are based on
expressions discussed by \citet{Li2007} and \citet{Li2013}, some of which we shall
rederive here for completeness.

Let $X^\mu = (ct, r,
\theta,\phi)$ denote the rest frame of the NS (or the ``lab frame'') with spherical coordinates and
Minkowski metric $g_{\mu\nu}=
\text{diag}(-1,1,r^2,r^2\sin^2\theta)$. Now consider the following
coordinate transformation $X^\mu \mapsto X'^\mu= (c\eta, \xi,
\theta,\phi)$, defined by
\begin{equation}
  t = \eta \cosh\xi,\mskip50mu r = c\eta\sinh\xi, \label{eq:t_r}
\end{equation}
which transforms the Minkowski metric into $g'_{\mu\nu} =
\text{diag}(-1,a^2(\eta),a^2(\eta)\sinh^2\xi,a^2(\eta)\sinh^2\xi\sin^2\theta)$,
with scale factor $a(\eta)=c\eta$, which is known as the metric of the
Milne universe (cf., e.g., Equation~(16.15) of
\citealt{Rindler2006}). For a test particle with constant radial
velocity $v$, we have $r = c\beta t$, where $\beta = v/c$. Using
Equation~\eqref{eq:t_r} this yields
\begin{equation}
  \xi = \arctanh\beta,\mskip50mu \eta = \gamma^{-1} t, \label{eq:eta_xi}
\end{equation}
where $\gamma = 1/\sqrt{1-\beta^2}$ is the Lorentz factor. Hence, $\eta$ is the proper time
of the particle and $\xi$ its
rapidity. At any time $t$, the ejecta shell can be assigned a Milne universe with
velocity $v=v_\text{ej}$ at $r=R_\text{ej}$ by appropriately rescaling
the lab frame time coordinate. For the ejecta shell thickness is
typically sufficiently small, i.e., $\Delta_\text{ej}\ll R_\text{ej}$,
at any given time we can describe the ejecta as a thin shell in this
Milne universe, which is what we call the comoving frame at time $t$.

First, we determine the three-acceleration $a$ in the lab frame in
terms of the corresponding quantity $\alpha$ in the comoving frame
(cf.~Equation~\eqref{eq:dvej_dt}). The four-acceleration of a particle in the
comoving frame is given by $A'^\mu = \rmn{d}^2X'^\mu/\rmn{d}\eta^2= (\gamma'^4\alpha v'/c, \gamma'^2(
\gamma'^2\alpha v' v'^i/c^2 + \alpha^i))$, where $v'^i =
(v'^\xi,v'^\theta,v'^\phi)$ is the three-velocity, $\alpha^i =
\rmn{d}v'^i/\rmn{d}\eta$, $v'=\sqrt{g'_{ij}v'^i v'^j}$, and $\alpha =
\rmn{d}v'/\rmn{d}\eta$.
For the fluid in the ejecta shell, $v'=v'^i =\alpha^\theta = \alpha^\phi = 0$. Hence,
\begin{eqnarray}
  A'^\mu &=& (0,\alpha^\xi,0,0), \label{eq:Apr_mu}\\
  v' &=& c\eta v'^\xi,\\
  \alpha &=& c\eta \alpha^\xi \label{eq:alpha_alphaxi}.
\end{eqnarray}
 Analogously, in the lab frame the four-acceleration is given by
 $A^\mu = (\gamma^4av/c, \gamma^2(\gamma^2 a vv^i/c^2 + a^i))$, where $v^i =
(v^r,v^\theta,v^\phi)$ is the three-velocity, $a^i =
\rmn{d}v^i/\rmn{d} t$, $v=\sqrt{g_{ij}v^i v^j}$, and $a =
\rmn{d}v/\rmn{d} t$. Using $v^\theta = v^\phi = a^\theta = a^\phi =
0$, we obtain
\begin{equation}
  A^\mu = (\gamma^4 a v/c, \gamma^4a,0,0).
\end{equation}
Moreover, using
Equations~\eqref{eq:Apr_mu},~\eqref{eq:alpha_alphaxi},~\eqref{eq:t_r},
and \eqref{eq:eta_xi}, we have $A^r = (\partial X^r/\partial X'^\xi)
A'^\xi = (\partial r/\partial\xi)\alpha^\xi =\gamma\alpha$ and thus we
obtain the desired expression (cf.~Equation~\eqref{eq:dvej_dt})
\begin{equation}
  a = \ddt{v} = \gamma^{-3}\alpha.
\end{equation}

Second, we determine the transformation of area, volume, and internal
energy. Henceforth, ``com'' refers to the comoving frame. The surface area of spheres of
radius $r$ and corresponding rapidity $\xi$ centered around
$r=\xi = 0$ are
invariant under the transformation between lab and comoving frames:
\begin{equation}
  S_\xi = c^2\eta^2\sinh^2\xi\int
  \sin\theta\rmn{d}\theta\rmn{d}\phi = 4\pi c^2\eta^2\sinh^2\xi = 4\pi
  r^2 = S_r. \label{eq:SxiSr}
\end{equation}
The corresponding spherical volumes transform as follows:
\begin{subequations} 
\label{eq:VcomV}
\begin{eqnarray}
 V_\text{com}(\xi) &=&
  c^3\eta^3\int\sin\theta\,\rmn{d}\theta\rmn{d}\phi\int_0^\xi\sinh^2\xi\,\rmn{d}\xi \\
  &=& \pi c^3\eta^3\int_0^{2\xi}(\cosh z -1) \,\rmn{d}z = \pi
      c^3\eta^3 (\sinh 2\xi - 2\xi) \label{eq:VcomVident}\\
  &=& \frac{4}{3}\pi r^3 \zeta = \zeta V(r),
\end{eqnarray}
\end{subequations}
where
\begin{equation}
  \zeta = \frac{3}{4}\frac{\sinh2\xi - 2\xi}{\sinh^3\xi} =
  \frac{3}{2}\frac{\gamma^2\beta - \arctanh\beta}{\gamma^3\beta^3}. \label{eq:zeta}
\end{equation}
Note that our definition of $\zeta$ differs from $\zeta(\beta)$
defined in \citet{Li2013} by a factor $\gamma^{-1} =
(\cosh\xi)^{-1}$. For the volume of the ejecta shell we thus find:
\begin{equation}
  V_\text{ej,com} = V_\text{com}(\xi_\text{ej}) -
  V_\text{com}(\xi_\text{n}) = \zeta V(R_\text{ej}) - \zeta
  V(R_\text{n}) = \zeta V_\text{ej}. \label{eq:VcomVshell}
\end{equation}
In order to compute the transformation of total energy of a fluid, let
$T_{\mu\nu} = (e + p)u^\mu u^\nu/c^2 + p g_{\mu\nu}$ denote the part
of the energy momentum tensor excluding kinetic and rest-mass energy,
with $e$ being the internal energy density, $p$ being the pressure,
$u^\mu$ being the four-velocity, and $g_{\mu\nu}$ being the metric of
flat spacetime. The total energy density as seen by
a normal (Eulerian) observer with four-velocity $n^\mu$ perpendicular
to the hypersurfaces of constant time $t$ in the Minkowski spacetime
is given by $e_\text{tot}=T_{\mu\nu}n^\mu n^\nu = \gamma^2 e +
(\gamma^2 - 1)p$. For a mixture
of ideal gas and radiation, we can rewrite $e_\text{tot}$ as
\begin{equation}
  e_\text{tot} = \frac{1}{3}(4\gamma^2 -1)e_\text{r} + [\gamma^2 +
  (\gamma^2-1)(\Gamma - 1)]e_\text{f} = \frac{1}{3}(4\gamma^2 -1)e, \label{eq:e_tot}
\end{equation}
where $e_\text{r}$ and $e_\text{f}$ refer to the internal energy
density of the radiation field and the gas, respectively. In the
second equality in Equation~\eqref{eq:e_tot} we have assumed an ideal
gas with adiabatic index $\Gamma = 4/3$ (as for the ejecta material we
consider). Consequently, the total energy of the fluid (apart from kinetic end
rest-mass energy) in a spherical volume as measured in the comoving
frame is given by (cf.~Equation~\eqref{eq:VcomVident})
\begin{equation}
  E_\text{com} = e_\text{tot,com} V_\text{com} = \pi
      c^3\eta^3 (\sinh 2\xi - 2\xi) e,
\end{equation}
whereas in the lab frame it is given by $E = 4\pi \int e_\text{tot}
r^2\,\rmn{d}r$, which after some manipulations can be rewritten as
\begin{equation}
  E = \frac{4}{3} \pi \eta^3 c^3 \sinh^3\xi\, e.
\end{equation}
Hence (cf.~Equations~\eqref{eq:zeta} and \eqref{eq:VcomV}),
\begin{equation}
  \frac{E_\text{com}}{E} = \zeta = \frac{V_\text{com}}{V}. \label{eq:EcomE}
\end{equation}
Thanks to Equation~\eqref{eq:VcomVshell}, this transformation also holds for
shell volumes.

Finally, we discuss how luminosities transform between the lab and the
comoving frame. Let $L$ denote the luminosity of thermal radiation
originating from the surface of the ejecta shell. Then using
Equation~\eqref{eq:SxiSr}, $L_\text{com} = S_{\xi_\text{ej}} \sigma
T_\text{eff,com}^4 = 4 \pi R_\text{ej}^2 \sigma T_\text{eff,com}^4$,
where $T_\text{eff,com} \simeq (16/3) T_\text{com}^4/(\Delta\tau_\text{com} + 1)$
is the effective temperature at optical depth $\Delta\tau_\text{com}$
in the comoving frame. The energy loss $\Delta E_\text{com}$
associated with this luminosity as measured in the comoving frame during a time
$\Delta \eta$ is given by $\Delta E_\text{com}=L_\text{com}\Delta\eta
= L_\text{com} \gamma^{-1}\Delta t$, where we have used
Equation~\eqref{eq:eta_xi} and $\gamma$ denotes the Lorentz factor of
the ejecta shell. The corresponding energy loss as measured in the lab
frame is then obtained as (cf.~Equation~\eqref{eq:EcomE}) $L\Delta t = \Delta E = \zeta^{-1}\Delta
E_\text{com} = \zeta^{-1} \gamma^{-1}L_\text{com}\Delta t $. Hence, we
arrive at
\begin{equation}
  L = \zeta^{-1} \gamma^{-1}L_\text{com},
\end{equation}
which is the desired result to motivate Equations~\eqref{eq:Lrad_phaseIII} and \eqref{eq:Lradin_phaseIII}.

\section{Synchrotron emission}
\label{app:synchrotron}
This appendix derives the relevant expressions needed to include
effects of synchrotron emission into our model of
the PWN in Phase III (Equations~\eqref{eq:gammadot_syn},
\eqref{eq:ndsyn}, and \eqref{eq:tau_syn}). The spontaneously emitted
spectral power at frequency $\nu$ of a single particle of charge $q$,
mass $m$, and Lorentz factor $\gamma$, accelerated by a magnetic field of strength $B$, is given by (\citealt{Rybicki_Lightman2004,Crusius1986})
\begin{equation}
  p(\nu,\theta,\gamma) =
  \frac{\sqrt{3}q^3B}{mc^2}\sin\theta\frac{\nu}{\tilde{\nu}_\text{c}}\int_{\nu/\tilde{\nu}_\text{c}}^\infty K_{\frac{5}{3}}(z)\,\rmn{d}z,
\end{equation}
where $c$ is the speed of light, $\theta$ is the angle between the magnetic field and the
direction of particle motion, $K_{\frac{5}{3}}$ denotes the modified
Bessel function of order $5/3$, and
$\tilde{\nu}_\text{c}=\nu_\text{c}\sin\theta = (3/4\pi)(qB\gamma^2/mc)
\sin\theta$. Assuming an isotropic distribution of particle velocities, we can
average over all possible angles for a given speed $\gamma$ to obtain
the average energy loss rate for a particle (cf.~\citealt{Crusius1986}):
\begin{equation}
  p(\nu,\gamma) = \frac{1}{4\pi}\int_0^{2\pi}\rmn{d}\phi\int_0^{4\pi}
  \rmn{d}\theta\,\sin\theta\, p(\nu,\theta,\gamma) = \frac{\sqrt{3}q^3B}{mc^2}\mathcal{R}(\nu/\nu_\text{c}),
\end{equation}
where
\begin{equation}
  \mathcal{R}(z) = \frac{\pi}{2}z[W_{0,\frac{4}{3}}(z) W_{0,\frac{1}{3}}(z) - W_{\frac{1}{2},\frac{5}{6}}(z) W_{-\frac{1}{2},\frac{5}{6}}(z)].
\end{equation}
Here, $W_{\lambda,\mu}(z)=\text{e}^{-\frac{1}{2}z}z^{\frac{1}{2}+\mu}
U(0.5+\mu-\lambda, 1+2\mu,z)$ is the Whittaker function and $U$
denotes the confluent hypergeometric function of second kind
(\citealt{AbramowitzStegun1972}, Section 13.1). Substituting $\nu$
with the dimensionless energy $x=h\nu/mc^2$, with $h$ being the Planck
constant, one has the relation $p(x,\gamma) = (mc^2/h) p(\nu,\gamma)$; accordingly,
we define $x_\text{c}=h\nu_\text{c}/mc^2$. Therefore we can write
the total cooling rate $\dot{\gamma}$ of a single particle as
\begin{equation}
  \dot{\gamma}(\gamma) = -\frac{1}{mc^2}\int p(x,\gamma)\,\rmn{d}x =
  -\frac{\sqrt{3}q^3B}{hmc^2}\int \mathcal{R}(x/x_\text{c})\,\rmn{d}x. \label{eq:app_gammadot_syn}
\end{equation}
Furthermore, the number density of photons emitted per unit time and
per unit normalized energy $x$ is given by
\begin{equation}
  \dot{n}(x)=\frac{1}{mc^2 x}\int N(\gamma)p(x,\gamma)\,\rmn{d}\gamma = \frac{\sqrt{3}q^3B}{hmc^2}
  \frac{1}{x}\int N(\gamma)
  \mathcal{R}(x/x_\text{c})\,\rmn{d}\gamma, \label{eq:app_ndsyn}
\end{equation}
where $N(\gamma)$ is the particle density per normalized energy
$\gamma$. Finally, assuming a uniform opacity coefficient the optical
depth $\Delta\tau$ to synchrotron self-absorption for a spherical volume of radius
$R$ is given by \citep{Ghisellini1988,Ghisellini1991}
\begin{equation}
  \Delta\tau(\nu) = \frac{R}{8\pi m
    \nu^2}\int\frac{N(\gamma)}{\gamma \alpha}
  \frac{\partial}{\partial t}[\gamma \alpha p(\nu,\gamma)]\, \rmn{d}\gamma, \label{eq:app_tau_syn_nu}
\end{equation}
where $\alpha = \sqrt{\gamma^2 -1}$. Using the information from above,
Equation~\eqref{eq:app_tau_syn_nu} can be rearranged to give
\begin{equation}
  \Delta\tau(x) = \frac{\sqrt{3}q^3h^2 B
    R}{8\pi
    m^4c^6}\frac{1}{x^2}\int
  N(\gamma)\left[\frac{\partial}{\partial\gamma}\mathcal{R}(x/x_\text{c}) +
    f(\gamma)\mathcal{R}(x/x_\text{c}) \right]\,\rmn{d}\gamma, \label{eq:app_tau_syn}
\end{equation}
where $f(\gamma) = (2\gamma^2 -1) / [\gamma(\gamma^2 -1)]$, with
$\lim_{\gamma \to 1}f(\gamma) =2$.

\section{Energy conservation in the nebula}
\label{app:energy_conservation}

This appendix demonstrates that Equations~\eqref{eq:LightmanI}
and \eqref{eq:LightmanII} conserve energy. Integrating
Equation~\eqref{eq:LightmanI} the total energy balance per unit
volume per unit time is given by
\begin{equation}
  \int \dot{n}_\text{esc} x\,\rmn{d}x = \int \dot{n}_0 x\,\rmn{d}x
  +\int \dot{n}_\text{A}x\,\rmn{d}x + \int
  \dot{n}_\text{C}^\text{NT}x\,\rmn{d}x + \int
  \dot{n}_\text{C}^\text{T}x\,\rmn{d}x + \int
  \dot{n}_\text{syn}x\,\rmn{d}x -\frac{c}{R_\text{n}}\int
  n\Delta\tau_\text{C}^\text{NT}x\,\rmn{d}x -\frac{c}{R_\text{n}}\int
  n\Delta\tau_{\gamma\gamma}x\,\rmn{d}x. \label{eq:ebalanceI}
\end{equation}
From Equation~\eqref{eq:N_gam} we have
\begin{eqnarray}
  -\int_1^{\gamma_\text{max}}
  \dot{\gamma}_\text{C,syn}(\gamma)N(\gamma)\,\rmn{d}\gamma &=&
  \frac{\sqrt{3}e^3B}{hmc^2}\int_1^{\gamma_\text{max}}\rmn{d}\gamma\int
  \rmn{d}x\,N(\gamma)\mathcal{R}(x/x_\text{c}) -
  \int_1^{\gamma_\text{max}}\dot{\gamma}_\text{C}N(\gamma)\,\rmn{d}\gamma
  \\
 &=& \int \dot{n}_\text{syn}x\,\rmn{d}x -\int_1^{\gamma_\text{max}}\dot{\gamma}_\text{C}N(\gamma)\,\rmn{d}\gamma,
\end{eqnarray}
and hence
\begin{equation}
   \int \dot{n}_\text{syn}x\,\rmn{d}x = \int_1^{\gamma_\text{max}}[Q(\gamma)
  + P(\gamma)](\gamma -1)\,\rmn{d}\gamma
  +\int_1^{\gamma_\text{max}}\dot{\gamma}_\text{C}N(\gamma)\,\rmn{d}\gamma. \label{eq:ndsyn_identitiy}
\end{equation}
Plugging this into Equation~\eqref{eq:ebalanceI} and noting that the
second term on the right hand side of
Equation~\eqref{eq:ndsyn_identitiy} cancels the third and sixth term
on the right hand side of Equation~\eqref{eq:ebalanceI} (see LZ87,
Appendix~A), we can proceed as in Appendix~A of LZ87 to conclude that
\begin{equation}
  \int \dot{n}_\text{esc} x\,\rmn{d}x = \int \dot{n}_0 x\,\rmn{d}x +
  \int_1^{\gamma_\text{max}} Q(\gamma)(\gamma -1)\,\rmn{d}\gamma.
\end{equation}
Therefore, the total injected power equals the total
power output, which shows that energy is conserved in steady state.

%%%%%%%%%%%%%%%%%%%%%%%%%%%%%%%%%%%%%%%%%
\end{document}